%% file: main.tex
\documentclass[12pt]{iopart}

\usepackage{iopams} 
\usepackage{braket}
\usepackage[dvipsnames]{xcolor}
\usepackage{bm}
\usepackage{dirtytalk}
\usepackage{hyperref}
\usepackage{graphicx}
\usepackage{comment}
\usepackage{array}
\usepackage{ragged2e}
\usepackage{etoolbox}
\usepackage{relsize}
\usepackage{float}

\makeatletter
\newrobustcmd{\fixappendix}{%
  \patchcmd{\l@section}{1.5em}{7em}{}{}%
  \patchcmd{\l@subsection}{2.3em}{7em}{}{}%
}
\makeatother

\newcommand{\of}[1]{\textcolor{Mulberry}{\textbf{(OF: #1)}}}

\newcommand{\change}[1]{#1}

\newcommand{\eg}{e.g$.$ }

\newcommand{\ie}{i.e$.$ }
\newcommand{\cf}{cf$.$ }

\newcommand{\wrt}{wrt$.$ }
\newcommand{\dd}{\ensuremath{\mathrm{d}}}

\newcommand{\secref}[1]{Section~\ref{sec:#1}}

\newcommand{\figref}[1]{Figure~\ref{fi:#1}}
\newcommand{\eqnref}[1]{Equation~\ref{eq:#1}}

\begin{document}

\title[Holographic phenomenology via overlapping degrees of freedom]{Holographic phenomenology via overlapping degrees of freedom}

\author[]{\parbox{\linewidth}{Oliver Friedrich$^{1,2\, *}$, ChunJun Cao$^{3,4,5}$, Sean M.\ Carroll$^{6,7}$, Gong Cheng$^{5,8}$, Ashmeet Singh$^9$}}

\address{$^1$ University Observatory, Faculty of Physics, Ludwig-Maximilians-Universität, Scheinerstr.\ 1, 81677 Munich, GER\textbackslash EU}
\address{$^2$ Excellence Cluster ORIGINS, Boltzmannstr.\ 2, 85748 Garching, GER\textbackslash EU}
\address{$^3$ Joint center for Quantum information and Computer science, University of Maryland, College Park, MD 20740, USA}
\address{$^4$ Institute for Quantum Information and Matter, California Institute of Technology, Pasadena, CA 91125, USA}
\address{$^5$ Department of Physics, Virginia Tech, Blacksburg, VA 24061, USA}
\address{$^6$ Departments of Physics \& Astronomy and Philosophy, Johns Hopkins University, Baltimore, MD 21218, USA}
\address{$^7$ Santa Fe Institute, Santa Fe, NM 87501, USA}
\address{$^8$ Maryland Center for Fundamental Physics, University of Maryland, College Park, MD 20740, USA}
\address{$^9$ Dept.\ of Physics, Whitman College, 345 Boyer Ave, Walla Walla, WA 99362, USA}
\ead{oliver.friedrich@physik.lmu.de, cjcao@vt.edu, seancarroll@gmail.com, cghope@terpmail.umd.edu, ashmeet@whitman.edu}

\begin{abstract}
The holographic principle suggests that regions of space contain fewer physical degrees of freedom than would be implied by conventional quantum field theory.
Meanwhile, in Hilbert spaces of large dimension $2^n$, it is possible to define 
$N \gg n$ Pauli algebras that are nearly anti-commuting (but not quite) and which can be thought of as ``overlapping degrees of freedom".
We propose to model the phenomenology of holographic theories by allowing field-theory modes to be overlapping, and derive potential observational consequences.
In particular, we build a Fermionic quantum field whose effective degrees of freedom approximately obey area scaling and satisfy a cosmic Bekenstein bound, and compare predictions of that model to cosmic neutrino observations. Our implementation of holography implies a finite lifetime of plane waves, which depends on the overall UV cutoff of the theory. To allow for neutrino flux from blazar TXS 0506+056 to be observable, our model needs to have a cutoff $\Lambda_{\mathrm{UV}} \lesssim 500\, \Lambda_{\mathrm{LHC}}\,$. This is broadly consistent with current bounds on the energy spectrum of cosmic neutrinos from IceCube, but high energy neutrinos are a potential challenge for our model of holography. We motivate our construction via quantum mereology, \ie using the idea that EFT degrees of freedom should emerge from an abstract theory of quantum gravity by finding quasi-classical Hilbert space decompositions. We also discuss how to extend the framework to Bosons. Finally, using results from random matrix theory we derive an analytical understanding of the energy spectrum of our theory.\ \ The numerical tools used in this work are publicly available within the \verb|GPUniverse| package, \url{https://github.com/OliverFHD/GPUniverse}~.
\end{abstract}
\hspace{2.5cm} {\small{${}^*$ corresponding}}

{\small\tableofcontents}


\markboth{Holographic phenomenology via overlapping degrees of freedom}{}

\section{Introduction \& summary of results}
\label{sec:intro}

\noindent The holographic principle has emerged as one of the most prominent conjectured properties of an eventual theory of quantum gravity \cite{'tHooft:1993gx,Susskind1995, Bousso2002, Pesci2022}. Its most precise formulation has been given by \cite{Bousso1999b} (generalising results of \cite{Fischler1998}), and it states that the entropy $S$ that can be accumulated on a light-sheet\footnote{In a $3+1$ dimensional spacetime a light-sheet emanating from a space-like surface $\mathcal{B}$ has been defined by \cite{Bousso1999b} as a 3 dimensional null hypersurface generated by a family of light-rays that are orthogonal to $\mathcal{B}$ and whose expansion $\theta$ is $\geq 0$ everywhere (i.e.\ the light-sheet stops where the expansion $\theta$ becomes negative).} emanating from a space-like surface $\mathcal{B}$ is bounded by a quarter of the area $|\mathcal{B}|$ of that surface (measured in units of the Planck area). In the weak gravity regime and for spherically symmetric, space-like volumes $\mathcal{R}$, a simpler version of the holographic principle states that the maximum entropy that can be localised within $\mathcal{R}$ equals the boundary area $|\delta\mathcal{R}|$ of that region divided by four times the Planck area,
\begin{equation}
\label{eq:Bekenstein}
    S(\mathcal{R}) \leq \frac{|\delta\mathcal{R}|}{4\ell_{\mathrm{P}}^2}\ .
\end{equation}
Such a bound is incompatible with standard quantum field theory, because the local degrees of freedom of the latter scale extensively with volume and not with area\footnote{In particular, in a cosmological context $S(\mathcal{R})$ would be the maximum entropy that can be localised on a light-sheet inside the cosmic horizon, which is strictly different from the cross-horizon entanglement entropy (which in fact does scale like area, \cf \cite{Srednicki1993, RT2006, HRT2007, QES2015}; the fact that those two different quantities are of the same order of magnitude remains to be explained, but is not addressed here).} \cite{Cohen1999, Thomas2000, Bousso2002, Yurtsever2003, Pesci2022}. The above bound also seems to imply that the Hilbert space of quantum gravity is locally finite \cite{Bao2017}, which calls the validity of bosonic quantum field theory into question even further \cite{Cao2019_essay, Friedrich2022}. Even within an AdS-framework, the emergence of sub-AdS locality remains an open problem where a large separation of the string scale and the AdS scale is needed~\cite{Heemskerk:2009pn,Belin_2017}. 

An area scaling of entropy requires the degrees of freedom of quantum gravity to be de-localised. But how is this reconciled with the fact that local degrees of freedom seem to be a ubiquitous part of nature? So far, the most concrete example of a mapping between local, bulk degrees of freedom and holographic, boundary degrees of freedom is given by the AdS/CFT correspondence \cite{Maldacena1999, Witten1998} which connects a theory of quantum gravity on $d+1$ dimensional anti-de~Sitter space to the degrees of freedom of a conformal field theory living on the boundary of that space. Within that correspondence, information present in the bulk theory can indeed be recovered from the boundary theory \cite{Hamilton2006, Jafferis2016, Dong2016}. \change{Concrete efforts to generalise results from AdS/CFT to more realistic spacetimes have for example attempted to establish a de~Sitter space / CFT correspondence. Initially, this was suggested to be a correspondence between quantum gravity on $d+1$ dimensional de~Sitter space and a $d$-dimensional CFT on the future conformal boundary of de~Sitter space (e.g.\ at future infinity), with time being an emergent coordinate - see \cite{Strominger2001} for the original proposal and e.g.\ \cite{Anninos2011, Anninos2017} for concrete implementations as well as \cite{Arkani_Hamed_2001, BANKS_2010} for potential phenomenological consequences. Alternatively, \cite{Kawamoto2023} have recently proposed a version of dS/CFT that is closer in spirit to AdS/CFT in that it proposes a duality between dS quantum gravity and a non-gravitating QFT at the \emph{time-like} boundary of part of de~Sitter space. In particular, that dual theory is supposed to display a (unitary) time evolution and can be viewed as a QFT on [($d$-1)+1]-dimensional de~Sitter space. \cite{Kawamoto2023} have found that such a duality is only possible, if the (emergent) bulk theory is over-counting the degrees of freedom present in the boundary theory\footnote{One way to understand why this problem does not appear in AdS/CFT is to look at discretised versions of anti-de~Sitter space in the form of tensor networks \cite{Pastawski2015, Cao2021}: the hyperbolic tessellation in such networks makes their boundary legs numerous enough to encode all information present in the bulk factors of the networks without the need for overcounting degrees of freedom.}. A potential duality between QFT in 3+1-dimensional Minkowski space and a boundary CFT has been exprored by \cite{deBoer2003}, and the celestial holography program \cite{Pasterski2017a, Pasterski2017b, Pasterski2021} is establishing such a duality for general gravitating, asymptotically flat spacetimes.}




The search for generalisations of AdS/CFT to more realistic spacetimes is one of the most promising routes to finding a holographic description of our own Universe. But it is worthwhile considering complementary test beds to study the consequences of holography. A constructive way to find such test beds is to modify quantum field theory in such a way that it obeys Equation~\ref{eq:Bekenstein} and its ramifications (such as local finite dimensionality). This has e.g.\ been explored by \cite{Yurtsever2003, Aste2005}, who simply removed the modes of a scalar field which upon excitation would form black holes, and more recently by \cite{Cao2019_essay, Friedrich2022}, who replaced the canonically conjugate field variables of a scalar field by finite-dimensional generalised Pauli operators \cite{Singh2018}. In this paper we present a new addition to this program \change{(though we discussion potential connections to the afore mentioned duality-based approaches in \secref{discussion})}: we approximate holography in quantum fields with the help of overlapping degrees of freedom. 

In quantum computing, it is customary to assume that $N$ independent qubits live in $N$ separate tensor factors of a $2^N$ dimensional Hilbert space. However, such independence is only approximate in practice. This is particularly pronounced, for instance, in various experimental setups involving long range interactions such as Rydberg systems where actions performed on one physical qubit influences other nearby qubits. To this end, Chao et. al. \cite{Chao2017} introduced the concept of overlapping qubits, such that Pauli operators $P_i, P_j$ supported on different qubits $i,j$ satisfy 
\begin{equation}
    ||[P_i,P_j]|\psi\rangle||<\epsilon.
\end{equation} 
Although these overlapping qubits are good approximations of independent qubits from the lens of commutation relations, they showed that the Hilbert space dimension associated with such systems can be as small as $poly(N)$. This surprising result follows from a measure concentration argument \cite{JohnsonLindenstrauss1984}  which states that one can identify exponentially many near-orthogonal vectors in a vector space (the Johnson-Lindenstrauss theorem). Related ideas have been applied to areas of computer science \cite{JLcs,review}, compressed sensing\cite{Baranuik,Ward2008}, and also lately in the context of quantum gravity (e.g. \cite{penington2020replica,Almheiri_2020,Marolf_2020,Akers:2021fut,Akers:2022qdl,balasubramanian2023microscopic,Cao:2023gkw}) to decrease the number of degree of freedom of the system. 

Here we want to apply the same idea as a simple model of potential holographic effects on quantum field theories and study its physical consequences. For instance, one can relax the causal locality condition $[O,O']=0$ for some spacelike separated operators $O,O'$ such that $[O,O'] \approx 0$. Although this may appear strange at first sight, \cite{Cao:2023gkw} suggests that similar behaviours are expected for gauge invariant operators coupled to long range interactions such as the electromagnetic or gravitational field \cite{donnelly+giddings2016_1,donnelly+giddings2016_2}, where the operator dressing renders an operator non-local. A similar relaxation for fermions is also possible, e.g. \cite{Nebabu-Qi}, where one considers the anti-commutator instead of the commutator. More generally, violation of microcausality is also expected at the non-perturbative level for quantum gravity \cite{Berglund_2023}.

\subsection{Summary of results}

For the impatient reader we are summarising our main results already here, with details and derivations given in subsequent sections as well as our appendices. We consider a massless, left-handed Weyl fermion which, when constrained to a box of finite side length $L$, can be decomposed into Fourier modes as
\begin{equation}
\label{eq:Weyl_decomposition_discrete_intro}
    \hat\psi(\bm{x}, t) = \sum_{\bm{p}}\frac{1}{(|\bm{p}| L^3)^{\frac{1}{2}}}\left\lbrace \hat c_{\bm{p}}(t)\ u(\bm{p})\ e^{i\bm{p}\bm{x}}+ \hat d_{\bm{p}}(t)^\dagger\ u(\bm{p})\ e^{-i\bm{p}\bm{x}} \right\rbrace\ ,
\end{equation}
where the sum is over all momenta $\bm{p} \in \lbrace\ 2\pi/L\cdot (n_1, n_2, n_3)\ |\ n_i \in \mathbb{Z} \ \rbrace$, and where we define the mode functions $u(\bm{p})$ in more detail in Section~\ref{sec:spinor_details}. The operator $\hat d_{\bm{p}}^\dagger$ can be thought of as creating an anti-spinor of momentum $\bm{p}$ while $\hat c_{\bm{p}}$ is destroying a spinor with momentum $\bm{p}$. In standard QFT these operators would satisfy the anti-commutation relations
\begin{eqnarray}
\label{eq:anti_commutation_intro}
0 = \lbrace \hat c_{\bm{p}}, \hat c_{\bm{q}} \rbrace = \lbrace \hat d_{\bm{p}}, \hat d_{\bm{q}} \rbrace = \lbrace \hat c_{\bm{p}}, \hat d_{\bm{q}} \rbrace = \lbrace \hat c_{\bm{p}}, \hat d_{\bm{q}}^\dagger \rbrace\\
\lbrace \hat c_{\bm{p}}, \hat c_{\bm{q}}^\dagger \rbrace = \lbrace \hat d_{\bm{p}}, \hat d_{\bm{q}}^\dagger \rbrace = \delta_{\bm{p},\bm{q}}\ .
\end{eqnarray}
These relations can only be satisfied if the individual momenta $\bm{p}$ represent independent Hilbert space factors $\mathcal{H}_{\bm{p}}$ (which further split into particles and anti-particles as $\mathcal{H}_{\bm{p}}^c\otimes \mathcal{H}_{\bm{p}}^d$). In particular, the overall Hilbert space in which the field $\hat\psi$ is defined needs to be given by
\begin{eqnarray}
\label{eq:non_overlapping_H_intro}
\mathcal{H} = \bigotimes_{\bm{p}}^{\mathrm{JW}} \mathcal{H}_{\bm{p}}\ ,
\end{eqnarray}
where the \say{JW} above the tensor product symbol indicates that operators of the individual Hilbert spaces have to be embedded into the product space via Jordan-Wigner strings in order to achieve anti-commutativity between different factors (as opposed to commutativity; \cf our \ref{app:JordanWigner}). We will explicitly break this product structure, and describe the field $\hat \psi$ using a Hilbert space that is strictly smaller than indicated by Equation~\ref{eq:non_overlapping_H_intro}. 

\begin{figure}
\centering
  \includegraphics[width=0.78\textwidth]{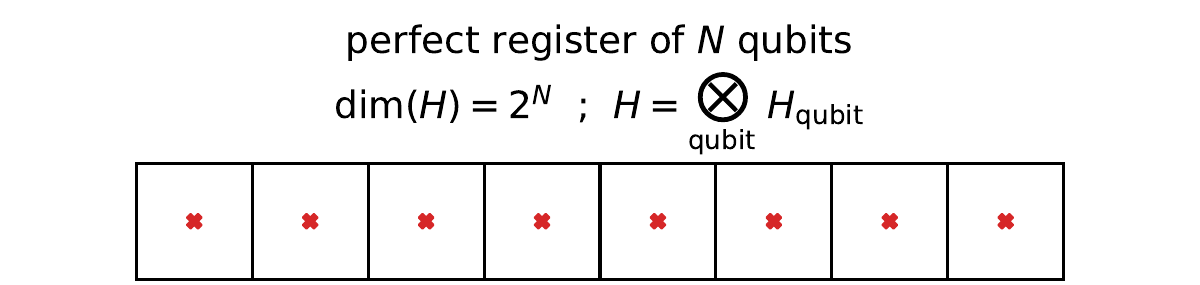}
  \noindent\rule[0.5ex]{0.8\textwidth}{1pt}
  \includegraphics[width=0.78\textwidth]{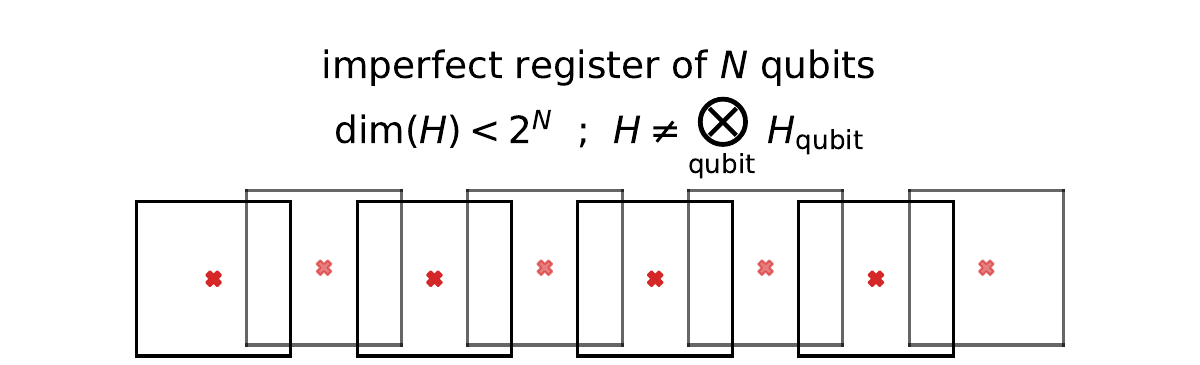}
   \caption{The upper sketch illustrates a perfect register of eight non-overlapping qubits. The Hilbert space of the entire register is the tensor product of the individual qubit Hilbert spaces. The lower sketch illustrates the situation when the total Hilbert space is strictly smaller than such a tensor product. In this case the qubits must be overlapping, by which we mean that the operator algebras defining the different qubits are not (anti-)commuting.}
  \label{fi:qubit_registers}
\end{figure}

We follow the lines of \cite{Chao2017}, who in the context of quantum computing have studied imperfect qubit registers that suffer from qubit overlap (cf.\ the sketch in our Figure~\ref{fi:qubit_registers} for an illustration). They considered $N$ qubits, but assumed them to be embedded in a Hilbert space of dimension $2^n < 2^N$. The Pauli algebra and the raising and lowering operators of an individual qubit can be represented in a Hilbert space of dimension $2^n$ as follows (though see our \secref{JL_embedding} for more details):
\begin{itemize}
    \item[1.] Choose generators $\bm{C}_1\, ,\, \dots\, ,\, \bm{C}_{2n}$ for the Clifford algebra in the Hilbert space of dimension $2^n < 2^N\, $.
    \item[2.] Choose a pair $\bm{v},\bm{w}$ of orthonormal vectors in $\mathbb{R}^{2n}\, $.
    \item[3.] Define
    \begin{eqnarray}
        \bm{\sigma}_x \equiv \sum_{j=1}^{2n} \braket{e_j|v} \bm{C}_j\ ,\ \bm{\sigma}_y \equiv \sum_{j=1}^{2n} \braket{e_j|w} \bm{C}_j\ ,\ \bm{\sigma}_z = -i\bm{\sigma}_x\bm{\sigma}_y\ ,\nonumber
    \end{eqnarray}
    where $\braket{e_j|v}$ and $\braket{e_j|w}$ are the $j$th components of $\bm{v},\bm{w}$ in some orthonormal basis $\lbrace \bm{e}_j\rbrace$ of $\mathbb{R}^{2n}\, $.
    \item[4.] Finally, define raising and lowering operators as
    \begin{eqnarray}
        \bm{c} = \frac{1}{2}\left(\bm\sigma_{x} + i\bm\sigma_{y}\right)\ ,\ \bm{c}^\dagger = \frac{1}{2}\left(\bm\sigma_{x} - i\bm\sigma_{y}\right)\ .\nonumber
    \end{eqnarray}
\end{itemize}
To embed $N$ qubits into the Hilbert space of dimension $2^n$, we simply choose $N$ orthonormal pairs $(\bm{v}_1,\bm{w}_1)\, ,\, \dots\, ,\, (\bm{v}_N,\bm{w}_N)$ at random. We then construct the operators defining the $j$th qubit with the help of the pair $\bm{v}_j,\bm{w}_j$ following the steps outlined above. This procedure raises (at least) two questions:
\begin{itemize}
    \item[A)] Since $n < N$, the operators we obtain for different qubits cannot satisfy the anti-commutation relations from Equations~\ref{eq:anti_commutation_intro}. How large are the deviations from those relations?
    \item[B)] The vector pairs $\bm{v}_j,\bm{w}_j$ are chosen at random. Where does this randomness come from? Which part of an underlying, more fundamental theory is represented by this randomness?
\end{itemize}
We discuss question B) in detail in \secref{mereology}~. To answer question A), let us look at the number $n$ and $N$ that would be required to describe the Weyl field.

\begin{figure}
\centering
  \includegraphics[width=0.6\textwidth]{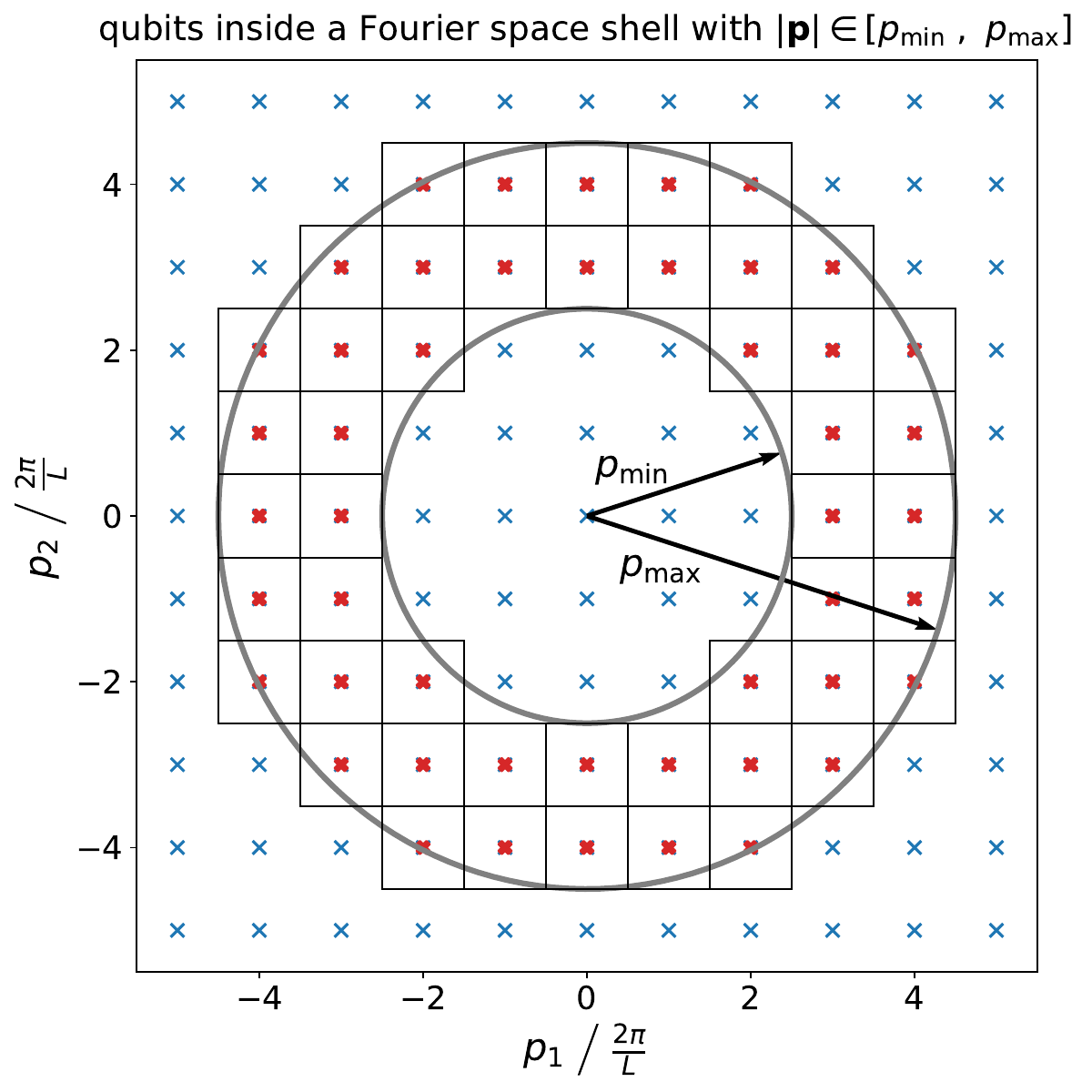}
  \noindent\rule[0.5ex]{0.8\textwidth}{1pt}
  \includegraphics[width=0.6\textwidth]{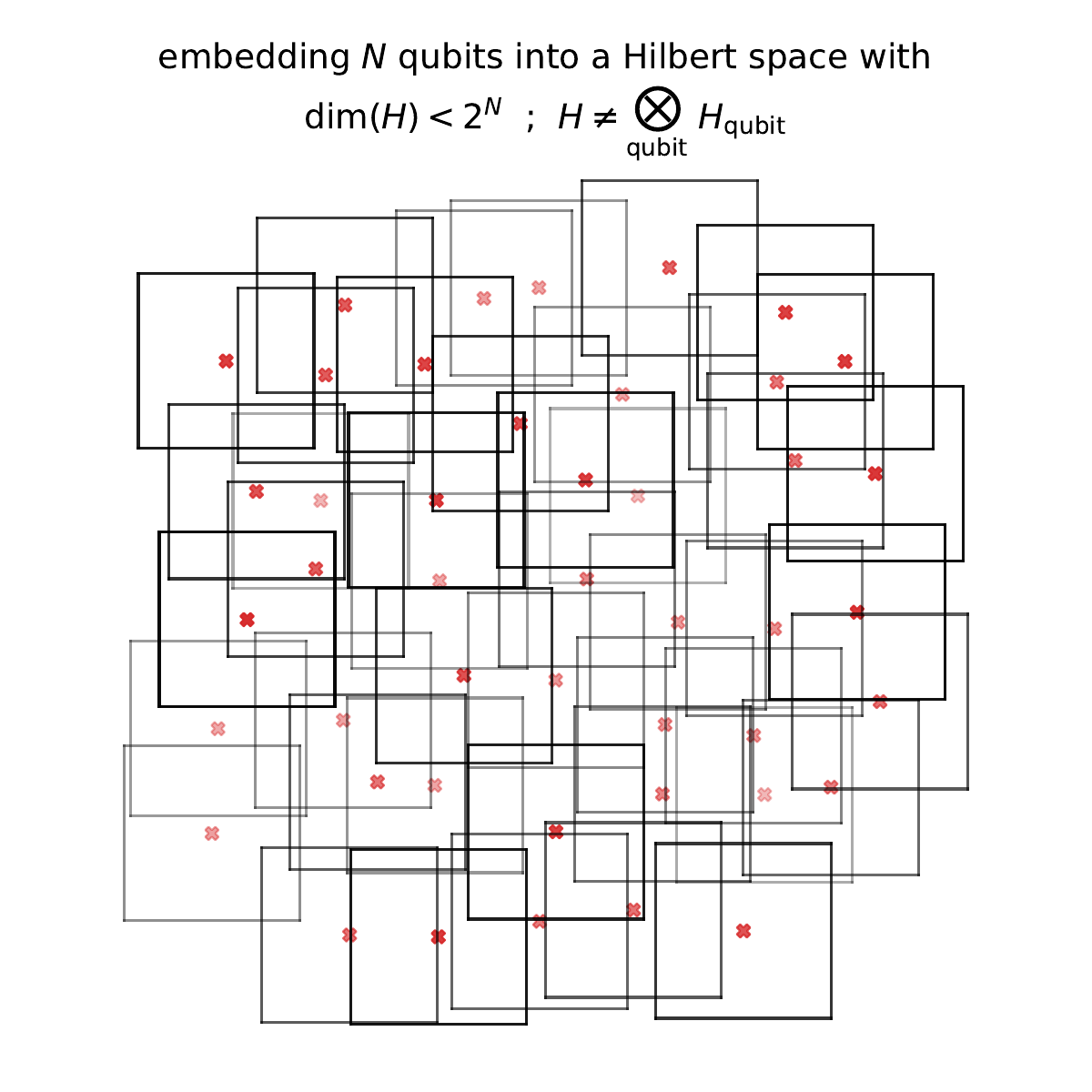}
   \caption{The ($c$-particle part of the) Weyl field can be decomposed into a collection of qubits, which are represented by points $\bm{p}$ in Fourier space (upper sketch). In 3 dimensions, the number of qubits in a small Fourier space shell $s$ (red crossed in upper sketch) scales as $k_s^2 \Delta_s$ where $k_s$ is the radius and $\Delta_s$ is the width of the shell. We want to embed these qubits into a Hilbert space that is strictly smaller then the tensor product of the individual qubit Hilber spaces (cf. Figure~\ref{fi:qubit_registers}) in order to achieve a holographic scaling of the effective degrees of freedom in our field.}
  \label{fi:sketch_with_tiles}
\end{figure}

We had already said that the Weyl field can be considered as a collection of qubits: one for each wave mode $\bm{p}$ in the particle sector and one for each wave mode $\bm{p}$ in the anti-particle sector (cf. Equation~\ref{eq:Weyl_decomposition_discrete_intro} as well as our sketch in Figure~\ref{fi:sketch_with_tiles}, which applies to either one of the two sectors). For a thin Fourier space shell $s$ of radius $k_s$ and width $\Delta_s$ the number density of modes $\bm{p}$ with $|\bm{p}|\in [k_s-\Delta_s/2, k_s+\Delta_s/2]$ per radius will scale as
\begin{eqnarray}
\label{eq:volume_scaling_intro}
    \frac{N_s}{\Delta_s} \propto k_s^2\ ,
\end{eqnarray}
where $N_s$ is the total number of modes inside shell $s$. The above scaling is the Fourier space manifestation of the volume extensive scaling of QFT degrees of freedom. We will now squeeze the qubits present at these modes into a Hilbert space of dimension $2^{n_s} < 2^{N_s}$ (cf.\ again Figure~\ref{fi:sketch_with_tiles}). To approximately achieve holography the radial number density of effective qubits in the shell should scale as \cite{Cao2019_essay} \footnote{According to \cite{Cohen1999, BanksDraper2019, BlinovDraper2021} the effective degrees of freedom should deplete even more quickly than this naive area scaling - \cf our comment in \secref{discussion}~.}
\begin{eqnarray}
    \frac{n_s}{\Delta_s} \propto k_s\ ,
\end{eqnarray}
where $n_s$ is now the effective number of qubit degrees of freedom in shell $s$. At a first glance this procedure may seem infeasible, because for high momenta $k_s$ it may lead to a vast overpopulation of the physical Hilbert space with our unphysical qubits. But the authors of \cite{Cao2017} have shown that - using the embedding procedure we outlined above - the physical Hilbert space can indeed be exponentially overcrowded with qubits (i.e.\ $n_s \sim \ln N_s$) while still keeping the operator norm of the anti-commutator of any pair of qubits smaller than some small parameter $\epsilon$.

Note that, in order to achieve holography, we don't even need this exponential overcrowding. We only need polynomial overcrowding,
\begin{eqnarray}
    N_s \propto n_s^2\ .
\end{eqnarray}
This leads us to our first result: in Section~\ref{sec:JL_embedding} we show that in order to make the Weyl field obey a holographic scaling of the effective degrees of freedom and also satisfy the Bekenstein bound for the box of size $L$, we need to allow overlaps of the creation and annihilation operators for different field modes that \change{are bounded by
\begin{eqnarray}
\label{eq:CJL_bound_intro}
    |\lbrace \hat c_{\bm{p}}^\dagger , \hat c_{\bm{q}} \rbrace|\, <\, \epsilon(k)\, &\approx\, \sqrt{\frac{48 \log\left(k_s L\right)}{\alpha \pi (k_sL)^2}}\, \frac{\Lambda_{\mathrm{UV}}}{\Lambda_{\mathrm{sp}}}\nonumber\\ 
    &\qquad\mathrm{for\ all}\ \ |\bm{p}| \approx k \approx |\bm{q}|,\ \bm{p}\neq\bm{q}\ .
\end{eqnarray}
Here $\alpha = \Delta_s / k_s$ is the relative width of the Fourier space shells, $\Lambda_{\mathrm{UV}}$ is a UV-cutoff of our theory, and $\Lambda_{\mathrm{sp}}$ is our analog of the so called species scale beyond which quantum gravitational effects are expected to become significant and different particle species present in the EFT of our Universe cannot be resolved anymore \cite{Veneziano2002, Han_and_Willenbrock_2005, Dvali2008, Dvali2010, Dvali2013, Castellano2022, Castellano2023}. In this paper we define $\Lambda_{\mathrm{sp}}$ in terms of the Planck scale $\Lambda_{\mathrm{Planck}}$ as
\begin{equation}
\label{eq:species_scale_intro}
    \Lambda_{\mathrm{sp}} \equiv \Lambda_{\mathrm{Planck}} \sqrt{N_{\mathrm{dof},\psi}/N_{\mathrm{dof,total}}}\ ,
\end{equation}
where $N_{\mathrm{dof,total}}$ are the total number of degrees of freedom present in the Universe and $N_{\mathrm{dof},\psi}$ are the degrees of freedom contributed by $\hat\psi$~. As we discuss in Sections~\ref{sec:Bekenstein_bound} and \ref{sec:N_dof_total}, both our definition of $\Lambda_{\mathrm{sp}}$ as well as the orders of magnitude we consider for the ratio $\Lambda_{\mathrm{sp}}/\Lambda_{\mathrm{Planck}}$ differ from typical discussions in the literature referenced above. In order to draw from the intuition built around the species
scale, and in order to simplify many of our expressions, we nevertheless go ahead with the definition in \eqnref{species_scale_intro}.}

\begin{figure}
\centering
  \includegraphics[width=0.92\textwidth]{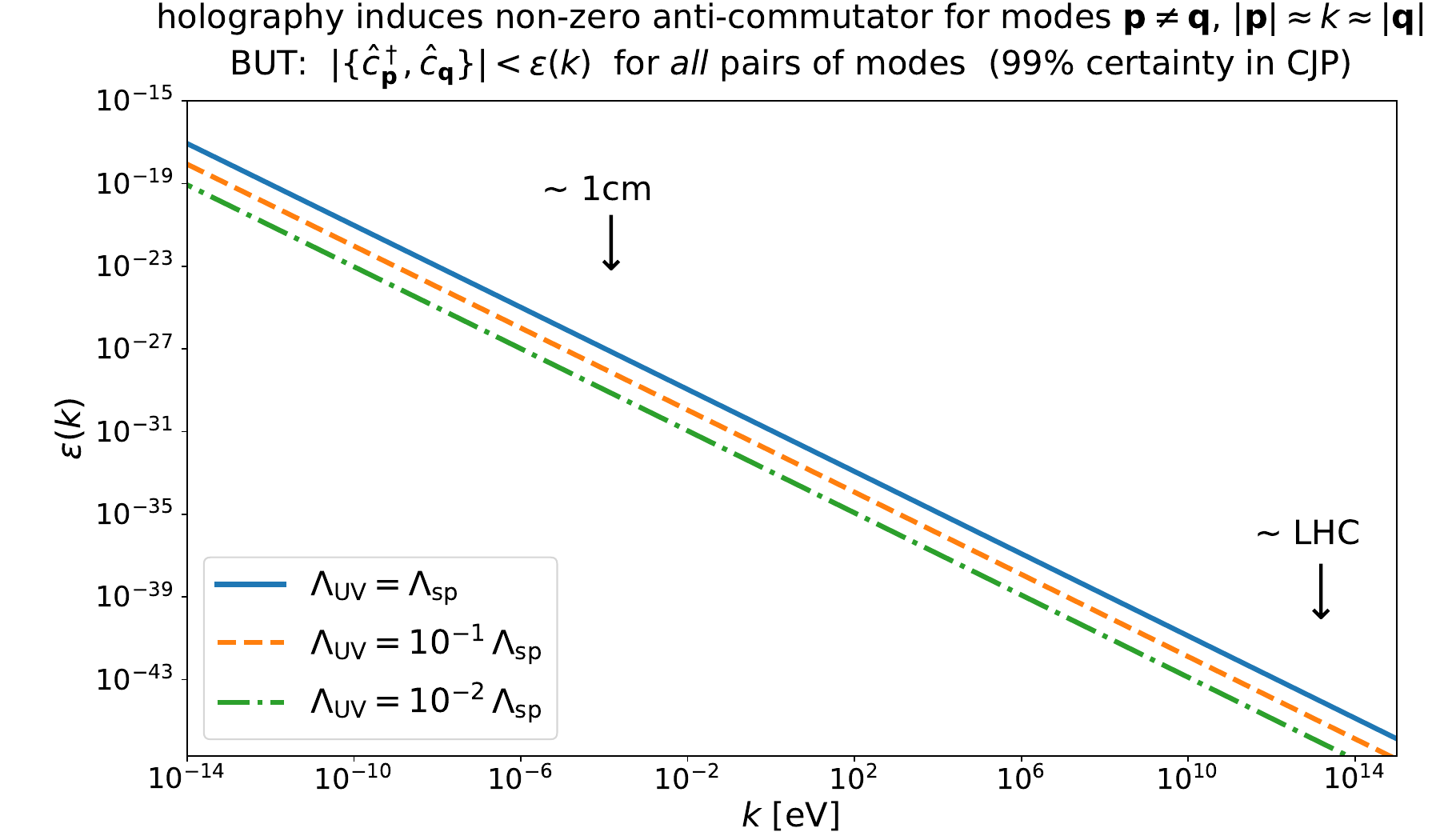}
   \caption{Bound on the anti-commutator $\lbrace \hat{c}_{\mathbf{p}}^\dagger , \hat{c}_{\mathbf{q}}\rbrace$ induced for modes $\mathbf{p}\neq\mathbf{q}$, $|\mathbf{p}|\approx k \approx |\mathbf{q}|$, due to squeezing the degrees of freedom of our field into too small Hilbert spaces via the Chao-Johson-Lindenstrauss embedding (cf. Section~\ref{sec:JL_embedding}). At energies probed by the Large Hadron Collider or at energies of everyday life (cf.\ small arrows) the bound is vastly smaller than the operator norms of $\hat{c}_{\mathbf{p}}, \hat{c}_{\mathbf{p}}^\dagger$, indicating that any pair of Fourier modes behave as almost independent degrees of freedom. \change{Different lines assume different ratios between the UV-cut of our theory and the (pseudo-)species scale $\Lambda_{\mathrm{sp}}$ defined in \eqnref{species_scale_intro}.}}
  \label{fi:epsilon_of_k}
\end{figure}

\change{Note that the bound of Equation~\ref{eq:CJL_bound_intro} is not an ansatz of ours, but a consequence of the Johnson-Lindenstrauss theorem together with our requirement that the effective number of field modes in each Fourier space shell scale like $n_s \propto k_s \Delta_s\,$ (\cf our discussion in \secref{JL_embedding}). We plot $\epsilon(k)$ for different values of the ratio $\Lambda_{\mathrm{UV}}\ /\ \Lambda_{\mathrm{sp}}$ in Figure~\ref{fi:epsilon_of_k}, fixing $\alpha=0.01\,$ and choosing $L$ to be today's physical size of the future particle horizon of our $\Lambda$CDM Universe (\ie the largest distance to which we will ever be causally connected; this is slightly larger than today's Hubble radius, \cf \cite{Friedrich2022}).} The bound $\epsilon(k)$ is in fact decreasing with increasing wave number $k$~, i.e.\ at higher energies the overlap between different modes becomes more and more suppressed. At energies probed by the Large Hadron Collider it falls down to $\sim 10^{-43}$, indicating that any pair of Fourier modes in that regime behave as almost independent degrees of freedom (though see the end of Section~\ref{sec:Bekenstein_bound} for a more nuanced discussion\change{, including a heuristic understanding of the scaling in Figure~\ref{fi:epsilon_of_k} using intuition from AdS/CFT}).

To investigate potential dynamical consequences of the non-vanishing Fourier mode overlaps, we consider a Hamiltonian for our field that takes the same form as for the standard non-overlapping Weyl field,
\begin{eqnarray}
\label{eq:Hamiltonian_intro}
\hat H &= \sum_{\bm{p}} |\bm{p}| \left\lbrace \left(\hat c_{\bm{p}}^\dagger \hat c_{\bm{p}} - \frac{1}{2}\right) + \left(\hat d_{\bm{p}}^\dagger \hat d_{\bm{p}} - \frac{1}{2}\right)\right\rbrace\nonumber \\
&\approx \sum_{\mathrm{shell}\ s} k_s \sum_{\bm{p}\in s} \left\lbrace \left(\hat c_{\bm{p}}^\dagger \hat c_{\bm{p}} - \frac{1}{2}\right) + \left(\hat d_{\bm{p}}^\dagger \hat d_{\bm{p}} - \frac{1}{2}\right)\right\rbrace\ ,
\end{eqnarray}
where in the second line we made explicit our decomposition of Fourier space into a set of shells. Within each shell the occupation number operators $\hat N_{\bm{p}} = \hat c_{\bm{p}}^\dagger\hat c_{\bm{p}}$, $\hat N_{\bm{q}} = \hat c_{\bm{q}}^\dagger\hat c_{\bm{q}}$ of different Fourier modes $\bm{p}$ and $\bm{q}$ will not commute anymore. This also means that $[\hat H , \hat N_{\bm{p}} ] \neq 0$ and the occupation number in a given Fourier mode is not conserved, despite \eqnref{Hamiltonian_intro} formally looking like the free field Hamiltonian. To determine the severity of this effect we first define plane wave states in our theory through the following steps:
\begin{itemize}
    \item consider the states $\ket{\psi(t)}$ which at an initial time $t_0$ start in the eigenspace of $\hat N_{\bm{p}}$ with eigenvalue $+1\,$, \ie $\hat N_{\bm{p}}\ket{\psi(t_0)} = +\ket{\psi(t_0)}$;
    \item in this subspace, find the states that minimize $\left|\left[\frac{\mathrm{d}^2}{\mathrm{d}t^2} \bra{\psi(t)}\hat N_{\bm{p}}\ket{\psi(t)}\right]_{t=t_0}\right|$ ;
    \item this condition is still satisfied by an entire subspace of states, so our final selection step is to choose the candidate state that initially has the lowest energy expectation value $\bra{\psi(t_0)}\hat H\ket{\psi(t_0)}$~.
\end{itemize}
In the following we will use the notation $\ket{\bm{p}} \equiv \ket{\psi(t_0)}\,$, where $\ket{\psi(t)}$ is chosen as explained above, and we will consider $\ket{\bm{p}}$ as the closest equivalent our theory has to the standard plane wave state $\hat c_{\bm{p}}^\dagger \ket{0}$ for the non-overlapping Weyl field. In the standard theory, plane wave states would be eigenstates of the Hamiltonian operator and thus have a trivial time evolution. In our holographic version of the Weyl field this is not the case anymore. Operators $\hat N_{\bm{p}}$ and $\hat N_{\bm{q}}$ for different modes $\bm{p},\bm{q}$ do not commute any longer, and thus the Hamiltonian (which is a sum over all the different occupation number operators) does not commute with any of the $\hat N_{\bm{p}}$ either. As a consequence, time evolution will move the states $\ket{\bm{p}}$ out of the $+1$-eigenspace of $\hat N_{\bm{p}}\,$. Note however that we have assumed different Fourier space shells to be non-overlapping, so the initial probability amplitude in $\ket{\bm{p}}$ will only be re-distributed with the shell that contains $\bm{p}\,$. Since in our fiducial construction the overlaps between different modes is independent of their relative direction, we interpret this modified time evolution as an isotropic scrambling of plane waves \change{(or equivalently, of the information stored in plane wave excitations)} within their Fourier space shells. This indicates a breaking of Lorentz symmetry (in particular, of momentum conservation), and we discuss implications of that in \secref{Lorentz}.

\begin{figure}
\centering
  \includegraphics[width=\textwidth]{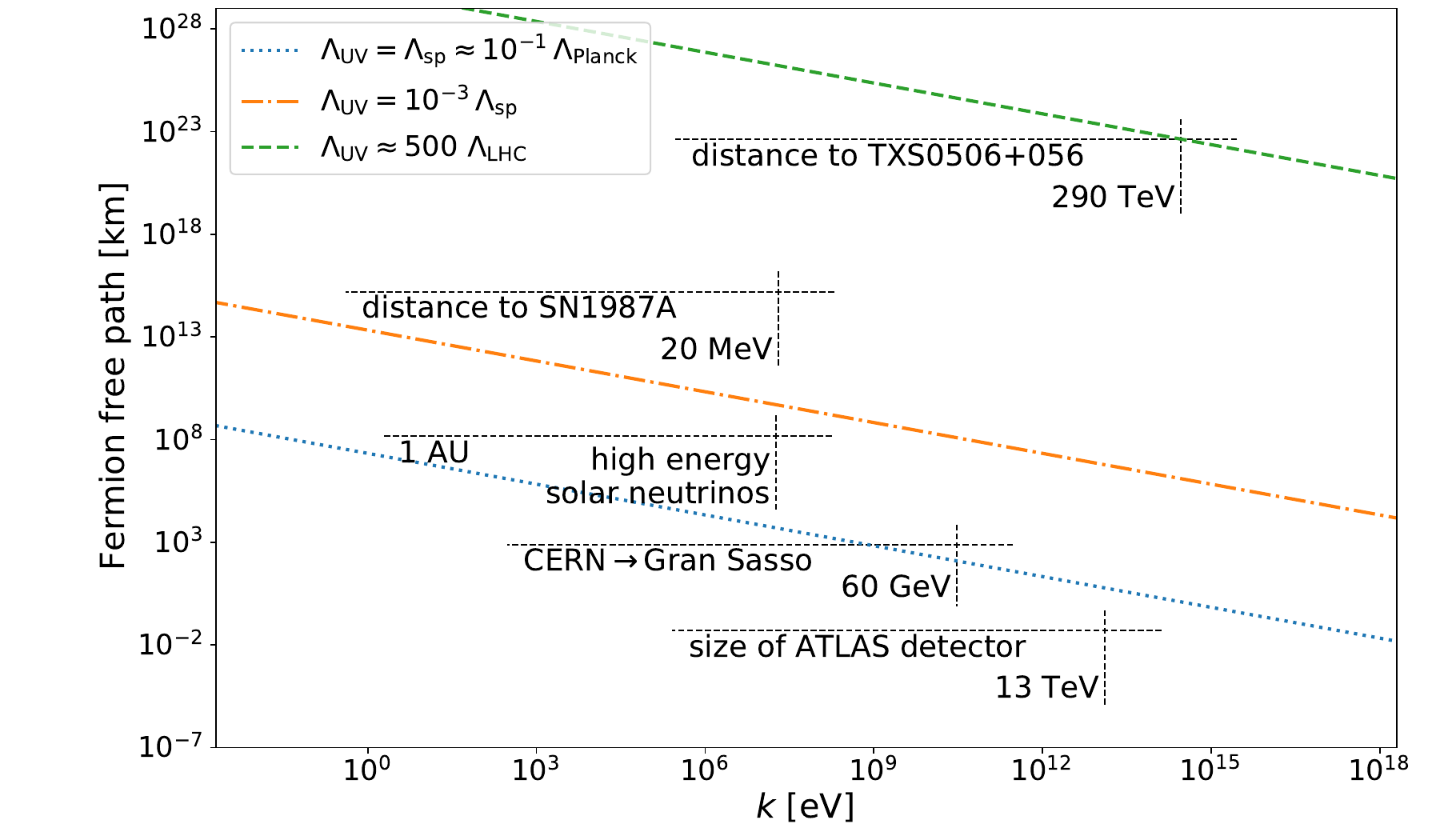}
   \caption{Lifetimes of plane waves in the holographic Weyl field depend on the UV-cut of our theory. We display the corresponding free path as a function of plane wave energy according to our calculations in \ref{app:lifetime} (see also \secref{lifetime}) for $\Lambda_{\mathrm{UV}} = \Lambda_{\mathrm{sp}}$ (blue, dotted line), for $\Lambda_{\mathrm{UV}} = 10^{-3}\, \Lambda_{\mathrm{sp}}$ (orange, dash-dotted line), and for $\Lambda_{\mathrm{UV}} = 470\, \Lambda_{\mathrm{LHC}}$ (green, dashed line), where $\Lambda_{\mathrm{LHC}} = 14\,$TeV represents the collision energies reached at the LHC and $\Lambda_{\mathrm{sp}} \approx 10^{-1}\, \Lambda_{\mathrm{Planck}}$ is the (pseudo-)species scale that appears in our construction (\cf our discussion in \secref{N_dof_total}). All of the displayed lines assume a static Universe and a light-like dispersion relation. We also indicate the energies and path lengths for a number of known and resolved sources of neutrino emission - cf.\ Table~\ref{tab:UV_cuts} for more details. To explain how all of these events could have been observed, we need to assume a break down of neutrino EFT at $\lesssim$ 500 times $\Lambda_{\mathrm{LHC}}$.}
  \label{fi:L_scramble_vs_k}
\end{figure}

To quantify the severity of the scrambling effect, we estimate a characteristic lifetime of plane waves as
\begin{eqnarray}
    \frac{1}{T_{\mathrm{scramble}}^2} = -\mathbb{E}\left\lbrace\frac{\mathrm{d}^2}{\mathrm{d}t^2}\bra{\bm{p}} \hat{N}_{\mathbf{p}} \ket{\bm{p}}\right\rbrace\ ,
\end{eqnarray}
where the expectation value $\mathbb{E}\lbrace \cdot\rbrace$ is taken with respect to the random vectors that are used in the CRSV-embedding to construct representations of the different field modes (\cf \secref{JL_embedding}). In \ref{app:lifetime} we demonstrate analytically that the following relation between $T_{\mathrm{scramble}}$ and the momentum $|\bm{p}|$ holds in the large-$n_s$ limit (cf.\ more details also in \secref{dynamics}):
\begin{eqnarray}
    \frac{\mathrm{d}^2}{\mathrm{d}t^2} \bra{\bm{p}}\hat N_{\bm{p}}\ket{\bm{p}} \propto \frac{|\bm{p}|^2}{|\bm{p}|L}\\
    \Rightarrow T_{\mathrm{scramble}} \propto \sqrt{\frac{L}{|\bm{p}|}}\ .
\end{eqnarray}
The exact amplitude of this scaling depends on the values of different parameters of our construction, and we explain in \secref{lifetime} that it is given by 
\begin{eqnarray}
\label{eq:T_scrample_intro}
    T_{\mathrm{scramble}} \approx 2\pi^2 \sqrt{\alpha} \left(\frac{\Lambda_{\mathrm{sp}}}{\Lambda_{\mathrm{UV}}}\right)^2 \sqrt{\frac{L}{|\bm{p}|}}\ .
\end{eqnarray}
In Sections~\ref{sec:Bekenstein_bound} and \ref{sec:N_dof_total} we argue for fiducial values of the IR-scale, $L\,$, and the relative width of holographically compressed Fourier space shells, $\alpha\,$. This makes the ratio $\Lambda_{\mathrm{UV}}/\Lambda_{\mathrm{sp}}$ the main unknown in the above expression. In particular, a higher $\Lambda_{\mathrm{UV}}$ seems to lead to a stronger scrambling. This can be understood as follows: a higher $\Lambda_{\mathrm{UV}}$ means that our field has more modes, which in turn requires larger mode overlaps in order to still satisfy the cosmic Bekenstein bound. Figuratively speaking, the mode overlaps generate a \say{cosmic fog} and that fog has to become thicker when $\Lambda_{\mathrm{UV}}$ is increased, \ie when more modes are allowed to exist.
In \figref{L_scramble_vs_k} we illustrate the impact of choosing different UV-cuts, fixing the other parameters to fiducial values detailed in Sections~\ref{sec:Bekenstein_bound} and \ref{sec:N_dof_total}. In that figure we also indicate the typical energies and travel distances of neutrinos created by a number of known and resolved sources (see also Table~\ref{tab:UV_cuts} for a summary; we assume that the neutrino dispersion relation is still close to light-like in order to translate lifetimes to distances). \change{The fact that we can meaningfully compare effects of holography to such low-energy observations (low-energy compared to the Planck scale) is indicative of a characteristic property of the theory of quantum gravity: it does not exhibit the UV/IR de-coupling that is present in e.g.\ effective field theories \cite{Cohen1999}.} We provide a more detailed discussion of the results of this comparison in \secref{dynamics}, but the highly energetic neutrino emission observed from blazar TXS 0506+056 requires $\Lambda_{\mathrm{UV}} < 470\, \Lambda_{\mathrm{LHC}}\,$, where we take $\Lambda_{\mathrm{LHC}} = 14\,$TeV. So electroweak theory should be modified only a few orders of magnitude above modern particle physics experiments. In \secref{CR_spectrum} we argue that this is still consistent with current measurements of the cosmic neutrino spectrum by IceCube (\cf \figref{The_CR_Spectrum_2023} in that section). The breakdown of our effective theory beyond the energy $\Lambda_{\mathrm{UV}}$ should happen either as a sharp cutoff or as a transition to \say{super-holographic} mode overlaps, because any modes present beyond $\Lambda_{\mathrm{UV}}$ need to be squeezed into a very small part of the total Hilbert space in order to still satisfy the Bekenstein bound. \change{Note that the fact that we find $\Lambda_{\mathrm{UV}} \ll \Lambda_{\mathrm{sp}}$ is in itself surprising, since usually $\Lambda_{\mathrm{sp}}$ is expected to be the scale of EFT-breakdown (see e.g.\ \cite{Dvali2008, Castellano2022}). We discuss this point further in \secref{discussion}.}

\begin{figure}
\centering
\includegraphics[width=0.9\textwidth]{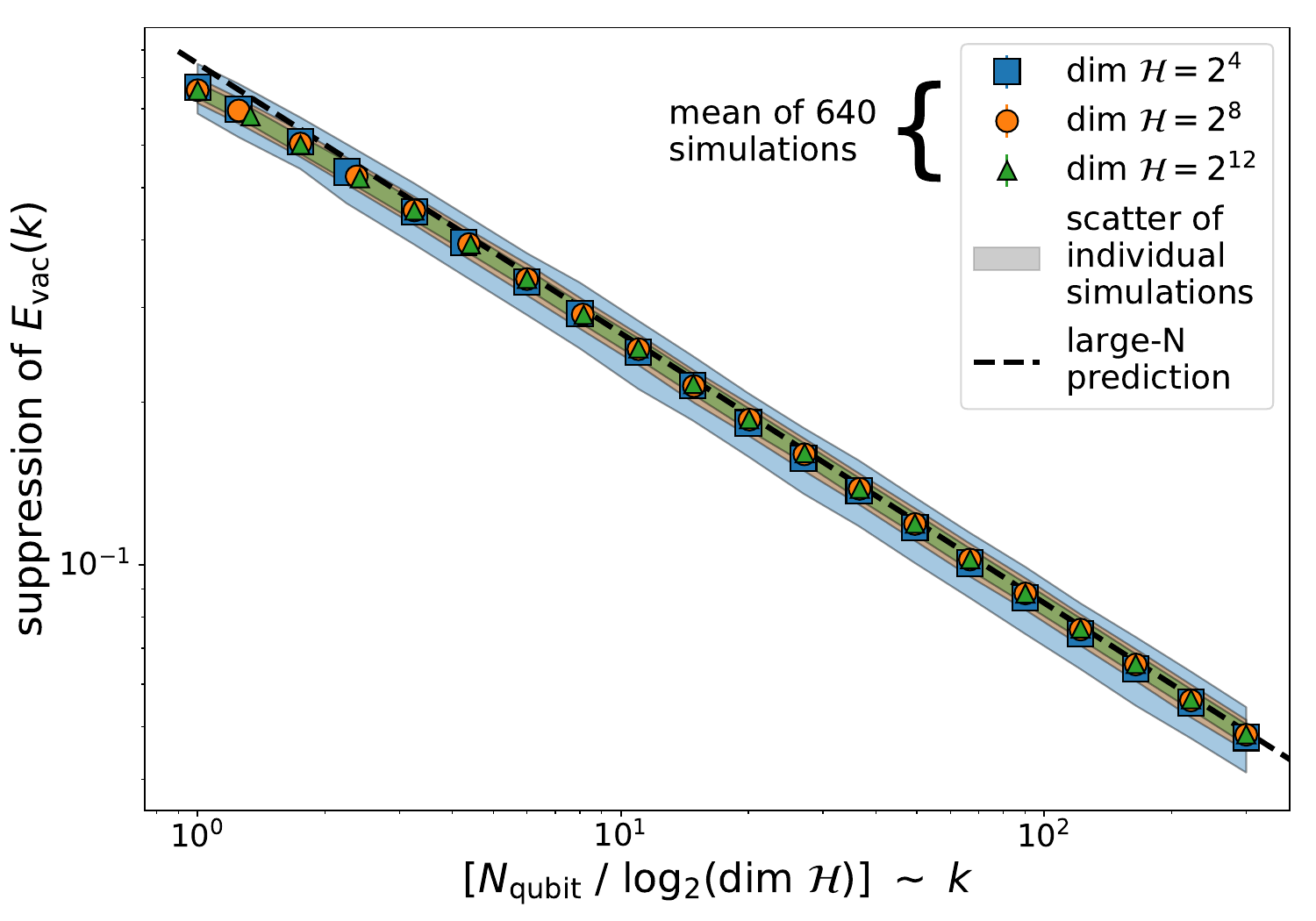}
   \caption{Suppression of minimum eigenvalue of the Hamiltonian (compared to the case of non-overlapping degrees of freedom) in a Fourier space shell $s$ of our holographic Weyl field as a function of its overcrowding $N_s/n_s$ ($= N_{\mathrm{eff}}/\log_2(\dim \mathcal{H})$ in the figure label). Note that $N_s \propto k_s^2\Delta_s$ and $n_s \propto k_s\Delta_s$, where $\Delta_s$ is the width of the shell and $k_s$ its radius. Thus, $N_s/n_s \propto k_s$, \ie the $x$-axis of the figure is proportional to shell radius. The different symbols represent measurements of vacuum energy suppression in simulated realisations of the Fourier space shells for $n_s = 4, 8, 12$. Overall $640$ independent realisations of those simulations have been averaged to obtain these measurements. The transparent bands indicate the standard deviation of the individual realisations. The dashed line displays our theoretical prediction for the suppression of vacuum energy, as derived in \ref{app:Heisenberg}. Note that due to numerical limitations we were only able to simulate very low dimensional shells. Our theoretical prediction will become even more precise at higher values of $n_s$ and $N_s$ because it describes an asymptotic, high-dimensional limit.}
\label{fi:Evac_suppression}
\end{figure}

Apart from the behaviour of plane waves, it is also possible to understand the dynamics of our field more generally. In \ref{app:energy_spectrum} we use results from random matrix theory to derive that the set of eigenvalues of the Hamiltonian in each individual shell of \eqnref{Hamiltonian_intro} is given by
\begin{eqnarray}
    \lbrace \lambda \rbrace &= \left\lbrace \left.\frac{k_s}{2} \sum_{i=1}^{n_s} \lambda_i s_i\ \right|\ \lbrace s_i \rbrace \in [-1, +1]^{n_s} \right\rbrace \ ,
\end{eqnarray}
where $\lambda_i$ are the positive eigenvalues of a generalised Wigner matrix. Using a rigidity property for the eigenvalues of Wiger matrices derived by \cite{Erdos2010b} the $i$th of these eigenvalues will be given with high accuracy by the $i/n_s$-quantile of a Wigner semi-circle distribution. In particular we are able to derive the vacuum energy in a Fourier space shell of width $\Delta_s$ and radius $k_s$ as
\begin{eqnarray}
\label{eq:minimum_E_intro}
    E_{\min , s} &\approx -\frac{N_s k_s}{2}\cdot\left(\frac{8}{3\pi} \sqrt{\frac{\displaystyle n_s}{\displaystyle N_s}}\right)\ .
\end{eqnarray}
In \figref{Evac_suppression} we have used simulated realisations of overlapping qubits to show that this result is accurate already for quite low numbers of field modes. At energies (and hence mode numbers) relevant for high-energy physics, it represents a strong suppression with respect to the non-overlapping Weyl field, whose vacuum energy in each shell would simply be $-N_s k_s/2$. When choosing $\Lambda_{\mathrm{UV}} = \Lambda_{\mathrm{Planck}}$ the total vacuum energy of the field is lowered by a factor of $\sim 10^{-30}$ (which is however not enough to alleviate the cosmological constant problem).

\begin{table}
\begin{tabular}{p{0.13\textwidth}|p{0.15\textwidth}|p{0.12\textwidth}|>{\raggedright}p{0.28\textwidth}||p{0.22\textwidth}}
Neutrino sources& characteristic energy scale & travel distance $d_{\mathrm{source}}$ & UV-scale where Fermion EFT must break down in order to ensure $d_{\mathrm{source}} < T_{\mathrm{scramble}}/c$ & \RaggedRight{vacuum energy suppression $\ \ \ \ $ (due to holography \& UV-cut)} \\
 &  &  &  &  \\
\hline
ATLAS (\cite{ATLAS2023}) & $\lesssim$ 14 TeV & 50 m & $1.04\, \Lambda_{\mathrm{Planck}}\, =\, 9.1\cdot 10^{14}\, \Lambda_{\mathrm{LHC}}$ & \RaggedRight{$1.4 \cdot 10^{-31}$ $\ \ \ \ \ \ \ $ (for $\Lambda_{\mathrm{UV}} = \Lambda_{\mathrm{Planck}}$)} \\
 &  &  &  &  \\
CERN $\rightarrow\ $ Gran Sasso (\cite{Adam2012}) & $\sim$ 30 GeV  & 731 km & $4.0\cdot 10^{-2}\, \Lambda_{\mathrm{Planck}}\, =\, 3.5\cdot 10^{13}\, \Lambda_{\mathrm{LHC}}$ & $1.9 \cdot 10^{-36}$ \\
 &  &  &  &  \\
Solar$\,\,\,\,\,\,$ (\eg \cite{Bellerive2004}) & $<$ 18 MeV & $1.5\cdot 10^{12}$ m & $1.8\cdot 10^{-4}\, \Lambda_{\mathrm{Planck}}\, =\, 1.6\cdot 10^{11}\, \Lambda_{\mathrm{LHC}}$ & $1.1 \cdot 10^{-44}$ \\
 &  &  &  &  \\
SN$\,\,$1987A $\ \ \ \ $ (\cite{SN1987A_1,SN1987A_2}) & $\lesssim$ 20 MeV &  $1.6\cdot 10^{18}$ m & $1.7\cdot 10^{-7}\, \Lambda_{\mathrm{Planck}}\, =\, 1.5\cdot 10^{8}\, \Lambda_{\mathrm{LHC}}$ & $2.9 \cdot 10^{-55}$ \\
 &  &  &  &  \\
TXS 0506+056 (\cite{blazarIceCube}) & $\sim$ 290 TeV  & $4.1\cdot 10^{25}$ m $\approx 1.3$~Gpc & $5.4\cdot 10^{-13}\, \Lambda_{\mathrm{Planck}}\, =\, 4.7\cdot 10^{2}\, \Lambda_{\mathrm{LHC}}$ & $1.7 \cdot 10^{-74}$ \\
\end{tabular}
    \caption{Known and resolved sources of neutrinos, their typical energies and their travel distances. In order for the travel time of these particles to be shorter than the plane wave scrambling time of our holographic construction, we need to impose UV cutoffs that are significantly below the Planck scale\change{ and, in fact, even below the species scale $\Lambda_{\mathrm{sp}}\,$, which for standard model degrees of freedom we assume to be $\approx 10^{-1}\, \Lambda_{\mathrm{Planck}}$ (cf.\ our discussion in \secref{N_dof_total})}. The most constraining UV-cut we derive is however still safely above energy scales probed at the LHC; cf.\ the 4th column. Holography and the UV cutoffs also combine to suppress vacuum energy density significantly compared to a standard Fermion with $\Lambda_{\mathrm{UV}} = \Lambda_{\mathrm{Planck}}$ (but not enough to solve the cosmological constant problem; cf.\ 5th column). The table employs our fiducial assumption that about $1\%$ of the Universes degrees of freedom are contributed by our holographic Fermion \change{(which is equivalent to $\Lambda_{\mathrm{sp}} \approx 10^{-1}\, \Lambda_{\mathrm{Planck}}$)}. Whether this is a realistic assumption depends on the degree to which different quantum fields overlap with each other (\cf again \secref{N_dof_total}).}
\label{tab:UV_cuts}
\end{table}

Finally, in Section~\ref{sec:real_space_propagator} we investigate the impact of the overlapping Fourier modes on the real space propagator $\lbrace i\psi_\alpha(\bm{x})^\dagger , \psi_\beta(\bm{y})\rbrace$~. After tracing out the spin degree of freedom labelled by $\alpha,\beta$ we obtain
\begin{equation}
\label{eq:propagator_intro}
\sum_\alpha \lbrace i\psi_\alpha(\bm{x})^\dagger , \psi_\alpha(\bm{y})\rbrace = 2i\delta_D(\bm{x}-\bm{y}) + 2i C(\bm{x}, \bm{y})\ .
\end{equation}
The first term on the right hand side of Equation~\ref{eq:propagator_intro} is the result one would obtain for the standard Weyl field and the second term is a correction induced by the mode overlap. Since we squeeze the degrees of freedom of our field into the true, physical Hilbert space of the field via a stochastic procedure, the function $C(\bm{x}, \bm{y})$ is also stochastic. It vanishes on average,  $\mathbb{E}[ C(\bm{x}, \bm{y}) ] = 0$ ~, but \change{it fluctuates with an amplitude
\begin{eqnarray}
\sqrt{\mathbb{E}\left[ |C(\bm{x},\bm{y})|^2 \right]} &= \frac{1}{\sqrt{64 \pi^5}}\left(\frac{\Lambda_{\mathrm{UV}}^2}{L}\right)\left(\frac{\Lambda_{\mathrm{UV}}}{\Lambda_{\mathrm{sp}}}\right)\ ,
\end{eqnarray}
where $L$ is again the IR scale corresponding to the size of the box to which we confine our field, $\Lambda_{\mathrm{UV}}$ is a UV cutoff and $\Lambda_{\mathrm{sp}}$ is the (pseudo-)species scale appearing in our construction.} Note that the amplitude of the correction term is in fact independent of $\bm{x}$, $\bm{y}$ and in particular their distance. So, within our construction, a holographic scaling of the effective field degrees of freedom manifests itself as a stochastic, long-range contribution to the field propagator. In particular, the correction can be interpreted as a source of non-locality because the field and its conjugate momentum do not anti-commute at non-zero, equal-time distances.

Since the correction term $2iC(\bm{x},\bm{y})$ in Equation~\ref{eq:propagator_intro} has a scale independent amplitude, it is tempting to assume that the standard term $2i\delta_D(\bm{x}-\bm{y})$ will always dominate the propagator at sufficiently small distances $\bm{x}-\bm{y}$~. To make this assumption more precise, and in order to test it, we smooth our field with a Gaussian test function of width $R$ (standard deviation). The resulting smoothed field $\hat\psi_R$ has the propagator
\begin{eqnarray}
    \sum_\alpha \lbrace i\psi_{R,\alpha}(\bm{x})^\dagger , \psi_{R,\alpha}(\bm{y})\rbrace\ &=\ 2iC_R(\bm{x},\bm{y}) + \frac{2i}{\sqrt{(4\pi)^3R^6}}\ e^{-\frac{1}{4R^2} |\bm{x}-\bm{y}|^2}\nonumber \\
    &\equiv\ 2i D_R(\bm{x},\bm{y})\ ,
\end{eqnarray}
where for simplicity we have again taken the trace over spin degrees of freedom, and where $C_R(\bm{x},\bm{y})$ is an appropriately smoothed version of $C(\bm{x},\bm{y})$.  We can then compare the average fluctuation of the long-range propagator (which is dominated by $C_R(\bm{x},\bm{y})$) to the average local propagator (which is given by the standard local propagator because $\mathbb{E}[ C_R] = 0$). In particular, we consider the ratio of these two and in Section~\ref{sec:real_space_propagator} \change{we obtain
\begin{eqnarray}
\label{eq:anticommutator_ratio_intro}
    \frac{\mathbb{E}\left[ |C_R(\bm{x},\bm{y})|^2 \right]}{\mathbb{E}[ D_R(\bm{x},\bm{x}) ]\ \mathbb{E}[ D_R(\bm{y},\bm{y}) ] }\ =\ \frac{8}{\pi^2}\, \left(\frac{R}{ L}\right)^2 \left(\frac{\Lambda_{\mathrm{UV}}}{\Lambda_{\mathrm{sp}}}\right)^2\ .
\end{eqnarray}
The square root of this ratio is displayed in the lower panel of Figure~\ref{fi:relative_longrange_propagator} for different values of the ratio $\Lambda_{\mathrm{UV}} / \Lambda_{\mathrm{sp}}$~. Even for cosmologically large smoothing scales ($R\,\sim 100$~Mpc) the typical amplitude of the long-range propagator remains at a level of $< 1\%$ of the local anti-commutator.} So the non-locality induced by our holographic scaling of the effective field degrees of freedom seems to become significant only on very large scales. In a sense, this is just the real space equivalent of our result regarding the mode anti-commutators $\lbrace \hat{c}_{\mathbf{p}}^\dagger , \hat{c}_{\mathbf{q}}\rbrace$ from \figref{epsilon_of_k}.

\subsection{Summary of assumptions and modelling choices}

\change{We discuss the assumptions and caveats of our analysis in detail in \secref{discussion}, but we already want to give a brief summary of our most crucial approximations and modelling choices here.
\begin{itemize}
    \item[A)] We construct our holographic field by covering the space of Fourier modes of the field with non-overlapping shells of constant relative width. The modes in each shell are then made to overlap with each other in an isotropic way that ignores how close any two modes are in Fourier space (though we do consider alternatives to our fiducial construction in \secref{alternative_choices}).
    \item[B)] The mode overlaps are stochastic, following the algorithm proposed by \cite{Chao2017}. At first glance, this may seem peculiar, but we argue in \secref{mereology} that this stochasticity is to be expected in a framework where quasi-classical degrees of freedom emerge from an ensemble of abstract quantum theories.
    \item[C)] We assume our holographic Weyl field to have a sharp UV-cutoff. This causes Lorentz invariance violations, which we discuss in \secref{Lorentz}.
    \item[D)] We discretise our field on a rectangular grid, which adds to the Lorentz invariance violation mentioned above.
    \item[E)] Our model implements a naive equal-time version of holography. We argue in Sections~\ref{sec:Lorentz} and \ref{sec:discussion} that a more appropriate construction would quantise the Weyl field on a light-sheet, \ie it would implement the co-variant entropy bound \cite{Bousso2002}.
    \item[F)] We assume that the future, co-moving particle horizon serves as an IR-cutoff to our model. This is the largest distance in today's Universe, to which we will ever be causally connected \cite{Friedrich2022}.
    \item[G)] We assume that the species scale is determined purely by the standard model degrees of freedom. We also allow our (effective) theory to break down at energies far below the species scale. This is in fact needed in order for our results to be consistent with cosmic neutrino observations, as we demonstrate in Sections~\ref{sec:lifetime} and \ref{sec:CR_spectrum}.
    \item[H)] We assume that the Hamiltonian of the holographic Weyl field retains the usual form given in \eqnref{Hamiltonian_intro}. In \secref{mereology} we argue that this assumption should eventually be replaced with an algorithm that extracts local field degrees of freedom from abstract Hamiltonian ensembles.
    \item[I)] We ignore cosmic expansion when applying our model to the propagation of cosmic neutrinos (an estimate of the mistake made by that approximation is discussed in \secref{CR_spectrum}). Note also that we explicitly assume a spatially flat Universe. In particular, our calculations do no apply to AdS spacetimes or hyperbolic geometries in general. For example, the maximum entropy attainable by our field is much lower that the entropy of, say, half an equal-time slice of an AdS spacetime whose radius equals our IR-scale $L$.
\end{itemize}}
\vspace{0.3cm}

\noindent \change{The holographic quantum field we build based on the above modelling choices is characterised by the following parameters.
\begin{itemize}
    \item $\alpha$: the relative with of the shells covering the field's Fourier space. This can be though of as the degree to which we allow mode overlaps to mix modes of different energy.
    \item $\Lambda_{\mathrm{UV}}$: the UV-cut of our model.
    \item $\Lambda_{\mathrm{sp}}$: a (pseudo-)species scale defined as $\Lambda_{\mathrm{sp}} \equiv \Lambda_{\mathrm{Planck}} \sqrt{N_{\mathrm{dof},\psi}/N_{\mathrm{dof,total}}}$, where $N_{\mathrm{dof,total}}$ are the total number of degrees of freedom present in the Universe and $N_{\mathrm{dof},\psi}$ are the degrees of freedom contributed by $\hat\psi$~.
    \item $L$: the IR-cut of our model.
\end{itemize}
In Sections~\ref{sec:Bekenstein_bound} and \ref{sec:N_dof_total} we motivate fiducial values for all of these parameters except for $\Lambda_{\mathrm{UV}}$. Hence, $\Lambda_{\mathrm{UV}}$ (or equivalently the ratio $\Lambda_{\mathrm{UV}}/\Lambda_{\mathrm{sp}}$) is the only parameter we vary in our analysis.} 

\subsection{Structure of the paper}

The rest of this paper is structured as follows. In \secref{spinor_details} we review basics of the Weyl field and how to decompose it into a set of qubits. \secref{JL_embedding} then introduces modifications to make the degrees of freedom of this field obey an approximate area law scaling. We study the dynamics of our holographic Weyl field in \secref{dynamics}. In particular, results about its energy spectrum are presented in \secref{E_spectrum} and we discuss our definition of plane waves and their lifetime in \secref{lifetime}. Section \ref{sec:real_space_propagator} investigates the real space anti-commutator of our field, while \secref{alternative_choices} repeats some of our analysis for two possible modifications of our fiducial construction. In \secref{mereology} we put our construction into the context of quantum mereology and discuss why classical stochasticity like the one present in our qubit overlaps is likely to arise in a dynamical emergence of classical degrees of freedom from an abstract quantum theory. Finally, \secref{discussion} summarises our paper and discusses the (many) questions that are left open by our analysis.

Note that many of our technical results are derived in our appendices (in particular \ref{app:Heisenberg}) and only quoted in the main text.

\section{The Weyl field as a collection of qubits}
\label{sec:spinor_details}

For the sake of simplicity we consider a Weyl fermion, which corresponds to one of the two helicity sectors of a massless Dirac fermion. This section serves as a reminder of the Weyl formalism. We also make explicit in which sense the Weyl field can be decomposed into a collection of qubits. In Section~\ref{sec:JL_embedding} we will use this decomposition to define a modified version of the Weyl field, in which certain degrees of freedom are \say{overlapping}.

\subsection{Weyl field basics}

The (left-handed) Weyl spinor is a two-component field $\psi$ with the Lagrangian
\begin{equation}
    \mathcal{L} = i\psi^\dagger \sigma^\mu \partial_\mu \psi\ ,
\end{equation}
where $\sigma^0 \equiv \mathbf{1}$ and $\sigma^i$ are \eg the Pauli matrices (we are following here the notation of \cite{Mandl_Shaw_1994, Pal2011}). The above Lagrangian leads to the equations of motion
\begin{equation}
\label{eq:Weyl_equation}
    \sigma^\mu \partial_\mu \psi = 0\ .
\end{equation}
A general solution to these equations can be expressed as
\begin{eqnarray}
\label{eq:expansion_of_solution}
    \psi(\bm{x}, t) &= \int \frac{\dd^3p}{(2\pi)^{3} E_p}\ \left\lbrace a_{\bm{p}}(t) u(\bm{p})\ e^{i\bm{p}\bm{x}}+ b_{\bm{p}}^*(t) u(\bm{p})\ e^{-i\bm{p}\bm{x}} \right\rbrace\ ,
\end{eqnarray}
where the time evolution of the coefficients $a_{\bm{p}}$ and $b_{\bm{p}}^*$ is given by
\begin{equation}
\label{eq:time_dependence_a_and_b_usual}
    a_{\bm{p}}(t) = a_{\bm{p},0}\ e^{-iE_p t}\ ,\ b_{\bm{p}}(t)^* = b_{\bm{p},0}^*\ e^{iE_p t}
\end{equation}
with $E_p \equiv |\bm{p}|$, and where $u(\bm{p})$ are eigenvectors of the matrix $\sigma^j p_j$ with eigenvalues $+E_p$\footnote{Note that $u(-\bm{p})$ is then an eigenvector with eigenvalue $-E_p$. This is why only a single family of functions $u(\bm{p})$ appears in the expansion of \eqnref{expansion_of_solution}.}~, 
\begin{equation}
    \sigma^j p_j \cdot u(\bm{p}) = +E_p\ u(\bm{p})\ .
\end{equation}
Note that in the following we will keep the time dependence of $a_{\bm{p}}$ and $b_{\bm{p}}^*$ implicit in our notation. The reason is that this dependence will deviate from \eqnref{time_dependence_a_and_b_usual} once we consider overlapping degrees of freedom.

Normalising the $u(\bm{p})$ such that
\begin{equation}
    u(\bm{p})^\dagger \cdot u(\bm{p}) = E_p
\end{equation}
ensures that they are orthogonal \wrt the Lorentz invariant momentum space measure
\begin{equation}
    \dd \tilde p := \frac{\dd^3 p}{(2\pi)^3\ E_p}\ .
\end{equation}
Up to an irrelevant phase factor we can \eg choose $u(\bm{p})$ as \cite{Pal2011}
\begin{equation}
    u(\bm{p}) = \sqrt{E_p} \pmatrix{ e^{-i\phi} \sin \frac{\theta}{2} \cr \cos \frac{\theta}{2}}\ ,
\end{equation}
where $\bm{p} = (p \sin\theta\cos\phi, p \sin\theta\sin\phi, p \cos\theta)^T$~.

In the quantum version of the above field theory we consider the operator valued field
\begin{eqnarray}
    \hat\psi(\bm{x}, t) &= \int \dd \tilde p\ \left\lbrace \hat a_{\bm{p}}(t) u(\bm{p})\ e^{i\bm{p}\bm{x}}+ \hat b_{\bm{p}}(t)^\dagger u(\bm{p})\ e^{-i\bm{p}\bm{x}} \right\rbrace\ ,
\end{eqnarray}
where the operator $\hat b_{\bm{p}}^\dagger$ can be thought of as creating an anti-spinor of momentum $\bm{p}$ while $\hat a_{\bm{p}}$ is destroying a spinor with momentum $\bm{p}$. At equal times these operators satisfy the anti-commutation relations
\begin{eqnarray}
0 = \lbrace \hat a_{\bm{p}}, \hat a_{\bm{q}} \rbrace = \lbrace \hat b_{\bm{p}}, \hat b_{\bm{q}} \rbrace = \lbrace \hat a_{\bm{p}}, \hat b_{\bm{q}} \rbrace = \lbrace \hat a_{\bm{p}}, \hat b_{\bm{q}}^\dagger \rbrace\\
\lbrace \hat a_{\bm{p}}, \hat a_{\bm{q}}^\dagger \rbrace = \lbrace \hat b_{\bm{p}}, \hat b_{\bm{q}}^\dagger \rbrace = (2\pi)^3 E_p\ \delta_D(\bm{p}-\bm{q})\ .
\end{eqnarray}
This ensures that the field operators satisfy the equal-time anti-commutation relation
\begin{eqnarray}
    &\lbrace \hat\psi(\bm{x}), i\hat\psi(\bm{y})^\dagger \rbrace \\
    &\ \ \ = i\int \dd\tilde p\dd\tilde q\ \left[ \lbrace \hat a_{\bm{p}}, a_{\bm{q}}^\dagger\rbrace\ u(\bm{p})u(\bm{q})^\dagger + \lbrace \hat b_{-\bm{p}}^\dagger, b_{-\bm{q}}\rbrace\ u(-\bm{p})u(-\bm{q})^\dagger\right] e^{i\bm{p}\bm{x} - i\bm{q}\bm{y}}\nonumber \\
    &\ \ \ = i\int \frac{\dd^3p}{(2\pi)^{3} E_p}\ \left[ u(\bm{p})u(\bm{p})^\dagger + u(-\bm{p})u(-\bm{p})^\dagger\right] e^{i\bm{p}(\bm{x} - \bm{y})}\nonumber \\
    \nonumber \\
    &\ \ \ = i\ \mathbf{1}_{\mathrm{2D}}\ \delta_D(\bm{x}-\bm{y})\ ,
\end{eqnarray}
as is needed because $i\hat\psi^\dagger$ is the conjugate momentum of the field $\hat\psi$\ .

\subsection{Decomposition into qubits}

To make it explicit that the above field can be considered as a collection of qubits, let us constrict $\psi(\bm{x})$ to a box of finite size $L$~. This means that we have to perform the substitutions
\begin{equation}
    \dd^3 p \rightarrow \frac{(2\pi)^3}{L^3}\ ,\ \delta_D(\bm{p}-\bm{q}) \rightarrow \frac{L^3}{(2\pi)^3}\delta_{\bm{p},\bm{q}}\ ,
\end{equation}
and that integrals over momenta will be replaced by sums over the discrete grid $\bm{p} \in \lbrace\ 2\pi/L (n_1, n_2, n_3)\ |\ n_i \in \mathbb{Z} \ \rbrace$~. For convenience, we will also consider redefined mode operators
\begin{equation}
    \hat c_{\bm{p}} = \frac{\hat a_{\bm{p}}}{(E_pV)^{\frac{1}{2}}}\ ,\ \hat d_{\bm{p}} = \frac{\hat b_{\bm{p}}}{(E_pV)^{\frac{1}{2}}}\ ,
\end{equation}
such that the new operators satisfy the anti-commutation relations
\begin{eqnarray}
0 = \lbrace \hat c_{\bm{p}}, \hat c_{\bm{q}} \rbrace = \lbrace \hat d_{\bm{p}}, \hat d_{\bm{q}} \rbrace = \lbrace \hat c_{\bm{p}}, \hat d_{\bm{q}} \rbrace = \lbrace \hat c_{\bm{p}}, \hat d_{\bm{q}}^\dagger \rbrace\\
\lbrace \hat c_{\bm{p}}, \hat c_{\bm{q}}^\dagger \rbrace = \lbrace \hat d_{\bm{p}}, \hat d_{\bm{q}}^\dagger \rbrace = \delta_{\bm{p},\bm{q}}\ .
\end{eqnarray}
Our field can now be decomposed in terms of these operators as
\begin{equation}
\label{eq:Weyl_decomposition_discrete}
    \hat\psi(\bm{x}, t) = \sum_{\bm{p}}\frac{1}{(E_pV)^{\frac{1}{2}}}\left\lbrace \hat c_{\bm{p}}(t)\ u(\bm{p})\ e^{i\bm{p}\bm{x}}+ \hat d_{\bm{p}}(t)^\dagger\ u(\bm{p})\ e^{-i\bm{p}\bm{x}} \right\rbrace\ .
\end{equation}
Usually, each of the grid points $\bm{p}$ in the above sum represents a 4-dimensional Hilbert space factor
\begin{eqnarray}
\mathcal{H}_{\bm{p}} = \mathcal{H}_{\bm{p}}^c\otimes^{\mathrm{JW}} \mathcal{H}_{\bm{p}}^d
\end{eqnarray}
and the total Hilbert space is the tensor product over all these factors,
\begin{eqnarray}
\mathcal{H} = \bigotimes_{\bm{p}}^{\mathrm{JW}} \mathcal{H}_{\bm{p}}\ ,
\end{eqnarray}
where the superscript \say{JW} again indicates that operators in the individual Hilbert spaces need to be embedded into the product space via Jordan-Wigner-factors (\cf \ref{app:JordanWigner}) in order for them to anti-commute (as opposed to commute). The $\hat c_{\bm{p}}$, $\hat d_{\bm{p}}$ and their Hermitian conjugates act non-trivially only on the factors $\mathcal{H}_{\bm{p}}^c$ and $\mathcal{H}_{\bm{p}}^d$ respectively. On the factor $\mathcal{H}_{\bm{p}}^c$ (and similarly for $\mathcal{H}_{\bm{p}}^d$) we can define
\begin{eqnarray}
\hat\sigma_{x,\bm{p}}^c &= \hat c_{\bm{p}} + \hat c_{\bm{p}}^\dagger\\
\hat\sigma_{y,\bm{p}}^c &= i\left(\hat c_{\bm{p}} - \hat c_{\bm{p}}^\dagger\right)\\
\hat\sigma_{z,\bm{p}}^c &= -i \hat\sigma_{x,\bm{p}}^c \hat\sigma_{y,\bm{p}}^c = 2\hat c_{\bm{p}}^\dagger \hat c_{\bm{p}} - 1\ .
\end{eqnarray}
These operators constitute a Pauli algebra on the qubit Hilbert space $\mathcal{H}_{\bm{p}}^c$~. 
Note however, that the labels $x, y, z$ are simply notation, and not meant to indicate directions in physical space.
The Hamiltonian of the field can be expressed in terms of these operators as
\begin{eqnarray}
\hat H &= \sum_{\bm{p}} E_p \left\lbrace \left(\hat c_{\bm{p}}^\dagger \hat c_{\bm{p}} - \frac{1}{2}\right) + \left(\hat d_{\bm{p}}^\dagger \hat d_{\bm{p}} - \frac{1}{2}\right)\right\rbrace\nonumber \\
 &= \sum_{\bm{p}} \frac{E_p}{2} \left\lbrace \hat\sigma_{z,\bm{p}}^c + \hat\sigma_{z,\bm{p}}^d\right\rbrace\ .
\end{eqnarray}
So our field behaves like a set of non-interacting spins in a $\bm{p}$-dependent magnetic field. \change{Of course, the occupation of different Fourier modes $\bm{p}$ of the Weyl field does not measure the state of any actual spins, but rather the existence or non-existence of particles with momentum $\bm{p}\,$.}

\section{Fermion field with overlapping qubits}
\label{sec:JL_embedding}

\subsection{Chao-Reichert-Sutherland-Vidick embedding}
\label{sec:JL_embedding_basics}

We will now use the scheme of \cite{Chao2017} for embedding $N$ qubits into an Hilbert space of dimension $2^n < 2^N$ (Chao-Reichert-Sutherland-Vidick embedding, or CRSV-embedding for the rest of this paper) to create a version of the Weyl field whose effective degrees of freedom approximately follow an area law scaling. To describe our approach we will focus on the particle part of the field (the anti-particle part can be treated analogously). We start by covering the entirety of Fourier space with adjacent but non-overlapping shells $\lbrace s \rbrace$ (cf.\ our sketch in Figure~\ref{fi:sketch_with_tiles}), and we introduce the following notation.
\begin{itemize}
    \item $k_s\, ,\ \Delta_s$ :
    
    the central radius and width of Fourier space shell $s$.
    \item $N_s$ :\,
    
    number of qubits inside Fourier space shell $s$\ \ (i.e.\ the number of momenta $\bm{p} \in s$).
    \item $n_s$ :
    
    logarithm (to the base $2$) of the dimension into which the $N_s$ will be embedded.
\end{itemize}
For our regular grid in Fourier space we expect
\begin{eqnarray}
N_s\ \approx\ 4\pi k_s^2 \Delta_s\ \bigg/\ \frac{(2\pi)^3}{L^3}\ ,
\end{eqnarray}
where $k_s$ is the central radius of shell $s$. The scaling $N_s \propto k_s^2\Delta_s$ indicates a volume scaling of the independent degrees of freedom, but we will squeeze the qubits of shell $s$ into a Hilbert space that is too small for them to be independent. In particular we would like to obtain an area-scaling of the effective degrees of freedom, such that \cite{Cao2019_essay}
\begin{eqnarray}
\frac{n_s}{N_s} \propto \frac{k_s \Delta_s}{k_s^2\Delta_s} = \frac{1}{k_s}\ .
\end{eqnarray}
This will be achieved if
\begin{eqnarray}
n_s = B\cdot 2\pi k_s\Delta_s\ \bigg/\ \frac{(2\pi)^2}{L^2}\ ,
\end{eqnarray}
where $B$ is some constant, and where the factors of $L/2\pi$ are just for later convenience. Along the lines of \cite{Chao2017} we will now proceed as follows:
\begin{itemize}
    \item[A)] For each shell $s$ we choose a $2^{n_s}$-dimensional matrix representation of the Clifford algebra with generators $\bm{C}_1$, \dots , $\bm{C}_{2n_s}$~.
    \item[B)] In the space $\mathbb{R}^{2 n_s}$~, we draw $2N_s$ random vectors from a standard normal distribution.
    \item[C)] We collect these vectors into $N_s$ pairs (corresponding to the $N_s$ momenta $\bm{p}$ in the Fourier space shell $s$), and choose orthonormal bases $\lbrace \ket{v_{\bm{p}}}, \ket{w_{\bm{p}}} \rbrace$ for the $N_s$ 2-dimensional subspaces spanned by these vector pairs.
    \item[D)] For some orthonormal basis $\lbrace \ket{e_j} \rbrace$ of $\mathbb{R}^{2 n_s}$, we define
    \begin{eqnarray}
    \label{eq:qubit_definitions}
    \bm\sigma_{x,\bm{p}}^c \equiv \sum_{j=1}^{2n} \braket{e_j|v_{\bm{p}}} \bm{C}_j\ \ ,\ \ \bm\sigma_{y,\bm{p}}^c \equiv \sum_{j=1}^{2n} \braket{e_j|w_{\bm{p}}} \bm{C}_j\ \ ,\ \ \bm\sigma_{z,\bm{p}}^c &= -i \bm\sigma_{x,\bm{p}}^c \bm\sigma_{y,\bm{p}}^c\ ,\nonumber \\
    \end{eqnarray}
    and we set
    \begin{eqnarray}
    \bm{c}_{\bm{p}} = \frac{1}{2}\left(\bm\sigma_{x,\bm{p}}^c + i\bm\sigma_{y,\bm{p}}^c\right)\ .
    \end{eqnarray}
    (Note that this is slightly different from how \cite{Chao2017} had originally defined their embedding. The reason for this is that we want the creation and annihilation operators for different modes $\bm{p}_1,\,\bm{p}_2$ to approximately anti-commute, while \cite{Chao2017} wanted operators on different qubits to approximately commute.)
    \item[E)] The matrices $\bm{c}_{\bm{p}}^\dagger$ and $\bm{c}_{\bm{p}}$ can serve as creation and annihilation operators in the Hilbert space $\mathcal{H}_s^c$ corresponding to shell $s$. In principle we now only have to stitch together different shells to obtain the creation and annihilation operators $\hat{c}_{\bm{p}}^\dagger$ and $\hat{c}_{\bm{p}}$ in the whole Hilbert space
    \begin{equation}
        \mathcal{H}^c = \bigotimes_s \mathcal{H}_s^c\ .
    \end{equation}
    A naive attempt to do so would be to define $\hat{c}_{\bm{p}} = \left( \bigotimes_{\bm{p}\notin s} \mathbf{1}_s \right) \otimes \bm{c}_{\bm{p}}$~, i.e.\ to consider a tensor product of the operator $\bm{c}_{\bm{p}}$ (which is defined on the Hilbert space of the shell that contains $\bm{p}$) with the unit operators in all shells that do now contain $\bm{p}$. This naive approach would not work, because we need operators $\hat{c}_{\bm{p}}$ and $\hat{c}_{\bm{q}}$ that live on different shells to be anti-commuting (instead of commuting).

    To achieve anti-commutation between different shells, we need to insert the appropriate Jordan-Wigner strings into the above, naive construction. We sketch in \ref{app:JordanWigner} how this can be done in practice. But the details of this do not matter in the following, since they do not impact any anti-commutators that are relevant to our calculations.
    \item[F)] Finally, we duplicate the above construction for the $d$-particles and build the full Hilbert space as the tensor product of the spaces of particles and anti-particles (again, using an appropriate Jordan-Wigner construction to ensure anti-commutation between $c$- and $d$-operators).
\end{itemize}
Within the above construction the anti-commutator between the operators $\hat c_{\bm{p}}^\dagger$ and $\hat c_{\bm{q}}$, where $\bm{p}$ and $\bm{q}$ are in the same Fourier space shell $s$, becomes
\begin{eqnarray}
\lbrace \hat c_{\bm{p}}^\dagger , \hat c_{\bm{q}} \rbrace &= \frac{1}{4}\sum_{jl} \braket{v_{\bm{p}} + i w_{\bm{p}}|e_j}\braket{e_l|v_{\bm{q}} + i w_{\bm{q}}} \lbrace \bm{C}_j, \bm{C}_l\rbrace\nonumber \\
&= \frac{\braket{v_{\bm{p}}|v_{\bm{q}}}+\braket{w_{\bm{p}}|w_{\bm{q}}}+i\braket{v_{\bm{p}}|w_{\bm{q}}}-i\braket{w_{\bm{p}}|v_{\bm{q}}}}{2}\nonumber \\
&\equiv \frac{\braket{z_{\bm{p}}|z_{\bm{q}}}}{2}\ ,
\end{eqnarray}
where in the second line we have used that $\bra{v_{\bm{p}} + i w_{\bm{p}}} = \bra{v_{\bm{p}}} - i\bra{w_{\bm{p}}}\,$, and in the last line we have introduced the complex vectors $\ket{z_{\bm{p}}} = \ket{v_{\bm{p}}}+i\ket{w_{\bm{p}}}$~. Note that in general $\braket{z_{\bm{p}}|z_{\bm{q}}} \neq 0$ for $\bm{p}\neq\bm{q}$, because the vectors $\ket{v_{\bm{p}}}, \ket{w_{\bm{p}}}$ are more numerous than their dimension (and because we chose these vector pairs randomly). At the same time, it is easy to check that $\braket{z_{\bm{p}}|z_{\bm{p}}} = 2$ and hence $\lbrace \hat c_{\bm{p}}^\dagger , \hat c_{\bm{p}} \rbrace = 1$.

The vectors $\ket{v_{\bm{p}}}, \ket{w_{\bm{p}}}$ that are used in the above embedding procedure are chosen at random (cf.\ steps B and C). As a consequence, the anti-commutators $\lbrace \hat c_{\bm{p}}^\dagger , \hat c_{\bm{q}} \rbrace$ for $\bm{p}\neq\bm{q}$ become classical random variables. We demonstrate in \ref{app:random_angles} that, as a consequence of the Johnson-Lindenstrauss theorem (JL-theorem, \cite{JohnsonLindenstrauss1984, Engebretsen2002, Dasgupta2003, Chao2017, Burr2018}) the vectors $\lbrace \ket{v_{\bm{p}}}, \ket{w_{\bm{p}}} \rbrace$ can be drawn such that with a probability $> 1 - \delta$ \emph{all} anti-commutators in a shell $s$ simultaneously satisfy
\begin{eqnarray}
\label{eq:Chao_bound}
    |\lbrace \hat c_{\bm{p}}^\dagger , \hat c_{\bm{q}} \rbrace| &< \sqrt{\frac{16}{n_s} \ln\left(\frac{2 N_s}{\sqrt{\delta}}\right)}\nonumber \\
    &= \sqrt{\frac{32\pi \log\left(k_s^2 \Delta_s L^3 / \left(\pi^2\sqrt{\delta}\right)\right)}{k_s B L^2 \Delta_s}} \ .
\end{eqnarray}
To asses whether or not this is small, we need to specify values for $L$, $\Delta_s$ and $B$. Our fiducial choice for these parameters is explained in the following subsection.

\subsection{Fiducial parameters and appearence of a species scale}
\label{sec:Bekenstein_bound}

Our procedure to construct overlapping creation and annihilation operators for the Weyl field depends on the values of a number of parameters:
\begin{itemize}
    \item $\Delta_s$: the width of the Fourier space shells $\lbrace s \rbrace$;
    \item $B$: the parameter determining the dimension of the Hilbert space factor, into which the qubits in one Fourier space shell will be embedded;
    \item $L$: the IR scale that induces the discreteness in Fourier space;
    \item $\Lambda_{\mathrm{UV}}$: a UV-cutoff.
\end{itemize}
In the following, we will take the UV scale $\Lambda_{\mathrm{UV}}$ to be a free parameter of our construction. As we will see later, in order to explain the existence of a number of observed cosmic neutrino fluxes this scale needs to be significantly lower than the Planck energy scale $\Lambda_{\mathrm{Planck}}$. Generally, having a lower $\Lambda_{\mathrm{UV}}$ will decrease effects of holography (or rather: of the mode overlaps present in our construction) because is allows us to choose a higher value of $B$ (and thus have smaller mode overlaps) while still satisfying the Bekenstein bound on the total number of degrees of freedom present in our field. We will interpret $\Lambda_{\mathrm{UV}} < \Lambda_{\mathrm{Planck}}$ as an indication that effective field theories of elementary Fermions need to break down already at sub-Planckian energies.

To fix $L$, we adopt the approach of \cite{Friedrich2022} who chose their IR scale as the future co-moving particle horizon of a $\Lambda$CDM universe similar to our own Universe. This scale is finite in a $\Lambda$-dominated cosmos, and \change{it corresponds to the largest distance in today's Universe to which we will ever be causally connected. Assuming the best-fitting $\Lambda$CDM model of \cite{Planck2018_short} we obtain $L \approx 4.5\, c/H_0$, where $c/H_0$ is today's Hubble radius.}

For concreteness we will assume a constant relative shell width, i.e.\ $\Delta_s = \alpha k_s$ for some $\alpha \ll 1$~. This is a somewhat arbitrary modelling choice, and we will quote many of the results in later parts of this paper in terms of general shells widths $\Delta_s$ (or in terms of general values for $n_s$ and $N_s$) in order to keep these results accessible to generalisation. Generally, having wider shells leads to a stronger suppression of holographic effects. The reason for this can be found in the Johnson-Lindenstrauss theorem itself, which states that the qubit overlap $\epsilon$ can be chosen such that $\epsilon^2 \sim (\ln N_s)/ n_s \sim (\ln \Delta_s)/ \Delta_s$~, which decreases as $\Delta_s$ is increased. Demanding $\alpha \ll 1$ seems natural because is leads to an approximately smooth radial mode density function in Fourier space, and in our fiducial construction we will set $\alpha = 0.01\,$.

Finally, we can motivate a choice for $B$ as follows. The overall number of effective qubits in our construction is given by
\begin{eqnarray}
\label{eq:N_eff}
    N_{\mathrm{eff}}\ =\ 2\sum_{\mathrm{shell}\ s} n_{s}\ =\ \frac{BL^2}{\pi} \sum_{\mathrm{shell}\ s} k_s \Delta_s\  &\approx\ \frac{BL^2}{\pi} \int_0^{\Lambda_{\mathrm{UV}}} \dd k\ k\nonumber \\
    &=\ \frac{B(L\Lambda_{\mathrm{UV}})^2}{2\pi}\ ,
\end{eqnarray}
where the factor of $2$ in the first line takes into account that there are both particles and anti-particles. At the same time the IR scale $L$ will correspond to a real-space boundary area $A_{\mathrm{boundary}} \sim 4\pi L^2\,$. The Bekenstein bound on the maximum entropy attainable within such a boundary should then lead to
\begin{eqnarray}
\label{eq:B_boundary}
    \frac{B(L\Lambda_{\mathrm{UV}})^2}{2\pi} &\lesssim \pi (L\Lambda_{\mathrm{Planck}})^2\\
\label{eq:B_boundary_2}
    \Rightarrow B &\lesssim 2\pi^2 \left(\frac{\Lambda_{\mathrm{Planck}}}{\Lambda_{\mathrm{UV}}}\right)^2\ .
\end{eqnarray}
Note again that a lower UV-scale $\Lambda_{\mathrm{UV}}$ allows us to choose a higher value for $B$. Since in turn the qubit overlaps scale as $\epsilon \sim 1/\sqrt{B}\ $ this means that holographic effects will be less pronounced. If our field $\psi$ constituted the only degrees of freedom in the Universe, then the above bound would be saturated, such that $B\approx 2\pi^2 \left(\Lambda_{\mathrm{Planck}}/\Lambda_{\mathrm{UV}}\right)^2$ and $n_s \approx \pi L^2 k_s \Delta_s \left(\Lambda_{\mathrm{Planck}}/\Lambda_{\mathrm{UV}}\right)^2$. We will instead take into account the possibility (or rather: certainty) that there are other quantum fields in the Universe. Assuming that the degrees of freedom of all types of fields (including spacetime degrees of freedom) individually satisfy an area law scaling leads to
\begin{eqnarray}
\label{eq:B_boundary_full}
    B\, \approx\, 2\pi^2\, \left(\frac{\Lambda_{\mathrm{Planck}}}{\Lambda_{\mathrm{UV}}}\right)^2\, \frac{N_{\mathrm{dof},\psi}}{N_{\mathrm{dof,total}}}\ ,
\end{eqnarray}
where $N_{\mathrm{dof,total}}$ are the total number of degrees of freedom present in the Universe and $N_{\mathrm{dof},\psi}$ are the degrees of freedom contributed by the field $\psi$~. \change{Note that the combination $\Lambda_{\mathrm{Planck}} \sqrt{N_{\mathrm{dof},\psi}/N_{\mathrm{dof,total}}}$ appearing in \eqnref{B_boundary_full} is reminiscent of the so called \emph{species scale} $\Lambda_{\mathrm{Planck}} / \sqrt{N_{\mathrm{species}}}\,$, where $N_{\mathrm{species}}$ is the number of particle species present in the effective field theory of our Universe. Beyond this scale the individual particle species cannot be resolved anymore and quantum gravitational effects are expected to become relevant, such that the species scale can be thought of as the maximal scale of validity of the EFT  \cite{Veneziano2002, Han_and_Willenbrock_2005, Dvali2008, Dvali2010, Dvali2013, Castellano2022, Castellano2023}. As we explain in \secref{N_dof_total}, $\Lambda_{\mathrm{Planck}} \sqrt{N_{\mathrm{dof},\psi}/N_{\mathrm{dof,total}}}$ is not perfectly analogous to usual definitions of the species scale, because Bosonic and Fermionic species contribute differently to $N_{\mathrm{dof,total}}$. But in order to draw from the intuition built around the species scale (and in order to simplify our expressions) we nevertheless define the scale
\begin{equation}
\label{eq:species_scale}
    \Lambda_{\mathrm{sp}} \equiv \Lambda_{\mathrm{Planck}} \sqrt{N_{\mathrm{dof},\psi}/N_{\mathrm{dof,total}}}\ .
\end{equation}
Note especially, that throughout the rest of our paper $\Lambda_{\mathrm{Planck}}\,$, $N_{\mathrm{dof},\psi}$ and $N_{\mathrm{dof,total}}$ always appear in the above combination. For example, \eqnref{B_boundary_full} then simplifies to
\begin{eqnarray}
\label{eq:B_boundary_species}
    B\, \approx\, 2\pi^2\, \left(\frac{\Lambda_{\mathrm{sp}}}{\Lambda_{\mathrm{UV}}}\right)^2\ .
\end{eqnarray}}
With our above choices for $\Delta_s$ and $B$ the bound on qubit overlap from Equation~\ref{eq:Chao_bound} now becomes
\begin{eqnarray}
\label{eq:Chao_bound_2}
    |\lbrace \hat c_{\bm{p}}^\dagger , \hat c_{\bm{q}} \rbrace| &<\, \sqrt{\frac{32\pi \log\left(k_s^2 \Delta_s L^3 / \left(\pi^2\sqrt{\delta}\right)\right)}{k_s B L^2 \Delta_s}}\nonumber \\
    &=\, \sqrt{\frac{16\left[3\ln\left(k_s L\right) + \ln\left(\alpha / \pi^2\sqrt{\delta}\right)\right]}{\alpha \pi (k_sL)^2}}\, \frac{\Lambda_{\mathrm{UV}}}{\Lambda_{\mathrm{sp}}}\nonumber \\
    &\approx\, \sqrt{\frac{48 \log\left(k_s L\right)}{\alpha \pi (k_sL)^2}}\, \frac{\Lambda_{\mathrm{UV}}}{\Lambda_{\mathrm{sp}}}\ \ \ \ \mathrm{for}\ k_s \gg L\ .
\end{eqnarray}
As we already noted in Section~\ref{sec:intro}, this bound is decreasing with increasing wave number $k$~. This means that at higher energies the overlap between different modes becomes more and more negligible. In Figure~\ref{fi:epsilon_of_k} we display the bound \change{for different values of the ratio $\Lambda_{\mathrm{UV}}\ /\ \Lambda_{\mathrm{sp}}$ in Figure~\ref{fi:epsilon_of_k}, fixing $\alpha=0.01\,$.} At everyday scales $k \sim \mathrm{cm}^{-1}$ the anti-commutators $\lbrace \hat c_{\bm{p}}^\dagger , \hat c_{\bm{q}} \rbrace$ have absolute values $\lesssim 10^{-27}$, and at energies probed by the Large Hadron Collider they drop to $\lesssim 10^{-43}$, indicating that any pair of Fourier modes in this energy range behaves as almost perfectly independent degrees of freedom. 

\change{At energy scales $k_s\sim \Lambda_{\mathrm{UV}}$ the above bound on $|\lbrace \hat c_{\bm{p}}^\dagger , \hat c_{\bm{q}} \rbrace|$ is suppressed by a factor $\sim 1/(\Lambda_{\mathrm{sp}}L)\,$. This may indicate that $\Lambda_{\mathrm{sp}}L$ plays a similar role as the parameter $N$ in AdS/CFT implementations of holography (\ie as the size $N$ of the gauge group of the boundary CFT). Note however that even below energies $\sim \Lambda_{\mathrm{UV}}$ the typical anti-commutator $\lbrace \hat c_{\bm{p}}^\dagger , \hat c_{\bm{q}} \rbrace$ remains small (compared to the operator norms of $c_{\bm{p}}^\dagger ,\, \hat c_{\bm{q}}$) as long as $k_sL \gg 1\,$.} In general, one can interpret the suppression of mode overlaps displayed in Figure~\ref{fi:epsilon_of_k} as a consequence of the JL-theorem: At fixed $\epsilon$ it predicts an exponential scaling $N_s \sim \exp[n_s\epsilon^2]$, but in order to achieve holography we only need polynomial scaling between $N_s$ and $n_s$! This allows us to choose a running $\epsilon$ instead, that quickly decreases with increasing energy. \change{The decrease in $\epsilon$ needed to achieve holography can also be heuristically understood using intuition from AdS/CFT\footnote{We thank an anonymous referee for this interpretation.}: High energy modes have smaller wavelengths, and can therefore fit into smaller entanglement wedges if pushed towards the AdS/CFT boundary. Thus there will be fewer entanglement wedges with which such modes have significant overlap, so that their non-locality is suppressed. In particular, modes near the boundary are quasi-local in AdS/CFT. Of course, one caveat to this explanation is the fact that we never actually define a boundary theory (though we potentially get close to it, \cf the discussion below as well as in \secref{discussion}). Hence, it is not entirely evident whether our model should benefit from such an AdS/CFT-inspired reasoning.}

\change{Note that the above construction of an approximately holographic Weyl field with overlapping degrees of freedom explicitly reduces the Hilbert space dimension of the field beyond just the usual reduction caused by imposing UV and IR cutoffs to a field theory. We thus do not expect that the phenomenology we derive in later sections of this paper can be fully captured by a standard, 3+1 dimensional effective field theory. In \ref{app:Heisenberg} we re-formulate our model in terms of non-overlapping degrees of freedom (whose number follows an area law by construction) and we explicitly derive the Hamiltonian of that dual model (assuming that the Hamiltonian in terms of overlapping degrees of freedom still has the standard Weyl form). That dual Hamiltonian turns out to be highly non-local in the new degrees of freedom. We think that this suggests the following viewpoint: a 2+1 dimensional theory that is highly non-local can be interpreted as a close-to-local 3+1 dimensional field theory, if we allow the mode operators of the latter to overlap slightly (see also the corresponding discussion in \secref{mereology}).}

\subsection{Degrees of freedom of other quantum fields and the value of \texorpdfstring{$\Lambda_{\mathrm{sp}}$}{}}
\label{sec:N_dof_total}

The standard model has 90 Fermionic and 28 Bosonic field degrees of freedom \cite{Husdal2016}. Neglecting Bosonic fields, and assuming that all Fermion degrees of freedom behave in a way similar to our holographic Weyl field, it seems that $N_{\mathrm{dof},\psi}\ /\ N_{\mathrm{dof,total}}\ <\ 1/90\,$. Of course, Bosonic degrees of freedom will account for vastly more effective qubit degrees of freedom, so a much smaller ratio $N_{\mathrm{dof},\psi}\ /\ N_{\mathrm{dof,total}}$ may seem likely. Technically, Bosons should contribute infinitely many qubits, but in the light of the holographic principle it has been argued that this is unphysical \cite{Bao2017, Cao2019_essay, Friedrich2022}. Let us assume that Bosonic degrees of freedom are approximated in the real Universe with generalised Pauli operators \cite{Singh2018, Friedrich2022} of dimension $\sim 1000\,$. This number is admittedly ad hoc, but the number of qubits contributed by such approximate Bosonic degrees of freedom depends only logarithmically on the dimension of the generalised Pauli operators. We would then expect $N_{\mathrm{dof},\psi}\ /\ N_{\mathrm{dof,total}}\ <\ 1/(90 + 28\log_2 1000)\ \approx\ 0.003\,$.

So, if we attempt to model \eg a neutrino of the standard model with our holographic version of the Weyl field, then it seems that we should use $N_{\mathrm{dof},\psi}\ /\ N_{\mathrm{dof,total}} \lesssim 0.003\,$. This however assumes that all the different fields in the standard model are independent and that there is no overlap between those fields \change{(\ie that field operators between different species perfectly commute / anti-commute)}. We consider this unlikely, and thus take the somewhat larger value of
\begin{eqnarray}
    N_{\mathrm{dof},\psi}\ /\ N_{\mathrm{dof,total}} \approx 0.01
\end{eqnarray}
to be our fiducial choice in the rest of this paper. \change{Correspondingly, our fiducial species scale will be
\begin{eqnarray}
\label{eq:Lambda_S_repeat}
    \Lambda_{\mathrm{sp}} = \sqrt{\frac{N_{\mathrm{dof},\psi}}{N_{\mathrm{dof,total}}}}\ \Lambda_{\mathrm{Planck}} \approx 0.1\, \Lambda_{\mathrm{Planck}}\ .
\end{eqnarray}
This choice for $\Lambda_{\mathrm{sp}}$ differs from the species scale that is e.g.\ considered by \cite{Dvali2010, Dvali2013, Castellano2022, Castellano2023} for two reasons. First, we only take into account standard model fields whereas string theory approaches to quantum gravity can predict significantly more particle species. Note however that the number of EFT-fields emerging from string theory doesn't seem to be as large as originally thought \cite{Dvali2010b, Dvali2010c} and is in fact bounded by $1/g_s^2$, where $g_s$ is the string coupling. The value of $g_s$ that is realised in our Universe is as of yet unknown, and we will thus stick to the known standard model degrees of freedom.}

\change{A second reason why our definition of $\Lambda_{\mathrm{sp}}$ is unusual is the fact that the ratio $N_{\mathrm{dof,total}}\ /\ N_{\mathrm{dof},\psi}$ does not simply count the number of particle species, but in fact up-weighs Bosonic degrees of freedom by $\ln d\,$, where $d$ is the dimension of the Hilbert space in which the Bosonic field modes are approximated (e.g.\ via generalised Pauli operators \cite{Singh2018, Friedrich2022}; see the explanations at the beginning of this sub-section). This will however only have an effect of $\lesssim\ln d$ on $\Lambda_{\mathrm{sp}}\,$, so we will interpret our $\Lambda_{\mathrm{sp}}$ as \emph{the} species scale for the remainder of this paper.}

\subsection{Breaking of Lorentz symmetry}
\label{sec:Lorentz}

\noindent Our construction breaks Lorentz invariance for a number of reasons:
\begin{itemize}
    \item[A)] we assume hard UV and IR cutoffs;
    \item[B)] we have employed a naive equal-time version of holography instead of imposing area scaling on a light-sheet;
    \item[C)] we have used non-local mode overlaps to build our holographic Fermion field.
\end{itemize}
We leave it to future work to address point B) by \eg quantizing our field directly on a cosmic light-sheet and then imposing holographic mode overlaps on that sheet. This may also address our need for an IR-cut, since the light-sheet naturally ends at the cosmic horizon.

In \secref{lifetime} we show that it is a consequence of point C) that plane waves in our field have a finite lifetime. In particular, we show that over time the power in a wave with wave vector $\bm{p}$ is isotropically scrambled to other modes $\bm{q}$ of similar absolute wave number $|\bm{q}| \approx |\bm{p}|\,$. Such a breaking of Lorentz symmetry is likely related to ideas of modified dispersion relations for relativistic particles, which have been proposed as a generic feature of quantum gravity before (\eg \cite{Amelino1998, Amelino2023}). At the same time, we demonstrate that the plane wave lifetime can be made cosmologically large by choosing a UV-cut that is far below the Planck scale, but still reasonably above the energies of current particle physics experiments.

This turns point A) into the main cause of our Lorentz symmetry breaking. It has been argued that quantum fields can be UV-regularized in a Lorentz invariant manner \cite{tHooft:1972tcz, Martin2012} via dimensional regularisation. But this regularisation scheme has also been criticised as being a mere mathematical trick that lacks concrete implementations \cite{Mathur2020}. Instead of assuming a sharp cutoff at $\Lambda_{\mathrm{UV}}$ we could have modelled our UV-cut as a transition from a holographic mode density below $\Lambda_{\mathrm{UV}}$ to a \say{super-holographic} mode density above $\Lambda_{\mathrm{UV}}\,$. This means that instead of squeezing the $\sim k_s^2 \Delta_s$ modes of Fourier space shell $s$ into $\sim k_s \Delta_s$ physical modes one would impose a stronger squeezing when $k_s > \Lambda_{\mathrm{UV}}\,$. With a sufficiently strong super-holographic squeezing, this may still satisfy the cosmic Bekenstein bound with a similar value for $\Lambda_{\mathrm{UV}}\,$.

We leave it to future work to investigate the extent to which such a regularisation scheme together with a light-sheet quantisation would restore Lorentz symmetry. \change{But it is already clear, that even with such a revised model the restoration of Lorentz symmetry will not be perfect. And as soon as we try to describe several, interacting particle species within our framework this residual breaking of Lorentz invariance may be particularly problematic. It has been demonstrated by \cite{Collins2004} that standard schemes to regularise interacting quantum field theories lead to unacceptable Lorentz breaking effects at low energies, even if the regularisation scales are chosen at unobservably high energies. In single species QFTs such a low-energy Lorentz breaking can be absorbed into a renormalised speed of light, but as soon as multiple fields are interacting, a preservation of low-energy Lorentz invariance seems to require either supersymmetry or an unnaturally fine tuning of how different fields are regularised \cite{Cortes2017}. The authors of \cite{Gambini2011} have argued that these findings can be circumvented in non-perturbative approaches to QFT / quantum gravity, but \cite{Polchinski2012} has objected that their argument only applies in non-generic situations and relies on Euclidean symmetries that are not present in Lorentzian field theory. In particular, according to \cite{Polchinski2012} the need for fine tuning remains in more general scenarios. It is at the moment unclear, to which degree the above discussion regarding Lorentz-violating regularisation schemes maps onto our construction based on overlapping degree of freedom. But we consider it likely, that similar arguments apply to our situation, at least as soon as we consider interactions between different particle species.}

We would like to stress that there are reasons to believe that the breaking of Lorentz symmetry is physical \cite{Gijosa2004, Bao2017, Mathur2020}. In particular, if spacetime and local degrees of freedom emerge as approximations to a more abstract class of quantum theories - as is e.g.\ assumed by the \emph{quantum mereology program} \cite{Cao2017, Carroll_Singh_2018, CaoCarroll2018, Carroll_Singh2020} - then also the symmetries of the emergent spacetime should be approximate. \change{If Lorentz invariance is indeed broken, then \cite{Cortes2017} suggest that the requirement of fine tuning should not be viewed as an obstacle, but rather as a guide for finding the correct UV-theory. For the mereology program this would mean that the fine tuning needed to approximate Lorentz symmetry at low energies should be implicit in the algorithm that extracts local, quasi-classical degrees of freedom from abstract classes of quantum theories. In \secref{mereology} we describe the mereology program in more detail and we discuss the extent to which our model fits into that framework. In particular, we argue that the emergence of quasi-classical degrees of freedoms from abstract Hamiltonian ensembles would give rise to stochastic features like those present in our model via the random mode overlaps. Modifying previous attempts at mereology to include Lorentz-preserving fine tunings is however beyond the scope of this work.}

\change{Finally, note that models of quantum gravity around an AdS background seem to strongly respect Lorentz invariance (see e.g.\ \cite{Herderschee_Maldacena_2024}) while at the same time implementing holography via the AdS/CFT correspondence. To see whether our construction can recover this result we would have to extend it to AdS backgrounds and in particular to hyperbolic geometry. This would significantly alter Equations~\ref{eq:N_eff}-\ref{eq:B_boundary_2}, and we leave the task of studying such modifications for future work.}


\section{Dynamics of the holographic Weyl field}
\label{sec:dynamics}

Let us define
\begin{eqnarray}
\hat \psi^c(x) &= \sum_{\bm{p}} \frac{1}{(|\bm{p}|V)^{\frac{1}{2}}}\ \hat c_{\bm{p}}(t)\ u(\bm{p})\ e^{i\bm{p}\bm{x}}\ \ ;\nonumber \\ 
\hat \psi^d(x) &= \sum_{\bm{p}}\frac{1}{(|\bm{p}|V)^{\frac{1}{2}}}\ \hat d_{\bm{p}}^\dagger(t)\ u(\bm{p})\ e^{-i\bm{p}\bm{x}}\ ,
\end{eqnarray}
where we have expanded the field in terms of \emph{time dependent} ladder operators $\hat c_{\bm{p}}(t)$ and $\hat d_{\bm{p}}^\dagger(t)$ instead of writing explicitly what this time dependence is. This is necessary because our use of overlapping qubits may change time evolution and we may in general have
\begin{eqnarray}
    \frac{\dd \hat c_{\bm{p}}}{\dd t} \neq -i |\bm{p}| \hat c_{\bm{p}}\ \ ,\ \ \frac{\dd \hat d_{\bm{p}}^\dagger}{\dd t} \neq i |\bm{p}| \hat d_{\bm{p}}\ .
\end{eqnarray}
We will investigate one particular choice for the Hamiltonian operator of our Fermion, namely we will assume that it remains of the form
\begin{eqnarray}
\hat H &= \sum_{\bm{p}} |\bm{p}| \left\lbrace \left(\hat c_{\bm{p}}^\dagger \hat c_{\bm{p}} - \frac{1}{2}\right) + \left(\hat d_{\bm{p}}^\dagger \hat d_{\bm{p}} - \frac{1}{2}\right)\right\rbrace\nonumber \\
 &\approx \sum_{s} \frac{k_s}{2}  \sum_{\bm{p}\in s} \left\lbrace \hat\sigma_{z,\bm{p}}^c + \hat\sigma_{z,\bm{p}}^d\right\rbrace\nonumber \\
 &\equiv \sum_{s} \hat H_s\ ,
\end{eqnarray}
where in the second line we have made explicit that the field is a sum over independent (non-overlapping) Fourier space shells $s$, and where we assumed that all momenta $\bm{p}$ in one shell $s$ have similar absolute values $|\bm{p}| \approx k_s$~. Because of the overlap between the operators $\hat c_{\bm{p}},\, \hat d_{\bm{p}}$ in one shell, the spectrum of this operator is different from the energy spectrum of a standard Weyl field. Also, the time evolution induced by this Hamiltonian will differ from the standard, non-overlapping case. In the following, we study instances of both of these modifications.

\subsection{Energy spectrum and re-formulation as non-local Heisenberg model}
\label{sec:E_spectrum}

In \secref{JL_embedding} we defined the operator algebra representing an individual Fourier mode $\bm{p}$ in a Fourier space shell $s$ of the particle sector of our field as
\begin{eqnarray}
    \bm\sigma_{x,\bm{p}}^c \equiv \sum_{j=1}^{2n} \braket{e_j|v_{\bm{p}}} \bm{C}_j\ \ ,\ \ \bm\sigma_{y,\bm{p}}^c \equiv \sum_{j=1}^{2n} \braket{e_j|w_{\bm{p}}} \bm{C}_j\ \ ,\ \ \bm\sigma_{z,\bm{p}}^c &= -i \bm\sigma_{x,\bm{p}}^c \bm\sigma_{y,\bm{p}}^c\ .\nonumber
\end{eqnarray}
From these operators we could then define annihilation operators as
\begin{eqnarray}
    \bm{c}_{\bm{p}} = \frac{1}{2}\left(\bm\sigma_{x,\bm{p}}^c + i\bm\sigma_{y,\bm{p}}^c\right)\ .\nonumber
\end{eqnarray}
This way we constructed a potentially large number of overlapping qubits to make up our holographic Weyl field. But since the Hilbert space of Fourier space shell $s$ of our field has dimension $2^{n_s}$, we can also decompose our construction into a set of non-overlapping qubits. Let us e.g.\ define the creation and annihilation operators of these qubits as
\begin{eqnarray}
    \hat{A}_j \equiv \frac{1}{2}\left(\bm{C}_{2j-1} + i\bm{C}_{2j}\right)\ ,\ \hat{A}_j^\dagger \equiv \frac{1}{2}\left(\bm{C}_{2j-1} - i\bm{C}_{2j}\right)\nonumber\\
    \Rightarrow \bm{C}_{2j-1} = \hat{A}_j + \hat{A}_j^\dagger\ ,\ \bm{C}_{2j} = -i(\hat{A}_j - \hat{A}_j^\dagger)\ .\nonumber
\end{eqnarray}
In \ref{app:Heisenberg} we show that with these non-overlapping creation and annihilation operators the Hamiltonian $\hat H_s$ in each Fourier space shell can be re-written as
\begin{eqnarray}
\label{eq:Heisenberg_Hamiltonian_main_1}
    \hat H_s &= \frac{k_s}{2} \left(\hat{A}_1^\dagger,\ \dots\ ,\ \hat{A}_n^\dagger,\ \hat{A}_n,\ \dots\ ,\ \hat{A}_1\right) \bm{K} \pmatrix{\hat{A}_1 \cr \dots \cr \hat{A}_n \cr \hat{A}_n^\dagger \cr \dots \cr \hat{A}_1^\dagger}\ ,
\end{eqnarray}
where $\bm{K}$ is a stochastic coefficient matrix which can be characterised as a generalised Wigner matrix (with the additional property of anti-persymmetry, cf. \ref{app:persymmetric_matrices}). \eqnref{Heisenberg_Hamiltonian_main_1} can be considered as defining a non-local Heisenberg model with general 2-site interactions. In \ref{app:energy_spectrum} we show that the eigenspectrum $\lbrace \lambda_H \rbrace$ of this Hamiltonian can be deduced from the positive half of the eigenvalues $\lbrace \lambda_K \rbrace$ of that coefficient matrix via
\begin{eqnarray}
    \lbrace \lambda_H \rbrace &= \left\lbrace \left.\frac{k_s}{2} \sum_{i=1}^{n_s} \lambda_{K,i}\, s_i\ \right|\ \lbrace s_i \rbrace \in \lbrace -1, +1\rbrace^{n_s} \right\rbrace \ ,
\end{eqnarray}
i.e.\ the eigenvalues of $\hat H_s$ are proportional to sums over the positive eigenvalues of $\bm{K}$ with all possible sets of prefactors $\pm 1$. The reason why only positive eigenvalues of $\bm{K}$ appear in this result is that all eigenvalues $\lambda$ of $\bm{K}$ come in pairs $\pm \lambda$ (because of the above mentioned anti-persymmetry, cf. \ref{app:persymmetric_matrices}).

Up to one additional symmetry, the coefficient matrix $\bm{K}$ is a generalised Wigner matrix, and the eigenvalues of such matrix ensembles are very well understood (see e.g.\ \cite{Dasgupta2003, Erdos2010a, Erdos2010b}). In particular, a rigidity property derived by \cite{Erdos2010b} states that in the limit $n_s \rightarrow \infty$ the $i$th of these eigenvalues will be given with increasing accuracy by the $i/n_s$-quantile of a Wigner semi-circle distribution. In \ref{app:energy_spectrum} we use this result to derive that the vacuum energy in a Fourier space shell $s$ as a function of $k_s$, $N_s$ and $n_s$ is given by
\begin{eqnarray}
\label{eq:minimum_E_main}
    E_{\min , s} &\approx -\frac{N_s k_s}{2}\cdot\left(\frac{8}{3\pi} \sqrt{\frac{\displaystyle n_s}{\displaystyle N_s}}\right)\ .
\end{eqnarray}
In \figref{Evac_suppression} we have used simulated realisations of overlapping qubits to show that this result is already accurate for very low numbers of field modes (cf.\ \ref{app:simulations} for a description of those simulations). At energies (and mode numbers) relevant for high-energy physics, it will be even more accurate because of the asymptotic nature of random matrix theory results \cite{Erdos2010a}. Equation~\ref{eq:minimum_E_main} result represents a strong suppression \wrt the non-overlapping Weyl field, whose vacuum energy in each shell would simply be $-N_s k_s/2\, $. When choosing $\Lambda_{\mathrm{UV}} = \Lambda_{\mathrm{Planck}}$ the total vacuum energy of the field is lowered by a factor of $\sim 10^{-30}$ (which is however not enough to alleviate the cosmological constant problem).

\subsection{Lifetime of plane waves}
\label{sec:lifetime}

\begin{figure}
\centering
\includegraphics[width=0.8\textwidth]{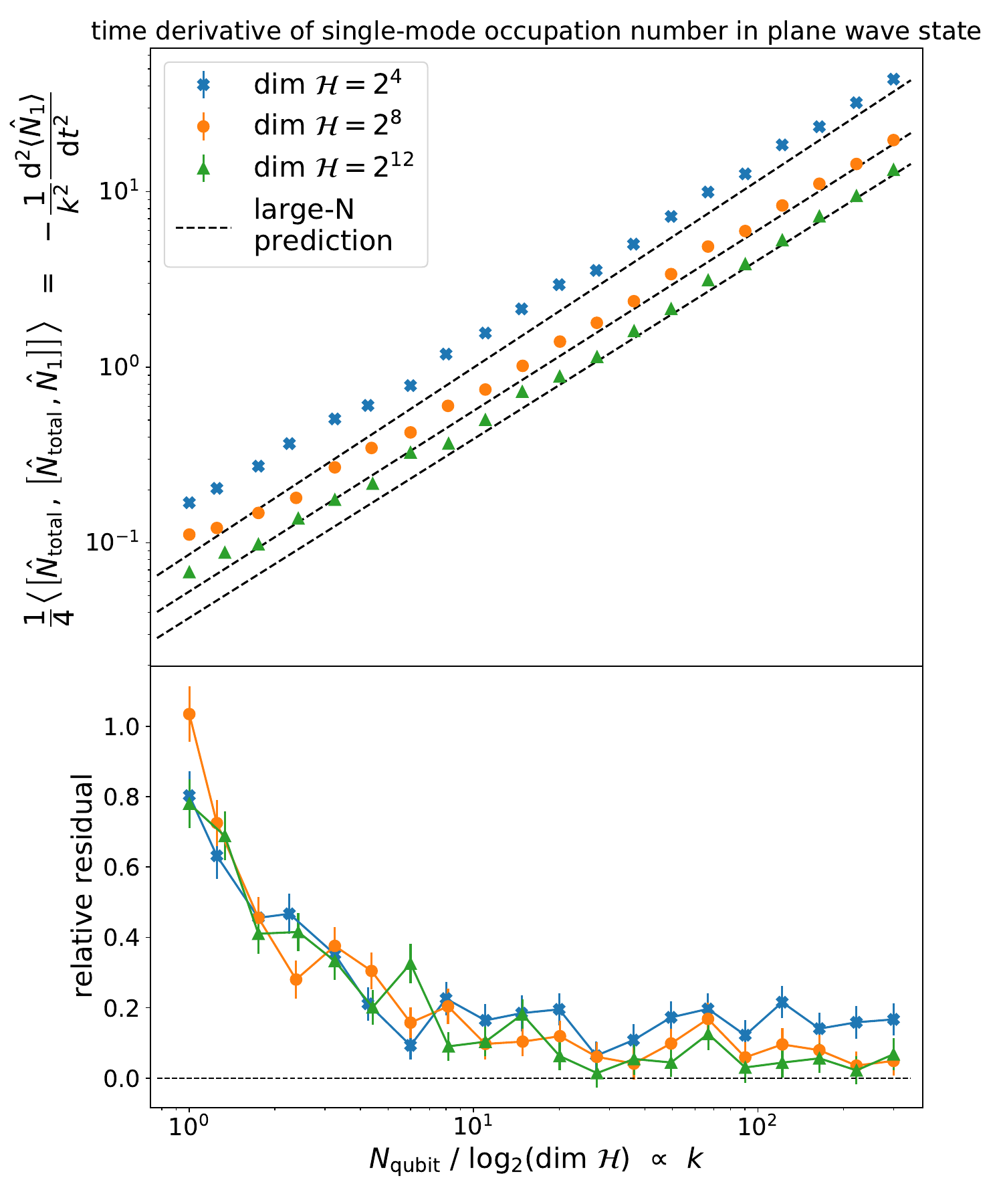}
   \caption{Comparing our analytical predictions for the lifetime of plane waves to results obtained from simulated realisations of low numbers of overlapping qubits. The upper panel displays the expectation value of the bi-commutator of the occupation number operator $\hat{N}_1$, representing e.g.\ a Fourier mode $\bm{k}_1$, with the total occupation number operator $\hat{N}_{\mathrm{total}}$ in the Fourier space shell to which $\bm{k}_1$ belongs, as a function of the ratio $N_s/n_s$ (which is proportional to the shell radius $k_s$). This bi-commutator is proportional to the second time derivative of the expectation value of $\hat{N}_1$, and can hence be used to define a characteristic lifetime for the plane wave state $\ket{\bm{k}_1}$, as we have done in \secref{lifetime}. The expectation values are taken \wrt the state $\ket{\bm{k}_1}$ and \wrt the CRSV-embedding.  Symbols represent results obtained from our simulations (\cf \ref{app:simulations}) and the dashed lines are the asymptotic results obtained in \ref{app:lifetime}. The lower panel shows relative residuals between the two.}
\label{fi:d2N_dt2}
\end{figure}

To investigate the impact of holography on the dynamics of our field further, we study the behaviour of plane wave states. Unfortunately, the vacuum state of our Hamiltonian is not an eigenstate of the occupation number operators $\hat N_{\bm{p}}$ of any of our overlapping Fourier modes $\bm{p}$. As a consequence, we cannot create a state in which a single mode is excited by acting on the vacuum state with a creation operator $\hat c_{\bm{p}}^\dagger\, $. Instead, we proceed as follows to define a state which approximates a plane wave excitation of our field with energy $\bm{p}$:
\begin{itemize}
    \item[A)] First we consider the subspace of states $\ket{\phi}$ with $\hat N_{\bm{p}}\ket{\phi} = +\ket{\phi}$. These are the states in which we would measure the occupation number of $\bm{p}$ to be 1.
    \item[B)] In this subspace, we find the states that minimize \smash{$\left|\frac{\mathrm{d}^2}{\mathrm{d}t^2} \bra{\phi}\hat N_{\bm{p}}\ket{\phi}\right|$}. These plane wave candidate states will be most stable in time. (Note that $\frac{\mathrm{d}}{\mathrm{d}t} \bra{\phi}\hat N_{\bm{p}}\ket{\phi} \equiv 0$ for all eigenstates of $\hat N_{\bm{p}}$.)
    \item[C)] Condition B) is still satisfied by an entire subspace of states (\cf \ref{app:lifetime}), and our final selection step is to choose the candidate state that has the lowest energy expectation value $\bra{\phi}\hat H\ket{\phi}$~.
\end{itemize}
We denote the state defined through conditions A) - C) with $\ket{\bm{p}}$ and we consider it as the closest equivalent our theory has to the standard plane wave state $\hat c_{\bm{p}}^\dagger \ket{0}$ for the non-overlapping Weyl field. We then estimate a characteristic time scale for the stability of this plane wave via the equation
\begin{eqnarray}
    \frac{1}{T_{\mathrm{scramble}}^2} \equiv -\mathbb{E}\left\lbrace\frac{\mathrm{d}^2}{\mathrm{d}t^2}\bra{\psi_{\mathbf{p}}} \hat{N}_{\mathbf{p}} \ket{\psi_{\mathbf{p}}}\right\rbrace\ .
\end{eqnarray}
Here the expectation value $\mathbb{E}\lbrace \cdot\rbrace$ is with respect to the random vectors $\bm{v}_{\bm{p}}$ and $\bm{w}_{\bm{p}}$ that were used to define the Pauli algebra of the mode $\bm{p}$ in the JLC-embedding (\cf \secref{JL_embedding}). Note that the different Fourier space shells in our construction do not overlap. Hence, the total occupation number within each individual shell is conserved, and the power that is lost in the mode $\bm{p}$ over time is isotropically scrambled to other modes in the same Fourier space shell $s$. In \ref{app:lifetime} we derive analytically that $T_{\mathrm{scramble}}$ for the mode $\bm{p}$ is given in terms of the parameters of our construction by
\begin{eqnarray}
\label{eq:T_scrample_main_1}
    T_{\mathrm{scramble}} \approx 2\pi^2 \sqrt{\alpha} \left(\frac{\Lambda_{\mathrm{sp}}}{\Lambda_{\mathrm{UV}}}\right)^2 \sqrt{\frac{L}{k_s}}\ ,
\end{eqnarray}
where the quantities appearing in this equation where introduced in \secref{JL_embedding}, and represent
\begin{itemize}
    \item \underline{$\alpha$:} relative width of the Fourier space shell $s$,
    \item \underline{$\Lambda_{\mathrm{UV}}/\Lambda_{\mathrm{sp}}$:} 
    
    UV-cut of our field in units of the species scale $\Lambda_{\mathrm{sp}} \approx 10^{-1}\, \Lambda_{\mathrm{Planck}}$ (cf.\ \secref{N_dof_total}),
    \item \underline{$L$:} IR-cutoff,
    \item \underline{$k_s$:} central radius of the Fourier space shell $s$.
\end{itemize}
Our analytical results are approximations that hold in the limit $n_s \rightarrow \infty$, $N_s \rightarrow \infty$. In \figref{d2N_dt2} we compare those analytical predictions to results obtained from low-dimensional simulations of overlapping qubits (\cf \ref{app:simulations}), and we find that they are already accurate for quite low numbers of qubits - see e.g.\ the green triangles in \figref{d2N_dt2}, which were obtained from simulating only $12$ overlapping qubits.

\eqnref{T_scrample_main_1} is the relation we have used to derive our results in Table~\ref{tab:UV_cuts} and \figref{L_scramble_vs_k}. In \figref{L_scramble_vs_k} we show $cT_{\mathrm{scramble}}$ for different UV-cuts. In that figure we also indicate the typical energies and travel distances of neutrinos emitted by a number of identified and resolved sources (see also Table~\ref{tab:UV_cuts} for a summary). The existence of solar neutrinos seems to demand that neutrino effective field theory already breaks down at $\Lambda_{\mathrm{UV}} \sim 10^{-3}\, \Lambda_{\mathrm{sp}}\ (\approx 10^{-4}\,\Lambda_{\mathrm{Planck}})$. Cosmic neutrino observations are even more constraining: the observation of a $\sim 290\,$TeV neutrino emitted from the blazar TXS 0506+056 \cite{blazarIceCube} requires $\Lambda_{\mathrm{UV}} < 470\, \Lambda_{\mathrm{LHC}}\,$, i.e. only a few orders of magnitude above modern particle physics experiments. As we discussed in \secref{Lorentz}, the breakdown of our effective theory beyond that energy should happen either as a sharp cutoff, or as a transition to super-holographic mode overlaps for modes with $|\bm{k}| > \Lambda_{\mathrm{UV}}\,$. This is because any modes present beyond $\Lambda_{\mathrm{UV}}$ need to be squeezed into a small part of the field's Hilbert space in order to still satisfy the cosmic Bekenstein bound. \change{The fact that we then need $\Lambda_{\mathrm{UV}} \ll \Lambda_{\mathrm{sp}}$ to be consistent cosmic neutrino observations is in surprising, because $\Lambda_{\mathrm{sp}}$ itself is usually expected to be the scale of EFT-breakdown (see e.g.\ \cite{Dvali2008, Castellano2022}). We discuss this point further in \secref{discussion}.}

Note again, that the above considerations assume a ratio of $N_{\mathrm{dof},\psi} / N_{\mathrm{total}} \sim 0.01$ between the degrees of freedom present in a single neutrino field and the total degrees of freedom present in the Universe (which is equivalent to assuming that $\Lambda_{\mathrm{sp}} \sim 0.1\, \Lambda_{\mathrm{UV}}$). In \secref{N_dof_total} we argued that this is a realistic estimate, assuming that Bosonic fields can be modelled via generalised Pauli operators (GPO \cite{Carroll_Singh_2018, Cao2019_essay, Friedrich2022}) with similar holographic mode overlaps.

\subsection{Consistency with the cosmic ray spectrum}
\label{sec:CR_spectrum}

\begin{figure}
\centering
  \includegraphics[width=0.98\textwidth]{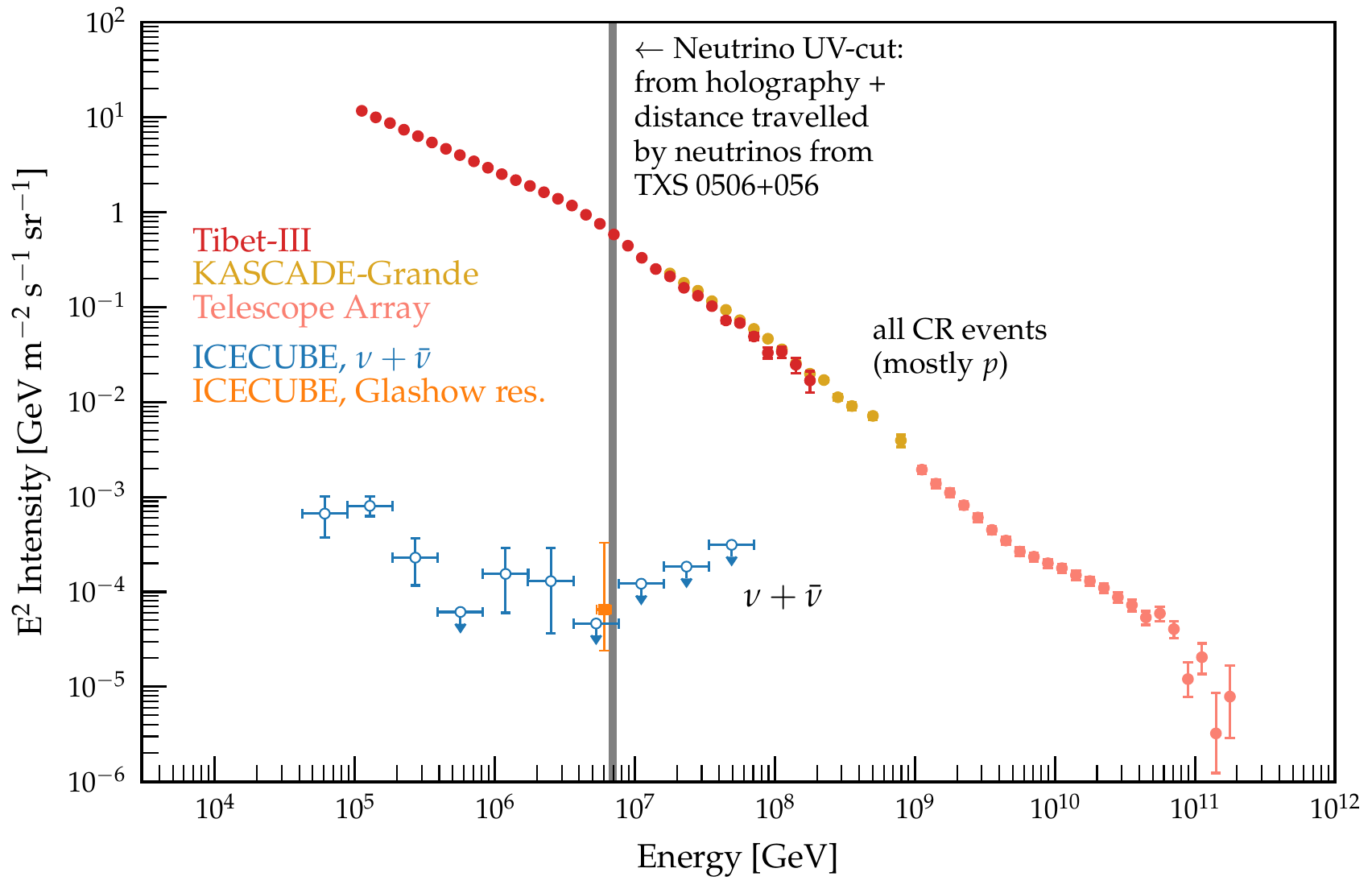}
  
  \includegraphics[width=0.96\textwidth]{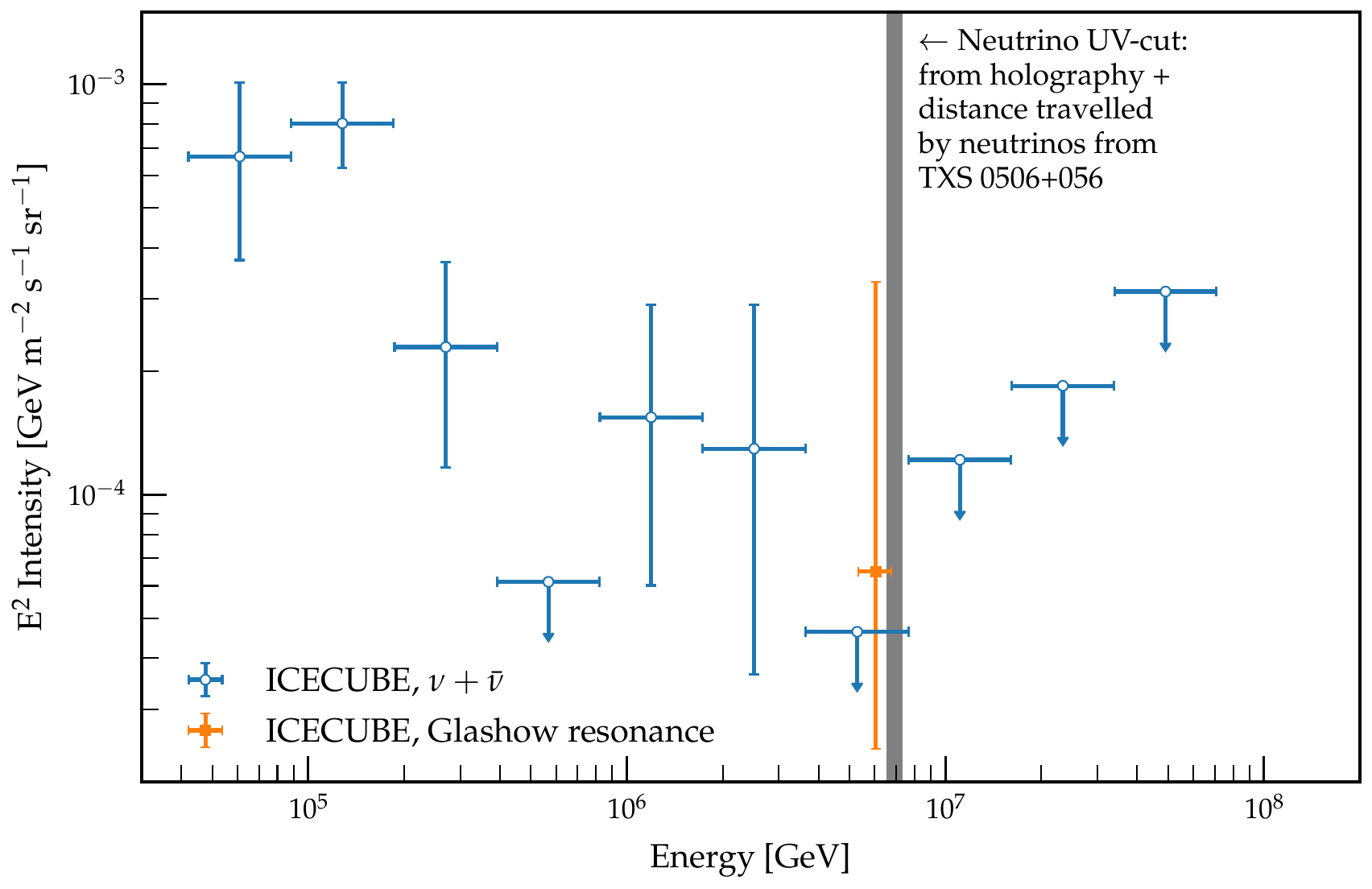}
   \caption{Top panel: Cosmic ray flux  (more concretely the $E^2$-Intensity, which is defined as $E^2\, \mathrm{d}N/\mathrm{d}E\mathrm{d}A\mathrm{d}t\mathrm{d}\Omega$) as a function of particle energy; using data and tools compiled by \cite{The_CR_spectrum} as well as observations reported by \cite{Glashow_res_2021_short}. Lower Panel: Zoom into the neutrino contributions to the cosmic ray spectrum. The grey vertical band indicates the neutrino UV-cut we have derived in \secref{lifetime}. The width of the band is an estimate of the mistake we make by assuming a static Universe. But note that a number of other simplified assumptions may also significantly impact our results.}
   \label{fi:The_CR_Spectrum_2023}
\end{figure}

As we discussed in the previous subsection, when we interpret our construction of the holographic Weyl field as a model for neutrinos then the fact that a $290\,$TeV neutrino has been observed from the far away blazar TXS 0506+056 indicates that neutrino physics should have a UV-cut of $\Lambda_{\mathrm{UV}} < 470\, \Lambda_{\mathrm{LHC}}$ (where we again take $\Lambda_{\mathrm{LHC}} = 14\,$TeV). Above that threshold we expect a rather sharp cutoff or at least a \say{super-holographic} behaviour of the neutrino fields, because whatever modes are present above $\Lambda_{\mathrm{UV}}$ need to be squeezed into a very small part of the total Hilbert space in order not to violate the Bekenstein bound.

Of course, our calculation has a number of caveats and relies on a number of modelling choices. We list several of these in \secref{discussion}, but the two main shortcomings of our model are that it assumes a static background space time (whereas the cosmos has significantly expanded in the time it took for neutrinos from TXS 0506+056 to arrive on earth), and that we implement a naive, equal-time version of holography instead of quantizing our field on a light-sheet. The UV-cut $\Lambda_{\mathrm{UV}} < 470\, \Lambda_{\mathrm{LHC}}\,$ may change once these shortcomings are corrected. But our current result does at least seem to be consistent with current intensity limits for the cosmic neutrino flux. 

In \figref{The_CR_Spectrum_2023} we have used tools and data compiled by \cite{The_CR_spectrum} (which includes data from the databases \cite{Maurin2014, DiFelice2017, Haungs2018}) as well as observations reported by \cite{Glashow_res_2021_short} to display current measurements of the intensities observed in different cosmic ray species as a function of particle energy. To date the IceCube experiment \cite{IceCube_2020_short} has only established upper limits for neutrino flux beyond the UV-cut we derived - \cf the blue, open circles in both panels of \figref{The_CR_Spectrum_2023}, which represent direct neutrino detections and the vertical grey band which indicates our result from the previous subsection. So far, the event that comes closest to our limit is an indirect detection of an excitation of the Glashow resonance which presumably took place in Earth's atmosphere and which has produced a shower of secondary particles which was observed in IceCube (\cf orange square; \cite{Glashow_res_2021_short}). This event indicates that electroweak theory is valid until shortly before the energy where our holographic theory for neutrinos needs to break down. Again, this comparison hinges on a number of assumptions, part of which we already know to be rather simplistic. To obtain an estimate of the impact of assuming a static space time, we re-derive our previous UV-cut when taking the travel distance from TXS0506+056 to be the distance between us and the blazar at the time when the observed high energy neutrino was emitted. This indeed moves the UV-cut to higher energies, and the width of the grey band in \figref{The_CR_Spectrum_2023} represents that shift. Of course, we could also directly calculate the travel time of neutrinos from TXS0506+056 and use that to derive a UV-cut. But that would also be a simplification, because we still wouldn't properly quantize our field on an expanding background. So for now, we stick with the above rough estimate of the impact of cosmic expansion. Another obstacle for further analysis is the stark Lorentz symmetry breaking of our theory close the the UV-cut \secref{Lorentz}. A better approximation of Lorentz invariance may be achieved by quantizing the field on light-sheets and by replacing the sharp UV cutoff with a transition to super-holographic mode overlaps. We leave this for future work.

In the upper panel of \figref{The_CR_Spectrum_2023} we also display measurements of cosmic ray flux that are attributed to cosmic protons (using data from the Tibet-III \cite{Amenomori_2008_short}, KASCADE-Grande \cite{Bertaina2016} and Telescope Array \cite{Ivanov2015} experiments; as compiled by \cite{The_CR_spectrum}). The measurements extend to much higher energies than the holographic UV-cut we derived for neutrinos, which seems to indicate that in QCD the breakdown of holography happens at higher energies. There is a number of subtleties which at the moment prevent us from drawing more concrete conclusions from this high energy proton flux. First, protons are composite particles and we did not investigate how to describe such composite states within our framework. Quark confinement, and the fact that quarks have fractional electric charges, will plausibly prevent individual proton constituents from scrambling in the same way that individual neutrino excitations do. The energy of a proton also does not directly translate to similar energies for the proton constituents. Finally, it is a possibility that holography indeed breaks down at different energies for different parts of the standard model. A direct derivation of a QCD UV scale from the travel distances of highly energetic protons (as we did in the previous subsection for neutrinos) is complicated by the fact that high energy protons scatter with the cosmic microwave background \cite{Kenneth1966, Zatsepin_Kuzmin_1966}. This mimics to some extent the plane wave scrambling we have derived here. Though in our construction that scrambling would not reduce proton energy, but instead create a background of diffuse, highly energetic particles.
For these reasons, we do not take the existence of proton cosmic rays above our neutrino cutoff as a significant problem for our model.

Finally, note that some of the high-energy neutrinos detected by IceCube may be associated with distant gamma ray bursts (GRBs). E.g.\ a neutrino with an energy of $\sim 60\,$TeV may be associated with a GRB that was located to be at redshift $z=1.38$ \cite{Amelino2023}. If we assume this association to be correct, then we could also use this pairing to derive a UV-cut for our holographic model of neutrinos. Estimating our systematic uncertainty due to assuming a static Universe in the same way as above would lead to an interval $390\, \Lambda_{\mathrm{LHC}} \lesssim \Lambda_{\mathrm{UV}} \lesssim 601\, \Lambda_{\mathrm{LHC}}\,$, which encompasses our fiducial result $\Lambda_{\mathrm{UV}} \approx 500\, \Lambda_{\mathrm{LHC}}$. The authors of \cite{Amelino2023} also report an extreme event of a $\sim 1800\,$TeV neutrino that may be associated with a GRB at a redshift of $z=3.93\,$. Our way of estimating the impact of cosmic expansion is likely inaccurate for a source at such a high redshift. If we nevertheless employ that estimate again, and if we assume that the association of said neutrino with the high redshift GRB is indeed correct, we arrive at $128\, \Lambda_{\mathrm{LHC}} \lesssim \Lambda_{\mathrm{UV}} \lesssim 283\, \Lambda_{\mathrm{LHC}}\,$. Such a range would indeed be in tension with the neutrino spectrum displayed in \figref{The_CR_Spectrum_2023}, but we deem our calculations unreliable in this situation. To adequately apply our framework to high-redshift sources, we would not just need to reformulate it on an expanding background, but we would also have to implement the covariant entropy bound (i.e.\ area scaling on light-sheets) instead of the equal-time area scaling present in our current model. As mentioned above, we leave this for future work.

\section{Real-space anti-commutator}
\label{sec:real_space_propagator}

\begin{figure}
\centering
  \includegraphics[width=0.92\textwidth]{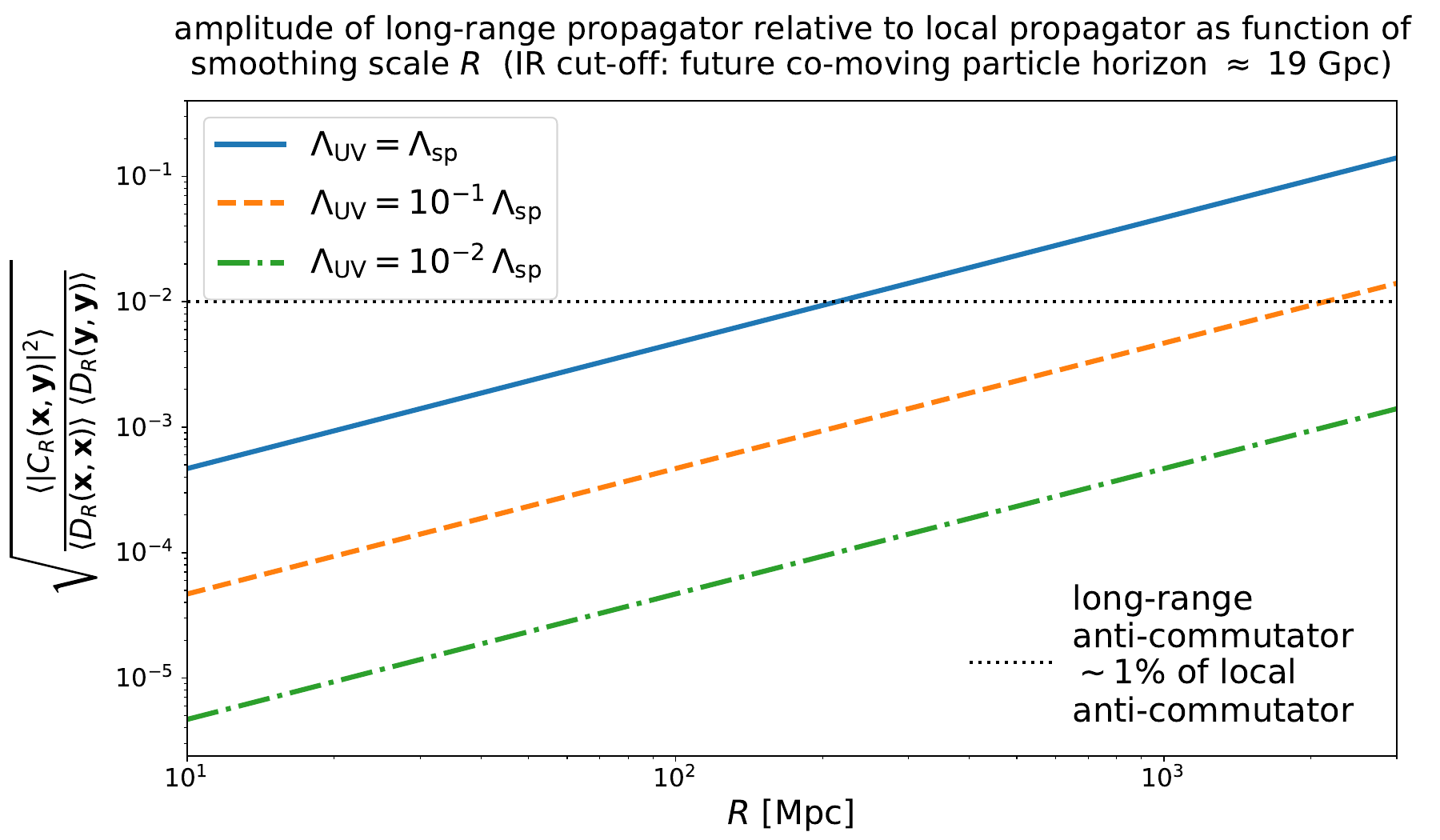}
   \caption{The overlap of Fourier modes generates a non-zero propagator at large equal-time distances in real space. We show the average amplitude of this long-range anti-commutator relative to the local anti-commutator when smoothing the field by different scales $R$~. Even for cosmological smoothing scales ($R\,\sim 100$ Mpc) this ratio remains at a level of $< 1\%$ (cf.\ Section~\ref{sec:real_space_propagator}).}
  \label{fi:relative_longrange_propagator}
\end{figure}

Let us rewrite ($\hat c$-part of) the real-space field as
\begin{eqnarray}
\hat \psi^c(x) &\approx \sum_{\mathrm{shell}\ s} \frac{1}{(k_s V)^{\frac{1}{2}}} \sum_{\bm{p}\in s} \hat c_{\bm{p}}(t)\ u(\bm{p})\ e^{i\bm{p}\bm{x}}\ ,
\end{eqnarray}
where we again made explicit that the field is a sum over independent (non-overlapping) Fourier space shells $s$, and where we assumed that all momenta $\bm{p}$ in one shell $s$ have similar absolute values $|\bm{p}| \approx k_s$~.

To asses how much the qubit overlap changes the behaviour of our field \wrt the non-overlapping case, let us consider the equal-time anti-commutator of $\psi^c$ and the $\hat c$-part of the momentum field, $i\psi_\alpha^c(\bm{x})^\dagger$,
\begin{eqnarray}
    \lbrace i\psi_\alpha^c(\bm{x})^\dagger , \psi_\beta^c(\bm{y})\rbrace &= i\sum_{\mathrm{shell}\ s} \frac{1}{k_s V}\sum_{\bm{p},\bm{q}\in s} \lbrace \hat c_{\bm{p}}^\dagger , \hat c_{\bm{q}} \rbrace\ u_\alpha(\bm{p})^\dagger u_\beta(\bm{q})\ e^{-i\bm{p}\bm{x} + i\bm{q}\bm{y}}\nonumber \\
    &= i\sum_{\mathrm{shell}\ s} \frac{1}{k_s V}\sum_{\bm{p},\bm{q}\in s, \bm{p}\neq\bm{q}} \lbrace \hat c_{\bm{p}}^\dagger , \hat c_{\bm{q}} \rbrace\ u_\alpha(\bm{p})^\dagger u_\beta(\bm{q})\ e^{-i\bm{p}\bm{x} + i\bm{q}\bm{y}}\ +\nonumber \\
    &\ \ + \frac{i}{V} \sum_{\bm{p}}\ \frac{u_\alpha(\bm{p})^\dagger u_\beta(\bm{p})}{|\bm{p}|}\ e^{-i\bm{p}(\bm{x}-\bm{y})}\ .
\end{eqnarray}
In particular, let us look at the trace of this \wrt the spinor components, 
\begin{eqnarray}
\label{eq:trace_of_anticommutator}
    \sum_\alpha \lbrace i\psi_\alpha^c(\bm{x})^\dagger , \psi_\alpha^c(\bm{y})\rbrace &= \frac{i}{2}\sum_{\mathrm{shell}\ s} \frac{1}{k_s V}\sum_{\bm{p},\bm{q}\in s, \bm{p}\neq\bm{q}} \braket{z_{\bm{p}}|z_{\bm{q}}}\ u(\bm{p})^\dagger u(\bm{q})\ e^{-i\bm{p}\bm{x} + i\bm{q}\bm{y}}\nonumber \\
    & \nonumber \\
    &\ \ + i \delta_D(\bm{x}-\bm{y})\nonumber \\
    &\nonumber \\
    &\equiv iC(\bm{x},\bm{y}) + i \delta_D(\bm{x}-\bm{y})\ .
\end{eqnarray}
The second term in the last line of Equation~\ref{eq:trace_of_anticommutator} would be the standard result in the case of independent qubits, while the first term is a correction due to qubit overlap. Because of the stochastic way by which we have embedded the individual qubit operators $\hat c_{\bm{p}},\hat c_{\bm{p}}^\dagger$ into the physical Hilbert space, the correction $C(\bm{x},\bm{y})$ is in fact a (classical) random variable. It can be shown to have vanishing expectation value, and its variance is given by
\begin{eqnarray}
    \mathbb{E}\left[ |C(\bm{x},\bm{y})|^2 \right] &= \frac{1}{4}\sum_{\mathrm{shell}\ s} \frac{\epsilon_s^2}{k_s^2V^2}\sum_{\bm{p},\bm{q}\in s, \bm{p}\neq\bm{q}} |u(\bm{p})^\dagger u(\bm{q})|^2 \ ,
\end{eqnarray}
where $\epsilon_s^2 = \mathbb{E}(|\langle z_{\bm{p}}|z_{\bm{q}}\rangle |^2)$ is the variance of the overlap between two qubits in the same shell. As we show in \ref{app:z_overlap}, \change{this variance is given by
\begin{equation}
    \epsilon_s^2 = \frac{2}{n_{s}} = \frac{2}{\pi L^2 k_s \Delta_s}\left(\frac{\Lambda_{\mathrm{UV}}}{\Lambda_{\mathrm{sp}}}\right)^2\ ,
\end{equation}
where $n_{s}$ is again the effective number of qubits within the physical Hilbert space of shell $s$ (\cf Section~\ref{sec:JL_embedding}) and $\Lambda_{\mathrm{sp}}$ is the species scale we defined in \eqnref{species_scale}.} Recall also, that the number of actual (overlapping) qubits within shell $s$ is given by
\begin{eqnarray}
    N_{s} = \frac{4\pi k_s^2 \Delta_s V}{(2\pi)^3}\ .
\end{eqnarray}
We can hence re-write the variance of $C(\bm{x},\bm{y})$ as
\begin{eqnarray}
    \mathbb{E}\left[ |C(\bm{x},\bm{y})|^2 \right] &= \frac{1}{4}\sum_{\mathrm{shell}\ s} \frac{\epsilon_s^2}{ V^2} N_{s}^2 \frac{1}{N_{s}^2}\sum_{\bm{p},\bm{q}\in s, \bm{p}\neq\bm{q}} \frac{|u(\bm{p})^\dagger u(\bm{q})|^2}{k_s^2} \nonumber\\
    &\approx \frac{1}{4}\sum_{\mathrm{shell}\ s} \frac{\epsilon_s^2}{V^2}\left(\frac{4\pi k_s^2 \Delta_s  V}{(2\pi)^3}\right)^2 \int \frac{\dd\Omega_{\bm{p}}\dd\Omega_{\bm{q}}}{(4\pi)^2} \frac{|u(\bm{p})^\dagger u(\bm{q})|^2}{k_s^2} \nonumber\\
    &\equiv \frac{A}{16 \pi^4}\sum_{\mathrm{shell}\ s} k_s^4 \epsilon_s^2 \Delta_s^2 \ .
\end{eqnarray}
Here we have introduced
\begin{eqnarray}
    A &\equiv \frac{1}{(4\pi)^2}\int \dd\Omega_{\bm{p}}\dd\Omega_{\bm{q}} \frac{|u(\bm{p})^\dagger u(\bm{q})|^2}{k_s^2}\nonumber \\
    &= \frac{1}{(4\pi)^2}\int \dd\theta_1 \sin\theta_1 \dd\phi_1 \int \dd\theta_2 \sin\theta_2 \dd\phi_2\ \left|\cos\frac{\theta_1}{2}\cos\frac{\theta_2}{2} + e^{i(\phi_2-\phi_1)}\sin\frac{\theta_1}{2}\sin\frac{\theta_2}{2}\right|^2\nonumber \\
    &= \frac{1}{2}\ .
\end{eqnarray}
We can hence conclude that
\begin{eqnarray}
    \mathbb{E}\left[ |C(\bm{x},\bm{y})|^2 \right] &= \frac{1}{16 \pi^5 L^2}\left(\frac{\Lambda_{\mathrm{UV}}}{\Lambda_{\mathrm{sp}}}\right)^2\sum_{\mathrm{shell}\ s} k_s^3 \Delta_s \nonumber\\
    &\approx \frac{1}{16 \pi^5 L^2} \left(\frac{\Lambda_{\mathrm{UV}}}{\Lambda_{\mathrm{sp}}}\right)^2 \int_0^{\Lambda_{\mathrm{UV}}} \dd k\ k^3 \nonumber \\
    &= \frac{\Lambda_{\mathrm{UV}}^4}{64 \pi^5 L^2} \left(\frac{\Lambda_{\mathrm{UV}}}{\Lambda_{\mathrm{sp}}}\right)^2 \\
    \Rightarrow \sqrt{\mathbb{E}\left[ |C(\bm{x},\bm{y})|^2 \right]} &= \frac{1}{\sqrt{64 \pi^5}}\frac{\Lambda_{\mathrm{UV}}^3}{\Lambda_{\mathrm{sp}}L}\ .
\end{eqnarray}
Note that - while the propagator $C$ does depend on $\bm{x}$ and $\bm{y}$ - it's average amplitude between two points is completely independent of those points and their separation. So our construction induces a long-range correlation in the spinor field $\psi$~.

To asses the severity of this effect let us introduce a finite smoothing scale to the real space field, i.e. we consider
\begin{eqnarray}
\hat \psi_R^c(x) &= \sum_{\mathrm{shell}\ s}\sum_{\bm{p}\in s}\ \frac{e^{-\frac{1}{2}(|\bm{p}|\cdot R)^2}}{(|\bm{p}| V)^{\frac{1}{2}}}\ \hat c_{\bm{p}}(t)\ u(\bm{p})\ e^{i\bm{p}\bm{x}}\ ,
\end{eqnarray}
and similarly for the $\hat d$-part of the field. This will change the trace of the anti-commutator to
\begin{eqnarray}
    \sum_\alpha \lbrace i\psi_{R,\alpha}^c(\bm{x})^\dagger , \psi_{R,\alpha}^c(\bm{y})\rbrace\ &=\ iC_R(\bm{x},\bm{y}) + \frac{i}{\sqrt{(4\pi)^3R^6}}\ e^{-\frac{1}{4R^2} |\bm{x}-\bm{y}|^2}\nonumber \\
    &\equiv\ i D_R(\bm{x},\bm{y})\ ,
\end{eqnarray}
where the variance of $C_R$ is now given by
\begin{eqnarray}
    \mathbb{E}\left[ |C_R(\bm{x},\bm{y})|^2 \right] &\approx \frac{1}{16 \pi^5 L^2} \left(\frac{\Lambda_{\mathrm{UV}}}{\Lambda_{\mathrm{sp}}}\right)^2 \int_0^{\Lambda_{\mathrm{UV}}} \dd k\ k^3\ e^{-k^2R^2}\nonumber \\ 
    &\approx \frac{1}{8\pi^5 L^2 R^4} \left(\frac{\Lambda_{\mathrm{UV}}}{\Lambda_{\mathrm{sp}}}\right)^2
\end{eqnarray}
as long as $R \gg 1/\Lambda_{\mathrm{UV}}$~. We can then compare the long-range anti-commutator $C_R(\bm{x},\bm{y})$ to the local anti-commutators $D_R(\bm{x},\bm{x})$ and $D_R(\bm{y},\bm{y})$ \change{by considering the ratio
\begin{eqnarray}
\label{eq:anticommutator_ratio}
    \frac{\mathbb{E}\left[ |C_R(\bm{x},\bm{y})|^2 \right]}{\mathbb{E}\left[ D_R(\bm{x},\bm{x}) \right]\ \mathbb{E}\left[ D_R(\bm{y},\bm{y}) \right]}\ =\ \frac{8}{\pi^2}\, \left(\frac{R}{ L}\right)^2 \left(\frac{\Lambda_{\mathrm{UV}}}{\Lambda_{\mathrm{sp}}}\right)^2\ .
\end{eqnarray}
So the qubit overlap only significantly impacts the anti-commutation behaviour of the field on scales $R$ that are comparable to the IR scale $L$ multiplied by the ratio $\Lambda_{\mathrm{sp}}/\Lambda_{\mathrm{UV}}$~. Choosing $L$ again as the co-moving size of our future particle horizon (which is finite in a dark energy dominated Universe), one would expect an effect of $\sim 1\%$ on scales $R\sim 200\, (\Lambda_{\mathrm{sp}}/\Lambda_{\mathrm{UV}})$~Mpc, \cf \figref{relative_longrange_propagator}.}

\section{Alternative modelling choices}
\label{sec:alternative_choices}

\begin{figure}
\centering
  \includegraphics[width=\textwidth]{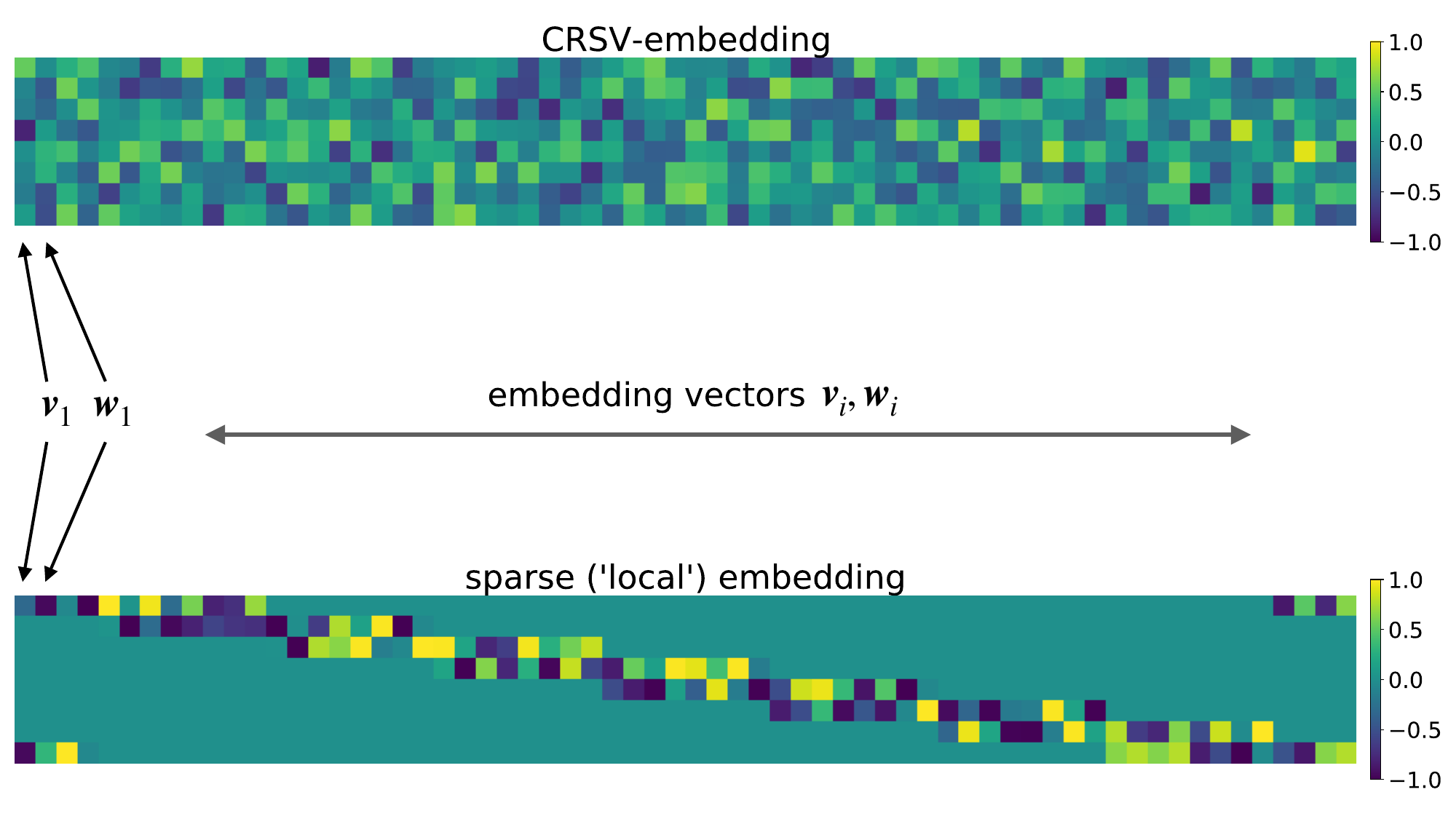}
   \caption{Sketch comparing our fiducial embedding scheme ("CRSV-embedding") to an alternative scheme investigated in \secref{alternative_choices}. See main text of \secref{non_isotropic} for details.}
   \label{fi:embedding_sketch}
\end{figure}

\begin{figure}
\centering
  \includegraphics[width=0.8\textwidth]{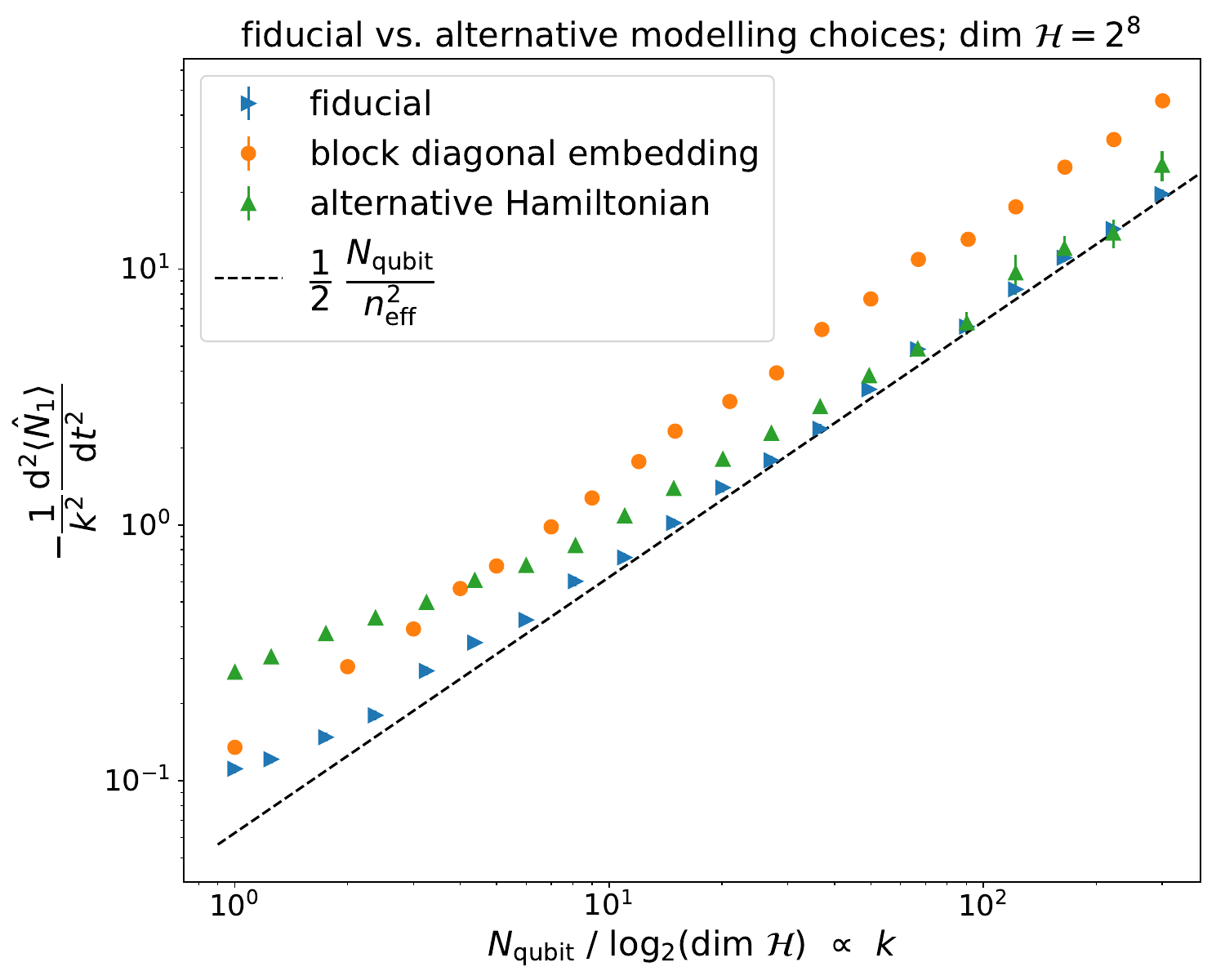}
   \caption{Same as \figref{d2N_dt2}, but comparing our fiducial model to alternative analysis choices. See main text of \secref{alternative_choices} for details. Note that due to numerical limitations we needed to reduce the amount of simulations run for our alternative Hamiltonian operator, such that for large ratios $N_s / n_s$ the results shown here are only averaged over $64$ (instead of $640$) realisations. This is why some of the green triangles display larger error bars than the simulation results obtained for the other modelling choices.}
   \label{fi:d2N_dt2_alt}
\end{figure}

\begin{figure}
\centering
  \includegraphics[width=\textwidth]{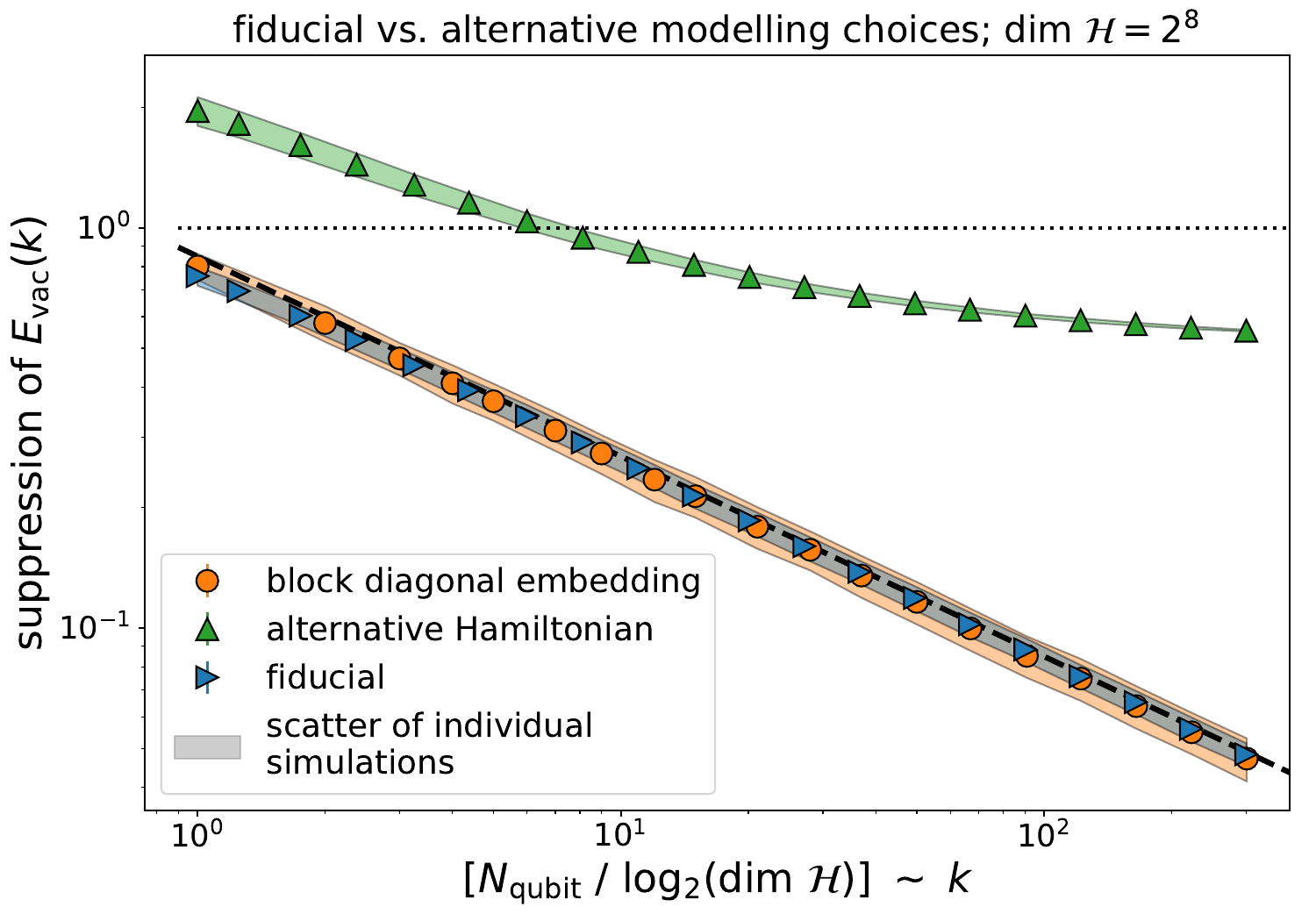}
   \caption{Same as \figref{Evac_suppression}, but comparing our fiducial model to alternative analysis choices. See main text of \secref{alternative_choices} for details.}
   \label{fi:Evac_suppression_alt}
\end{figure}

\subsection{non-isotropic overlaps}
\label{sec:non_isotropic}

Our fiducial method of embedding $N$ qubits into a Hilbert space of dimension $2^n,\ n < N$, relies on choosing orthonormal pairs of vectors $\bm{v}_i, \bm{w}_i\ (i = 1, \dots, N)$ randomly in $\mathbb{R}^{2n}$ and defining the Pauli algebra of the $i$th qubit from these vector pairs via Equations~\ref{eq:qubit_definitions}. The upper panel of \figref{embedding_sketch} displays the $2n\times 2N$ matrix whose columns consist of such such random vector pairs (in that instance, $n=4$ and $N=32$). In our fiducial construction each pair $\bm{v}_i, \bm{w}_i$ represents one Fourier mode of our quantum field in a thin Fourier space shell, and the scalar products between different vector pairs determine how much the corresponding Fourier modes overlap.

The above way of constructing operators for the different field modes leads to mode overlaps that are independent of how far away two wave vectors $\bm{k}$ and $\bm{p}$ are from each other in Fourier space. We want to investigate the impact of such a non-local overlapping structure on some of our results. To do so, we study an alternative embedding scheme in which the vectors $\bm{v}_i, \bm{w}_i$ form a matrix with a sparse structure, as is depicted in the lower panel of \figref{embedding_sketch}. Concretely, this structure is obtained by populating elements above and below the diagonal of the embedding matrix such that both horizontally and vertically only one quarter of the matrix elements are non-zero (and we include non-zero elements in the upper right and lower left corners to obtain cyclic symmetry). Such an embedding could e.g.\ represent a local overlapping structure in a one-dimensional string of field modes. This would in principle not be appropriate in our case since our overlapping modes live in Fourier space shells which can be thought of as approximately two-dimensional. But with this alternative embedding we can nevertheless qualitatively test for differences between the statistically isotropic overlaps of our fiducial model and a more local scheme of overlaps.

In \figref{d2N_dt2_alt} we show how this new embedding scheme impacts the lifetime of plane waves. Modulo a multiplicative factor, this figure displays $\bra{\bm{k}} \dd^2 \hat N_{\bm{k}} / \dd t^2 \ket{\bm{k}}$, i.e.\ the expectation value of the second time derivative of the occupation number operator $\hat N_{\bm{k}}$ in a single field mode (modulo a multiplicative factor) as a function of the energy $|\bm{k}|$ of the plane wave state $\ket{\bm{k}}$ \wrt which the expectation value is taken (also modulo a multiplicative factor, since $N_s/n_s \sim k_s$). In \secref{lifetime} we had used this second derivative as a definition of $1/T_{\mathrm{scramble}}^2$, where we took $T_{\mathrm{scramble}}$ to represent a characteristic time scale within which the plane wave will be completely scrambled within its Fourier space shell. The orange circles in \figref{d2N_dt2_alt} represent values for that second time derivative obtained from averaging $640$ simulated realisations of the new embedding scheme, while the blue triangles use our fiducial scheme and the dashed line represents the analytical prediction derived in \ref{app:lifetime}. In the limit of large ratios $N_s/n_s$ the second time derivative of $\hat N_{\bm{k}}$ is larger in the alternative embedding scheme compared to the fiducial scheme by more than a factor of $2$, indicating a plane wave lifetime that is shorter by about a factor of $1.5$. This seems to indicate that the statistically isotropic CJL-embedding leads to more stable plane waves than a more local embedding scheme. It is however possible that this picture would reverse if we were to consider wave packets instead of plane waves. Many plane waves will be superposed in such a packet in a way that is local in Fourier space, and it may be that local mode overlaps cause such a superposition to become more stable. At this point neither our analytical nor our simulation tools are powerful enough to test this hypothesis in a meaningful way, and such an analysis needs to be carried out in follow-up work.

Finally, we also investigate the impact of the new embedding scheme on the vacuum energy of our field. \figref{Evac_suppression_alt} compares how that energy is suppressed in the different embedding schemes, and surprisingly we find that that suppression is almost identical in both approaches. For the CJL-embedding we had found in \ref{app:energy_spectrum} that the vacuum energy suppression only depends on the ratio $N_s/n_s$. Since our alternative embedding cuts down the non-zero elements of the embedding matrix by the same factor both horizontally and vertically (cf.\ our sketch in \figref{embedding_sketch}), we speculate that the more local overlap structure retains the same \say{effective} ratio of $N_s/n_s$, and hence retains the same amount of vacuum energy suppression. We however do not repeat our calculations of \ref{app:energy_spectrum} for the new embedding scheme, and a more thorough investigation is again postponed to follow-up work.

\subsection{alternative Hamiltonian}

We also want to consider an alternative, more general form for the Hamiltonian of our field. In this section we thus take the Hamiltonian in one Fourier space shell $s$ to be given by
\begin{eqnarray}
\hat H_s &= \sum_{\bm{q}, \bm{k}\in s} A_{\bm{q}\bm{k}} \left\lbrace \left(\hat c_{\bm{q}}^\dagger \hat c_{\bm{k}} - \frac{1}{2}\right) + \left(\hat d_{\bm{q}}^\dagger \hat d_{\bm{k}} - \frac{1}{2}\right)\right\rbrace\ ,
\end{eqnarray}
where we choose the ansatz $A_{\bm{q}\bm{k}} = \mathcal{N}_s \braket{z_{\bm{k}}|z_{\bm{q}}}$, such that the coupling of any two Fourier modes in the Hamiltonian is proportional to the overlap of these modes within the CRSV-embedding. The time evolution of $\hat c_{\bm{k}}$ with this modified Hamiltonian is given by
\begin{eqnarray}
\label{eq:dcdt_new_Hamiltonian}
    \frac{\dd \hat c_{\bm{p}}}{\dd t} &= i \left[\hat H_s, \hat c_{\bm{p}}\right]\nonumber \\
    &= \frac{i\mathcal{N}_s}{2} \sum_{\bm{q}, \bm{k}\in s} \braket{z_{\bm{k}}|z_{\bm{q}}} \left\lbrace \braket{z_{\bm{k}}^*|z_{\bm{p}}} \hat c_{\bm{q}}^\dagger  - \braket{z_{\bm{q}}|z_{\bm{p}}} \hat c_{\bm{k}} \right\rbrace\ .
\end{eqnarray}
Now remember that we had represented the annihilation operators $\hat c_{\bm{q}}$ by matrices
\begin{equation}
\label{eq:representation_of_cq}
    \hat{c}_{\bm{q}} = \frac{1}{2} \sum_{j=1}^{2n} \braket{e_j|z_{\bm{q}}} \bm{C}_j\ ,
\end{equation}
where $\bm{C}_j$ are generators of the Clifford algebra in $2n$ dimensions and $\ket{e_j}$ are an orthonormal basis in $\mathbb{R}^{2n}$. As we had already discussed in \secref{JL_embedding}, Equation~\ref{eq:representation_of_cq} constitutes a stochastic embedding, because the vectors $\ket{z_{\bm{q}}}$ are chosen stochastically in the CRSV-algorithm. But at least on average we would like the right-hand side of Equation~\ref{eq:dcdt_new_Hamiltonian} to equal $- i|\bm{p}| \hat c_{\bm{p}}$, which would be the time derivative of the standard, non-overlapping annihilation operators. This can be achieved by properly choosing the normalisation factor $\mathcal{N}_s\, $. First, let us calculate the average operator that would appear on the right-hand side of Equation~\ref{eq:dcdt_new_Hamiltonian}. To do so, we will keep the embedding of the operator $\hat c_{\bm{p}}$ fixed, i.e.\ we fix the  coefficients $\braket{e_j|z_{\bm{p}}}$, and we will average over the embedding of the other operators, i.e.\ over the coefficients $\braket{e_j|z_{\bm{q}}}$, $\braket{e_j|z_{\bm{k}}}$ with $\bm{q} \neq \bm{p} \neq \bm{k}$. It can then be shown that
\begin{eqnarray}
    \mathbb{E}\left[\frac{\dd \hat c_{\bm{p}}}{\dd t}\right]_{\braket{e_j|z_{\bm{p}}}\ \mathrm{fixed}} &= -\frac{i\mathcal{N}_s}{2}\ \hat c_{\bm{p}}\ \sum_{\bm{q}\in s} \mathbb{E}\left[|\langle z_{\bm{q}}|z_{\bm{p}}\rangle|^2\right]_{\braket{e_j|z_{\bm{p}}}\ \mathrm{fixed}} \nonumber \\
    &= - 2i \mathcal{N}_s \hat c_{\bm{p}}\ \left\lbrace \frac{(N_s - 1)}{n_s} + 1\right\rbrace\nonumber \\
    &\approx - 2i \mathcal{N}_s\ \hat c_{\bm{p}}\ \left\lbrace 1 + \frac{k_s L}{2 \pi^3} \right\rbrace\ .
\end{eqnarray}
So in order for the average time evolution to equal that of the standard Weyl field, we need to choose
\begin{eqnarray}
    \mathcal{N}_s = \frac{k_s}{2\left(\frac{(N_s - 1)}{n_s} + 1\right)}\ . 
\end{eqnarray}
Such a normalisation factor was not necessary in our fiducial Hamiltonian, because there $\mathbb{E}[\mathrm{d} \hat c_{\bm{p}} / \mathrm{d}t] = -i|\bm{p}| \hat c_{\bm{p}}$ was already satisfied to begin with (we do not explicitly show this here, but it is straight forward to demonstrate).

The green triangles in \figref{d2N_dt2_alt} show $\bra{\bm{k}} \dd^2 N_{\bm{k}} / \dd t^2 \ket{k}$ as a function of $N_s / n_s$ in simulated realisations of overlapping qubits for the alternative Hamiltonian discussed above. For large ratios $N_s / n_s$ this second time derivative seems to approach the same asymptotic behaviour as plane waves with our fiducial Hamiltonian (blue triangles, and the dashed line which represents our analytical result from \ref{app:lifetime}). One may speculate that this is a result of our choice for the normalisation constant $\mathcal{N}_s$ which ensures that the average dynamics of the field modes is the same for our fiducial and alternative Hamiltonian. We however do not extend our analytical calculations from \ref{app:lifetime} to the more general Hamiltonian and cannot confirm this speculation at this point. The fact that the lifetime of plane waves doesn't seems to be significantly altered between our fiducial Hamiltonian and a more general one nevertheless lends a degree of robustness to one of our key results. Of course, we have only considered a rather restricted generalisation and a much more general class of Hamiltonians should be investigated in follow-up work. Note in particular that (as we discuss in \secref{mereology}) the ansatz of our paper needs to be more thoroughly motivated by concepts of quantum mereology, in which a given Hamiltonian dictates which degrees of freedom are good candidates for semi-classical field degrees of freedom and thus how the Hamiltonian itself looks like in terms of these degrees of freedom.

Finally, while plane waves behave quite similarly with our alternative Hamiltonian, the energy eigenspectrum seems to be significantly altered. The green triangles in \figref{Evac_suppression_alt} show the suppression of vacuum energy with our alternative Hamiltonian, and it is much less severe than that present with our fiducial Hamiltonian. Note also, that the scatter of vacuum energy (\wrt different realisations of the CRSV-embedding) is much reduced for the alternative Hamiltonian. Again, we do not attempt to extend our calculations of \ref{app:energy_spectrum} to the more general Hamiltonian and postpone a more thorough analysis to follow-up work. In particular, at this point it is not clear which form of the Hamiltonian would be motivated by a more thorough integration of our work into a quantum mereology framework.

\section{Why these random overlaps? Context within quantum mereology}
\label{sec:mereology}

In this section, we would like to embed our construction into the context of \emph{quantum mereology} \cite{Carroll_Singh2020}. We will argue, that our investigation is in a sense \say{mereology done the wrong way around}. This will also shed light on a peculiar feature of our model of a holographic Weyl field: the fact that it contains a stochastic element in the form of randomly chosen embeddings of the field modes into the smaller, physical Hilbert space.

Let us again consider a thin shell $s$ in Fourier space with radius $k_s$ and width $\Delta_s \ll k_s$. The particle-part of our field (and similarly, the anti-particle-part) consists of 
\begin{eqnarray}
    N_s\, \approx\, \frac{4\pi k_s^2 \Delta_s}{(2\pi/L)^3} \,\propto\, k_s^2 \Delta_s
\end{eqnarray}
Fourier modes $\bm{p}\in s$, where $L$ is the IR-cut of our theory. In the usual Weyl field, each of these modes corresponds to one independent qubit in the overall Hilbert space. This would however lead to a volume scaling of the field's number of degrees of freedom. To achieve an (approximate) area scaling, we instead sought to represent the field's Fourier space shell $s$ in a Hilbert space of dimension $2^{n_s} < 2^{N_s}$ with
\begin{eqnarray}
    n_s \propto k_s \Delta_s\ .
\end{eqnarray}
We did so by choosing $2n_s$ Clifford generators $\bm{C}_j$ in the Hilbert space of dimension $2^{n_s}$ and then representing the Pauli algebra of each field mode $\bm{p}$ (for the particle-, or $c$-part of the field) as
\begin{eqnarray}
    \bm\sigma_{x,\bm{p}}^c \equiv \sum_{j=1}^{2n} \braket{e_j|v_{\bm{p}}} \bm{C}_j\ \ ,\ \ \bm\sigma_{y,\bm{p}}^c \equiv \sum_{j=1}^{2n} \braket{e_j|w_{\bm{p}}} \bm{C}_j\ \ ,\ \ \bm\sigma_{z,\bm{p}}^c &= -i \bm\sigma_{x,\bm{p}}^c \bm\sigma_{y,\bm{p}}^c\nonumber\ ,
\end{eqnarray}
where $\braket{e_j|v_{\bm{p}}}$ and $\braket{e_j|w_{\bm{p}}}$ are the $j$th components of two orthonormal vectors $\bm{v}_{\bm{p}}$ and $\bm{w}_{\bm{p}}$ that are randomly chosen in $\mathbb{R}^{2n_s}$. The Johnson-Lindenstrauss theorem (\cf our \ref{app:JL_theorem}) together with the results of \cite{Chao2017} (\cf our \secref{JL_embedding_basics}) then guarantees that the field modes for two different $\bm{p}\neq \bm{q}$ in the same shell $s$ almost anti-commute (up to a small $\epsilon$, \cf our \eqnref{Chao_bound}). We have referred to the deviations from perfect anti-commutation as the \say{overlap} between the modes $\bm{p}$ and $\bm{q}$. That overlap is a stochastic quantity itself, since it is proportional to the scalar products between the random vectors $\bm{v}_{\bm{p}},\bm{w}_{\bm{p}}$ and $\bm{v}_{\bm{q}},\bm{w}_{\bm{q}}$~.

\begin{figure}
\centering
  \includegraphics[width=\textwidth]{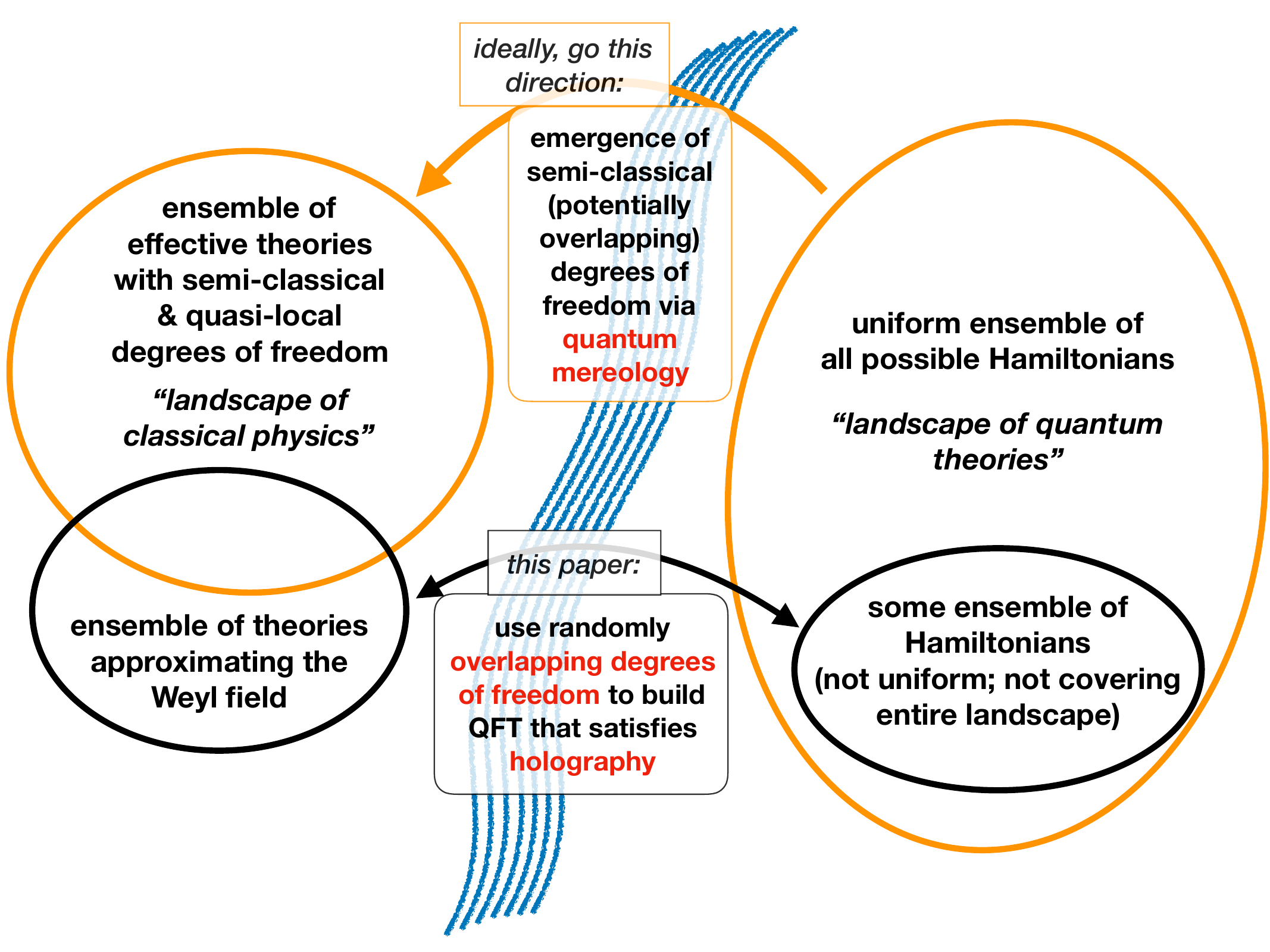}
   \caption{Putting our work into the context of the quantum mereology / quantum first program \cite{Carroll_Singh_2018, Carroll_Singh2020, Cao2017, CaoCarroll2018, Singh2018}. The latter conjectures, that classical physics and in particular classical, local degrees of freedom emerge from an abstract quantum theory (specified through a Hamiltonian operator in some Hilbert space) by decomposing the Hilbert space of that theory into factors that satisfy certain classicality conditions (upper, orange part of the sketch). Finding and implementing such a mereology algorithm is challenging \cite{Carroll_Singh2020}, and our work (lower, black part of the sketch) does not quite tackle this task (\cf our discussion in \secref{mereology}). But our insights and the tools we developed can assist future implementations of the quantum mereology program. And as an additional product, we managed to derive a concrete phenomenology of the holographic principle.}
  \label{fi:mereology_sketch}
\end{figure}

Why these random overlaps? Why should the true, low-energy EFT of quantum gravity be described by a model with stochastic ingredients, and what is the mechanism to choose the random vectors $\bm{v}_{\bm{p}},\bm{w}_{\bm{p}}$~? We conjecture that the answer to these questions can be found in the concept of quantum mereology. In \cite{Carroll_Singh2020} the authors consider general quantum theories, which are defined by an arbitrary Hilbert space $\mathcal{H}$ as well as a Hamilton operator $\hat{H}$ on that Hilbert space. The ensemble $\lbrace \mathcal{H}, \hat{H}\rbrace$ can be considered as the \say{landscape} of all possible quantum theories. An arbitrarily picked point in that landscape is not equipped with any pre-defined notion of \say{degrees of freedom}, i.e.\ the Hilbert space $\mathcal{H}$ does not have any a priori decomposition $\mathcal{H} = \mathcal{H}_1\otimes \mathcal{H}_2 \otimes \dots$ into factors that represent individual degrees of freedom. The authors of \cite{Carroll_Singh2020} also assume that there are no pre-defined observables other then the Hamiltonian $\hat{H}$. Instead, they argue that such operators as well as a preferred Hilbert space factorisation should \emph{emerge}, by demanding that - under the evolution dictated by $\hat{H}$ - the emergent factors and operators should behave in a quasi-classical way. More concretely, they propose and implement the following algorithm for quantum mereology:
\begin{itemize}
    \item[1.] Choose a candidate factorisation $\mathcal{H} = \mathcal{H}_1\otimes \mathcal{H}_2 \otimes \dots$~.
    \item[2.] Among all observables of the form $\hat{O} = \hat{O}_1\otimes \hat{O}_2 \otimes \dots$~, find the one that is most compatible with the Hamiltonian $\hat{H}$ (i.e.\ whose commutator with $\hat{H}$ has the smallest operator norm).
    \item[3.] Consider a set of product states $\lbrace \ket{\psi} = \ket{\psi}_1\otimes \ket{\psi}_2 \otimes \dots \rbrace$ that correspond to \say{peaked} configurations in the observables $\hat{O}_i$~, i.e.\ the $\ket{\psi}_i$ are superpositions of the eigenstates of $\hat{O}_i$ that are concentrated around one of the eigenvalues of $\hat{O}_i$.
    \item[4.] Evaluate two measures of classicality for this candidate factorisation: A) how slowly does entanglement entropy grow when starting in the states $\lbrace \ket{\psi}\rbrace$, and B) how slowly do the peaked states $\lbrace \ket{\psi}\rbrace$ disperse (when viewed as wave packets in the eigenspectrum of the operators $\hat{O}_i$). These criteria have been interpreted by \cite{Carroll_Singh2020} as follows: A) measures the robustness of the observables $\hat{O}_i$ towards decoherence, and B) measures predictability in the sense of asking how long it takes for peaked states in the observables $\hat{O}_i$ to disperse.
    \item[5.] Find the factorisation $\mathcal{H} = \mathcal{H}_1\otimes \mathcal{H}_2 \otimes \dots$\ that maximises some joint measure of criteria A) and B). This is the emergent, semi-classical factorisation of $\mathcal{H}$. And the operators $\hat{O}_i$ measure observable degrees of freedom living on the factors $\mathcal{H}_i$~.
\end{itemize}
Originally, \cite{Carroll_Singh2020} implemented their mereology algorithm for the situation where $\mathcal{H}$ is only split into two factors (a \emph{system} and an \emph{environment}), but we want to consider their ideas here in the context of many degrees of freedom. In particular, we would like to interpret the $\hat{O}_i$ as local degrees of freedom in some effective field theory. To do that, we would need to complement the original algorithm by some locality criterion. But such a generalised version of mereology may be difficult to realise, because it is not guaranteed that an arbitrary Hamiltonian admits a factorisation in which dynamics are local. The authors of \cite{Cotler2019} have shown that such a factorisation rarely exists, and that it is essentially unique if it exists. Though the situation may be not as dire as these results indicate, because according to the findings of \cite{Loizeau2023} factorisations in which a given Hamiltonian appears approximately local do exist for quite general ensembles of Hamiltonians.

One way to increase the success probability of finding degrees of freedom with quasi-local dynamics when given an arbitrary Hamiltonian is to make the mereology algorithm more flexible: instead of insisting that the Hilbert space $\mathcal{H}$ be split into an exact factorisation and that the observables $\hat{O}_i$ have exactly vanishing (anti-)commutators, we may instead task mereology with finding operators $\hat{O}_i$ that are only quasi-commuting - similar to the mode operators $\hat c_{\bm{p}}, \hat d_{\bm{p}}$ of our holographic Weyl field. (Technically, $\hat c_{\bm{p}}, \hat d_{\bm{p}}$ are of course not local because they are the Fourier space degrees of freedom. But also the real space degrees of our field suffer random overlaps, as is demonstrated by our investigation of the real space anti-commutator in \secref{real_space_propagator}.) In the following, we will assume that the mereology algorithm is given this additional leeway and that, given a Hamiltonian $\hat{H}$, it identifies EFT degrees of freedom $\hat{O}_i$ that have local dynamics but that can have small non-zero overlaps.

Now note that we have started the mereology algorithm by randomly picking a point $(\mathcal{H}, \hat{H})$ in the landscape of quantum theories. Even the points on that landscape with the same Hilbert space $\mathcal{H}$ will still vary in their Hamiltonian $\hat{H}$, i.e.\ these points will result in different EFT degrees of freedom $\hat{O}_i$~. So the mereology algorithm we conjectured above will map an ensemble $\lbrace\hat{H}\rbrace$ of Hamiltonian operators to an ensemble $\lbrace \lbrace \hat{O}_1\, ,\, \hat{O}_2\, ,\, \dots\, \rbrace_{\hat{H}} \rbrace$ of sets of quasi-classical degrees of freedom, with each set $\lbrace \hat{O}_1\, ,\, \hat{O}_2\, ,\, \dots\, \rbrace_{\hat{H}}$ corresponding to a different quasi-local EFT. In \ref{app:Heisenberg}, we find evidence that our model of the Weyl field with overlapping degrees of freedom indeed fits into such a perspective. We show there that the Hamiltonian $\hat H_s$ in one Fourier space shell $s$ of our field can be written as
\begin{eqnarray}
\label{eq:Heisenberg_Hamiltonian_main_text}
    \hat H_s &\propto \sum_{j,k=1}^n \left( M_{jk} \hat{A}_j\hat{A}_k  + X_{jk} \hat{A}_j \hat{A}_k^\dagger  - X_{jk}^* \hat{A}_j^\dagger \hat{A}_k  - M_{jk}^* \hat{A}_j^\dagger \hat{A}_k^\dagger  \right)\ ,
\end{eqnarray}
where the $\hat A_j, \hat A_j^\dagger$ are non-overlapping creation and annihilation operators and $M_{jk}, X_{ij}$ are random coefficients that depend on the vectors $\bm{v}_{\bm{p}}, \bm{w}_{\bm{p}}$ from above. In particular, these coefficients have zero mean and identical variances (see \ref{app:Heisenberg} for more details). So the Hamiltonian above is that of a non-local Heisenberg model with 2-site interactions and randomly drawn interaction coefficients. This is a very general ensemble of theories! And nevertheless, with the help of overlapping degrees of freedom, this can be reformulated into an ensemble of theories that approximates the standard Weyl field. If we interpret this as an instance of quantum mereology, then the randomness of the embedding vectors $\bm{v}_{\bm{p}}, \bm{w}_{\bm{p}}$ simply appears to be a result of sampling from a more general landscape of quantum theories.

There are a number of caveats to this admittedly enthusiastic interpretation of our results, which we list in the following.
\begin{itemize}
    \item We have only shown in a quite limited way, that our model indeed approximates the Weyl field. We demonstrated that plane waves in our field can have cosmic life times (\cf \secref{dynamics}), that anti-commutators between Fourier modes are tiny (\cf \secref{Bekenstein_bound}), and that notable differences to the real-space propagator only appear at cosmological scales (\cf \secref{real_space_propagator}). We have not studied the full dynamics of our field, and have only limited reason to claim that the locality of that dynamics is preserved.
    \item We present a link between a somewhat general ensemble of quantum theories and an ensemble of theories approximating a quasi-local EFT, but we do not demonstrate that this link is made via any sort of mereology algorithm.
    \item While we did reduce our construction to a quite general ensemble of Heisenberg models, this is still far from sampling the space of all possible Hamiltonians. In particular, our results are restricted to Heisenberg models with only 2-site interactions, and we still assume different Fourier space shells to be uncoupled. (Though note that \cite{Loizeau2023} showed that Hamiltonians with 2-site interactions are approximately unitary equivalent to the ensemble of all possible Hamiltonians in $\mathbb{C}^{2^n}$ with exponentially small error in the limit $n \rightarrow \infty\,$.)
\end{itemize}
Part of these caveats are summarised by our sketch in \figref{mereology_sketch}. Ideally, the program of quantum mereology would be implemented by sampling from the landscape of all possible, abstract Hamiltonians (right hand-side of the sketch) and then applying a mereology algorithm to re-interpret the latter as an ensemble of theories with quasi-classical degrees of freedom and dynamics (left hand-side of the sketch). If we view the process of mereology as \say{bridging a river} between the landscape of abstract quantum theories and the landscape of potential quasi-classical physics, then the approach of this paper has rather been the following: we float down the river and throw ropes to either side of it, in the hopes that they stick somewhere (lower part of the sketch; and creating a concrete implementation of holography in the process). As a result we sample neither on the right nor on the left hand-side of the sketch quite the ensembles of theories we would like to sample. Nevertheless, we consider our findings as strong evidence, that the mereology program (and related ideas about the emergence of local degrees of freedom from random Hamiltonian ensembles, \cf \cite{AlbrechtIglesias2008, AlbrechtIglesias2015, Loizeau2023, Zanardi2024, Adil2024}) can work in principle. And the framework of overlapping degrees of freedom we have developed may assist in future implementations of that program.

\section{Discussion}
\label{sec:discussion}

Since \secref{intro} included a detailed overview of our results, we will summarise them here only very briefly, and instead focus on open questions. 
The essential results are:
\begin{itemize}
    \item[A)] We have introduced a modified version of the Weyl field whose effective degrees of freedom obey an approximate area scaling (exact area scaling in Fourier space), and that satisfies the Bekenstein bound for the total degrees of freedom inside the cosmic horizon.
    \item[B)] We find that each pair of Fourier modes of that field still behaves like a pair of almost independent, non-overlapping degrees of freedom ($|\lbrace \hat c_{\bm{p}}, \hat c_{\bm{q}} \rbrace| \sim 10^{-43}$ at LHC energies). The real space anti-commutator obtains a stochastic, non-local correction which however becomes significant only on cosmological scales.
    \item[C)] We define plane wave states in our holographic field and find that they have a finite lifetime. In particular, the power in a wave of energy $\bm{p}$ \change{(or equivalently, the information stored in the excitation of the mode $\bm{p}$)} is isotropically scrambling to modes of similar absolute value $|\bm{p}|$. We find that this is not in contradiction to observed cosmological sources of highly energetic neutrino emission as long as our effective theory breaks down at energies $\Lambda_{\mathrm{UV}} < 470\, \Lambda_{\mathrm{LHC}}\,$. We argued that this is still consistent with current bounds on the validity regime of electroweak theory from observations of the high-energy cosmic neutrinos. But a potential tension arises if we naively apply our calculations to a highly energetic neutrino that according to \cite{Amelino2023} may be associated with a gamma ray burst detected at a high redshift ($z\approx 3.93$).
    \item[D)] We have motivated our construction via quantum mereology, \ie using the idea that EFT degrees of freedom should emerge from an abstract theory of quantum gravity by finding quasi-classical Hilbert space decompositions.
    \item[E)] We also find that vacuum energy is significantly reduced in our construction, e.g. by a factor $\sim 10^{-30}$ if $\Lambda_{\mathrm{UV}} \sim \Lambda_{\mathrm{Planck}}$ and by $\sim 10^{-25}$ if $\Lambda_{\mathrm{UV}} \sim 470\, \Lambda_{\mathrm{LHC}}$ (which corresponds to an overall suppression of $\sim 10^{-74}$ compared to a naive calculation with $\Lambda_{\mathrm{UV}} = \Lambda_{\mathrm{Planck}}$ and no holography). Neither of this is enough to be relevant for the cosmological constant problem.
\end{itemize}
\vspace{0.3cm}
All of these findings depend on a number of parameters and modelling choices, and a large number of questions remain unaddressed. We list those in the following:
\begin{itemize}
    \item \change{Our construction explicitly reduces the Hilbert space dimension of the Weyl field beyond just the reduction that would be caused by imposing UV and IR cutoffs to a field theory. Hence, we speculated that the phenomenology we derived cannot be fully captured by a standard 3+1 dimensional effective field theory. In \ref{app:Heisenberg} we went on to re-formulate our model in terms of non-overlapping degrees of freedom (whose number follows an area law by construction) and we explicitly derived the Hamiltonian of that dual model. The dual Hamiltonian turns out to be highly non-local in the new degrees of freedom, and we argued that this suggests the following viewpoint: a 2+1 dimensional theory that is highly non-local can be interpreted as a close-to-local 3+1 dimensional field theory, if we allow the mode operators of the latter to overlap slightly (see also the corresponding discussion in \secref{mereology}). This point should be explored further and made more precise in follow-up analysis.}
    \item Several of our results assumed a ratio of $N_{\mathrm{dof},\psi} / N_{\mathrm{dof,total}} \sim 0.01$, \ie that our holographic field makes up $\sim 1\%$ of all degrees of freedom in the Universe. This is equivalent to assuming that the species scale that appears in our constrcution is given by $\Lambda_{\mathrm{sp}}\approx 10^{-1}\, \Lambda_{\mathrm{Planck}}\,$. In \secref{N_dof_total} we have only given a heuristic justification for that number. Our argument did take into account the fact that Bosonic fields may need to be modelled via generalised Pauli operators (as e.g.\ done by \cite{Carroll_Singh_2018, Cao2019_essay, Friedrich2022}) and that such fields would contribute significantly more degrees of freedom to the Universe's total budget than Fermionic fields. But a more thorough analysis, that e.g.\ more closely examines potential overlaps between different matter fields, is needed. Note also that in string theory models, many more particle species than those of the standard model are expected in the Universe's EFT \cite{Dvali2008, Dvali2010b, Castellano2022}.
    \item While we have considered one alternative to our fiducial Hamiltonian and found that both Hamiltonians lead to similar plane wave lifetimes, we have not demonstrated that there isn't any other alternative Hamiltonian for the holographic Weyl field that more closely preserves the classical, non-overlapping dynamics.
    \item As explained in \secref{mereology}, our analysis can be viewed as an incomplete and, in a sense, inverted version of the quantum mereology program, which aims to describe the emergence of classical degrees of freedom from abstract quantum theories (\cf \figref{mereology_sketch}). At the same time, the technology we have developed for describing overlapping degrees of freedom may facilitate more faithful implementations of quantum mereology (and of related ideas, \cf \cite{AlbrechtIglesias2008, AlbrechtIglesias2015, Loizeau2023, Zanardi2024, Adil2024}) in the future.
    \item We have studied the dynamics of plane waves, but particle propagation would more realistically be described via wave packets. The latter would be described by superpositions of plane waves, but the overlap between the different Fourier modes of our field may cause such superpositions to display non-trivial behaviour.
    \item To define the lifetime of plane waves we have only considered averages of the second time derivative of mode occupation numbers $\hat N_{\bm{p}}\,$. Significant scatter around those averages could lead to further effective changes to the dispersion relation of plane waves.
    \item We have investigated the suppression of vacuum energy of our field. But our results do in principle allow us to calculate the entire energy eigenspectrum of the field. A more detailed study of that spectrum may reveal even more observational signatures of holography.
    \item We have considered only one alternative to our fiducial approach of making degrees of freedom overlap in such a way that an approximate area scaling is reached. There maybe a multitude of alternatives, and some of those may preserve classical, non-overlapping dynamics more closely, especially when investigating those dynamics beyond plane waves.
    \item We have implemented a naive, equal-time version of holography instead of imposing a holographic scaling of degrees of freedom on a light-sheet. We have also quantized our field on a static, non-expanding background. Both of these simplification impair the applicability of our calculations to high-redshift sources such as the ones reported by \cite{Amelino2023}.
    \item We have imposed a sharp UV cutoff in our model. Instead, UV regularisation could also take place as a transition from holographic to super-holographic mode overlaps, such that all modes above the regularisation scale $\Lambda_{\mathrm{UV}}$ are squeezed into a small part of the overall Hilbert space (\cf the corresponding discussion in \secref{Lorentz}). Such a regularisation may decrease the severity Lorentz symmetry breaking, especially when paired with a quantisation that implements the covariant entropy bound on light-sheets instead of our naive equal-time area scaling.
    \item \change{It has been argued that the species scale $\Lambda_{\mathrm{sp}}$ should be the energy scale beyond which the EFT that describes the low energy limit of quantum gravity should break down \cite{Veneziano2002, Han_and_Willenbrock_2005, Dvali2008, Dvali2010, Dvali2013, Castellano2022, Castellano2023}. But we had to impose a separate UV-cut $\Lambda_{\mathrm{UV}} \ll \Lambda_{\mathrm{sp}}$ in order for the predictions of our model to agree with cosmic neutrino observations (\cf Sections~\ref{sec:lifetime} and \ref{sec:CR_spectrum}). Following the arguments of \cite{Cohen1999} it is not entirely unexpected that quantum gravity should modify quantum field theory even below the species scale (see also \cite{Castellano2022}, who explicitly point this out). However, according to \cite{Castellano2022} the holographic entropy bound should be satisfied right up until the species scale $\Lambda_{\mathrm{sp}}$, whereas we explicitly argued that the UV cut $\Lambda_{\mathrm{UV}} \ll \Lambda_{\mathrm{sp}}$ should be viewed as a scale where holography breaks down (e.g.\ via a transition to super-holographic mode overlaps, \cf the discussion above). Of course, the species scale could be much smaller than what we have assumed (\cf the discussion in \secref{N_dof_total}). But our results depend only on the ratio $\Lambda_{\mathrm{UV}}/\Lambda_{\mathrm{sp}}$, such that a lower $\Lambda_{\mathrm{sp}}$ would only force us to choose an even lower $\Lambda_{\mathrm{UV}}\,$.}
    \item The authors of \cite{BanksDraper2019, BlinovDraper2021} have investigated a number of observational consequences of EFT mode depletion which we have not considered. Our construction may serve as a concrete model that implements their ideas. To achieve that, our model overlaps would need to be modified such that the degrees of freedom underlying effective field theory deplete even more quickly with energy than the naive area scaling $n_s \sim k_s \Delta_s$ we have assumed. \change{
    \item Even though the independent degrees of freedom of our field satisfy an approximate area scaling, we have not worked out a theory that is dual to our construction and that lives on our construction's boundary. This obscures potential connections of our model to approaches that explicitly construct such dualities in situations more general than AdS/CFT such as e.g.\ \cite{Anninos2017, Kawamoto2023, Pasterski2017a, Pasterski2017b, Pasterski2021}. The shape of the real-space anti-commutator of our field (\cf \secref{real_space_propagator}) hints at differences to \cite{Anninos2017}. In the latter the authors also find modifications to the field algebra induced by holography, but the commutator between fields at different locations $\bm{x}, \bm{y}$ still retains the Heisenberg shape $\propto \delta(\bm{x} - \bm{y})\,$. This indicates an absence of field overlaps, in contrast to our construction. At the same time, our construction may be supported by the findings of \cite{Kawamoto2023}. The latter have proposed a duality between dS quantum gravity and a non-gravitating QFT at the \emph{time-like} boundary of part of de~Sitter space. They then show that such a duality is only possible, if the (emergent) bulk theory is over-counting the degrees of freedom present in the boundary theory - in line with our thoughts. Finally, since our current construction is set in flat space, its results should be related to existing results from the celestial holography program \cite{Pasterski2017a, Pasterski2017b, Pasterski2021}. An interesting avenue to work out such a relation would be to derive the soft-limit of the $S$-matrix of our theory. Note in particular that, while we chose the Hamiltonian of our field to have the same shape as that of the free Weyl field, our mode overlaps should nevertheless generate a non-trivial $S$-matrix. Note also, that we indeed did disentangle our mode overlaps in \ref{app:Heisenberg}, and the resulting collection of Heisenberg models (one for each shell) should at least be isomorphic to a potential boundary theory. Our $S$-matrix, together with the dictionary worked out within the celestial holography program may then tell us, what that isomorphism is. Making the Lorentz invariance violations of our construction less severe (e.g.\ via quantisation on light-sheets; \cf our discussion above) may be a necessary prerequisite to working out such connections. Also, we are at the moment still lacking a description of multi-particle states.}
\end{itemize}
The above points constitute a multitude of potential problems and open questions for our construction. Nevertheless, the tools we have developed here may serve to open a fruitful new pathway towards studying observational signatures of the holographic principle.

\ack
\textcopyright 2024. We we would like to thank George Efstathiou, Kristina Giesel, Daniel Gruen, Anik Halder, Luca Marchetti, Daniele Oriti and Rogerio Rosenfeld as well as our two anonymous referees for helpful comments and discussions. O.F.\ was supported by a Fraunhofer-Schwarzschild-Fellowship at Universitätssternwarte München (LMU observatory) and by DFG's Excellence Cluster ORIGINS (EXC-2094 – 390783311). C.C.\ acknowledges the support by the Air Force Office of Scientific Research (FA9550-19-1-0360), the National Science Foundation (PHY-1733907), and the Commonwealth Cyber Initiative. The Institute for Quantum Information and Matter is an NSF Physics Frontiers Center. A.S.\ acknowledges support by the M.\ J.\ Murdock Charitable Trust, and by Whitman College.  We are grateful for the invaluable work of the teams of the public python packages \verb|NumPy| \cite{NumPy}, \verb|SciPy| \cite{scipy}, \verb|CuPy| \cite{Okuta2017CuPyA} and \verb|Matplotlib| \cite{Matplotlib}.

\section*{Data availability} 

\verb|Python| tools to reproduce our results (analytical predictions and simulated realisations of overlapping qubits) are publicly available at \url{https://github.com/OliverFHD/GPUniverse}~. We also provide data measured in a suite of simulations of overlapping qubits in that repository.

\section*{References}

\input{journaldef}
\bibliographystyle{iopart-num}
\bibliography{main}

\appendix
\addtocontents{toc}{\fixappendix}

\section{Simulated realisations of the CRSV-embedding}
\label{app:simulations}

We have complemented the analytical calculations presented in \ref{app:Heisenberg} and in the main text with simulated realisations the the CRSV-embedding of $N$ qubits into a Hilbert space of dimension $2^n,\ n < N$. Comparing our analytical calculations to these simulations in particular allows us to determine how quickly the asymptotic limit in which these calculations are valid is reached.

To generate overlapping qubits, we follow very closely to the CRSV-procedure described in \secref{JL_embedding}. This requires us to
\begin{itemize}
    \item choose $2n$ generators of Clifford algebra in the Hilbert space $\mathbb{C}^{2^n}$; we do this via the iterative algorithm proposed by \cite{Hile_Lounesto_1990},
    \item generate $2N$ random vectors on the unit sphere in $\mathbb{R}^{2n}$,
    \item in the planes spanned by each pair of these vectors generate an orthonormal basis  by applying the Gram-Schmidt algorithm to the original pairs.
\end{itemize}
If $\bm{C}_j$ are the Clifford generators and $\bm{v},\bm{w}$ is one of the orthonormal vector pairs, a Pauli algebra that represents one qubit can be obtained via
\begin{eqnarray}
    \bm\sigma_{x}^c \equiv \sum_{j=1}^{2n} v^j \bm{C}_j\ \ ,\ \ \bm\sigma_{y,\bm{p}}^c \equiv \sum_{j=1}^{2n} w^j \bm{C}_j\ \ ,\ \ \bm\sigma_{z,\bm{p}}^c &= -i \bm\sigma_{x,\bm{p}}^c \bm\sigma_{y,\bm{p}}^c\ .
\end{eqnarray}
Each pair of orthonormal vector then generates one qubit, and as has been demonstrated by \cite{Chao2017} (see also the calculations we present in \secref{JL_embedding_basics}, \ref{app:real_angles}, \ref{app:JL_theorem}) the anti-commutator between the lowering and raising operators of different qubits will depend on the scalar products between the random vectors that were used to construct the Pauli operators corresponding to those qubits.

Due to computational constraints, we were only able to simulate quite low numbers of qubits. In the most extensive case we simulated $N=3600$ qubits squeezed into a Hilbert space of $\log_2$-dimension $n=12$ (i.e.\ $\dim \mathcal{H} = 2^{12} = 4096$). It however turns out that this is already enough to reach the asymptotic regime in which our calculations based on random matrix theory are valid - see e.g.\ \figref{Evac_suppression} and \figref{d2N_dt2} as well as their descriptions in the main text. Each simulation was run $640$ times and these figures present averages of our quantities of interest over these repeated runs.

To implement the above calculations we made use of the \verb|python| packages \verb|cupy| (package enabling computations with graphical processing units; see \url{https://cupy.dev}) as well as \verb|numpy| \cite{NumPy} and \verb|scipy| \cite{scipy}. Our simulation code is publicly available within our \verb|GPUniverse| package (\url{https://github.com/OliverFHD/GPUniverse}).

\section{Random vectors on the unit sphere and the Johnson-Lindenstrauss theorem}
\label{app:random_angles}

\subsection{Moments on the unit sphere}
\label{app:real_angles}

Let us consider two random vectors $\bm{v}_1, \bm{v}_2 \in \mathbb{R}^n$ that are normalised, $|\bm{v}_1|^2 = 1 = |\bm{v}_2|^2$~, and that are otherwise drawn from a uniform distribution on the unit sphere. Because of symmetry reasons (e.g.\ mirroring $\bm{v}_1$ wrt.\ the hyperplane orthogonal to $\bm{v}_2$) the expectation value $\mathbb{E}[\bm{v}_1\cdot\bm{v}_2]$ must vanish. In other words: on average the two vectors will be orthogonal.

To calculate the variance of $\bm{v}_1\cdot\bm{v}_2$ we then only have to compute the 2nd moment of $\cos \theta$, where $\theta$ is the angle of points on the $n$-dimensional unit sphere wrt.\ an arbitrary axis. This 2nd moment is given by
\begin{eqnarray}
    \frac{\int_0^\pi \dd \theta\ (\sin\theta)^{n-2}\ (\cos \theta)^2}{\int_0^\pi \dd \theta\ (\sin\theta)^{n-2}} = \frac{\frac{1}{2} \sqrt{\pi} \ \Gamma\left(\frac{n-1}{2}\right) / \Gamma\left(\frac{n}{2}+1\right)}{\sqrt{\pi} \ \Gamma\left(\frac{n-1}{2}\right) / \Gamma\left(\frac{n}{2}\right)}
\end{eqnarray}
So the variance of the scalar product between two random unit vectors in $\mathbb{R}^{n}$ is given by
\begin{eqnarray}
    \mathrm{Var}(\bm{v}_i\cdot\bm{v}_j) = \frac{\Gamma\left(\frac{n}{2}\right)}{2\Gamma\left(\frac{n}{2}+1\right)} &= \frac{1}{n}\ .
\end{eqnarray}
Now consider a single random vector $\bm{v}$ drawn uniformly on the unit sphere in $\mathbb{R}^n$. For some of the results of this paper we need to calculate moments like $\mathbb{E}[v_1^2 v_2^2]$, where $v_i$ it the $i$th component of $\bm{v}$. This amounts to computing
\begin{eqnarray}
    &\frac{\int_0^\pi \dd \theta\ \sin^{n-2}\theta\ \cos^2 \theta \sin^2 \theta \int_0^\pi \dd \phi\ \sin^{n-3}\phi\ \cos^2 \phi}{\int_0^\pi \dd \theta\ \sin^{n-2}\theta \int_0^\pi \dd \phi\ \sin^{n-3}\phi} \nonumber \\
    &= \frac{\frac{1}{2} \sqrt{\pi} \ \Gamma\left(\frac{n+1}{2}\right) / \Gamma\left(\frac{n+2}{2}+1\right)}{\sqrt{\pi} \ \Gamma\left(\frac{n-1}{2}\right) / \Gamma\left(\frac{n}{2}\right)} 
    \frac{\frac{1}{2} \sqrt{\pi} \ \Gamma\left(\frac{n-2}{2}\right) / \Gamma\left(\frac{n-1}{2}+1\right)}{\sqrt{\pi} \ \Gamma\left(\frac{n-2}{2}\right) / \Gamma\left(\frac{n-1}{2}\right)}\nonumber \\
    %
    %
    %
    &= \frac{1}{4} \frac{\Gamma\left(\frac{n}{2}\right)}{\Gamma\left(\frac{n}{2}+2\right)} = \frac{1}{n\left(n+2\right)}\ .
\end{eqnarray}
Similarly (but simpler) the moment $\mathbb{E}[v_1^4]$ requires calculation of
\begin{eqnarray}
    \frac{\int_0^\pi \dd \theta\ \sin^{n-2}\theta\ \cos^4 \theta}{\int_0^\pi \dd \theta\ \sin^{n-2}\theta} &= \frac{\int_0^\pi \dd \theta\ \left[\sin^{n-2}\theta\  - \sin^{n}\theta\right]\ \cos^2 \theta}{\int_0^\pi \dd \theta\ \sin^{n-2}\theta}\nonumber \\
    &= \frac{1}{n} - \frac{1}{2}\frac{\Gamma\left(\frac{n-1}{2}+1\right) / \Gamma\left(\frac{n}{2}+2\right)}{\Gamma\left(\frac{n-1}{2}\right) / \Gamma\left(\frac{n}{2}\right)}\nonumber \\
    &= \frac{1}{n} - \frac{n-1}{n\left(n+2\right)} = \frac{3}{n(n+2)}\ .
\end{eqnarray}
And the 6th moment $\mathbb{E}[v_1^6]$ is given by
\begin{eqnarray}
    \frac{\int_0^\pi \dd \theta\ \sin^{n-2}\theta\ \cos^6 \theta}{\int_0^\pi \dd \theta\ \sin^{n-2}\theta} &= \frac{\int_0^\pi \dd \theta\ \left[\sin^{n-2}\theta\  - 2\sin^{n}\theta + \sin^{n+2}\theta\right]\ \cos^2 \theta}{\int_0^\pi \dd \theta\ \sin^{n-2}\theta}\nonumber \\
    &= \frac{1}{n} + \frac{1}{2}\frac{\Gamma\left(\frac{n}{2}\right)}{\Gamma\left(\frac{n-1}{2}\right)}\left[\frac{\Gamma\left(\frac{n-1}{2}+2\right)}{\Gamma\left(\frac{n}{2}+3\right)} - 2\frac{\Gamma\left(\frac{n-1}{2}+1\right)}{\Gamma\left(\frac{n}{2}+2\right)}\right]\nonumber \\
    &= \frac{15}{n\left(n+2\right)\left(n+4\right)}\ .
\end{eqnarray}
We also need to calculate moments of the form $\mathbb{E}[v_1^4 v_2^2]$. This amounts to computing
\begin{eqnarray}
    &\frac{\int_0^\pi \dd \theta\ \left[\sin^{n}\theta\  - \sin^{n+2}\theta\right]\cos^2 \theta \int_0^\pi \dd \phi\ \sin^{n-3}\phi\ \cos^2 \phi}{\int_0^\pi \dd \theta\ \sin^{n-2}\theta \int_0^\pi \dd \phi\ \sin^{n-3}\phi} \nonumber \\
    &= \frac{1}{4}\left[\frac{\Gamma\left(\frac{n+1}{2}\right)}{\Gamma\left(\frac{n}{2}+2\right)} - \frac{\Gamma\left(\frac{n+3}{2}\right)}{\Gamma\left(\frac{n}{2}+3\right)} \right]
    \frac{\Gamma\left(\frac{n}{2}\right)}{\Gamma\left(\frac{n+1}{2}\right)}\nonumber \\
    %
    %
    &= \frac{3}{n\left(n+2\right)\left(n+4\right)}\ .
\end{eqnarray}
Finally, we need to calculate moments of the form $\mathbb{E}[v_1^2 v_2^2 v_3^2]$. This amounts to computing
\begin{eqnarray}
    &\frac{\int_0^\pi \dd \theta\ \sin^{n+2}\theta\ \cos^2 \theta \int_0^\pi \dd \phi\ \sin^{n-1}\phi\ \cos^2 \phi  \int_0^\pi \dd \chi\ \sin^{n-4}\chi\ \cos^2 \chi}{\int_0^\pi \dd \theta\ \sin^{n-2}\theta \int_0^\pi \dd \phi\ \sin^{n-3}\phi  \int_0^\pi \dd \chi\ \sin^{n-4}\chi\ \cos^2 \chi} \nonumber \\
    &= \frac{1}{8}\frac{\Gamma\left(\frac{n+3}{2}\right) / \Gamma\left(\frac{n+4}{2}+1\right)}{\Gamma\left(\frac{n-1}{2}\right) / \Gamma\left(\frac{n}{2}\right)}
    \frac{\Gamma\left(\frac{n}{2}\right) / \Gamma\left(\frac{n+1}{2}+1\right)}{\Gamma\left(\frac{n-2}{2}\right) / \Gamma\left(\frac{n-1}{2}\right)}
    \frac{\Gamma\left(\frac{n-3}{2}\right) / \Gamma\left(\frac{n-2}{2}+1\right)}{\Gamma\left(\frac{n-3}{2}\right) / \Gamma\left(\frac{n-2}{2}\right)}
    \nonumber \\
    %
    %
    %
    %
    %
    %
    %
    %
    %
    %
    %
    %
    &= \frac{1}{8}\frac{\Gamma\left(\frac{n}{2}\right)}{\Gamma\left(\frac{n}{2}+3\right)}
     = \frac{1}{n\left(n + 2\right)\left(n + 4\right)}\nonumber \\
\end{eqnarray}

\subsection{Moments of orthogonal vectors}
\label{app:moments_ortho}

Consider two random unit vectors $\bm{v},\bm{\tilde w}\in\mathbb{R}^n$. We can define a $\bm{w}\perp\bm{v}$ via
\begin{eqnarray}
    \bm{w} &= \bm{\tilde w} - (\bm{\tilde w}\cdot\bm{v})\, \bm{v}\nonumber \\
    \Rightarrow w_i &= \tilde w_i - v_i \sum_j v_j \tilde w_j\ .
\end{eqnarray}
Note that $\bm{w}$ is not yet normalised. We will address this missing normalisation later, but for now let us consider joint moments of $\bm{w}$ and $\bm{v}$. We start with
\begin{eqnarray}
    \mathbb{E}[ w_1^2 v_1^2 ] &= \mathbb{E}[ \tilde w_1^2]\,\mathbb{E}[ v_1^2] - 2\mathbb{E}[ \tilde w_1 v_1^3 \sum_j v_j \tilde w_j ] + \mathbb{E}[ v_1^4 \sum_{jk} v_jv_k \tilde w_j \tilde w_k ] \nonumber \\
    &= \mathbb{E}[ \tilde w_1^2]\,\mathbb{E}[ v_1^2] - 2 \mathbb{E}[ \tilde w_1^2]\, \mathbb{E}[ v_1^4] + \sum_j \mathbb{E}[ \tilde w_j^2]\, \mathbb{E}[ v_1^4 v_j^2 ]\nonumber \\
    &= \frac{1}{n^2} - \frac{6}{n^2(n+2)} + \frac{3n-3}{n^2\left(n+2\right)\left(n+4\right)} + \frac{15}{n^2(n+2)(n+4)}\nonumber \\
    &= \frac{1}{n^2} - \frac{3}{n^2\left(n+2\right)}\nonumber \\
    &= \frac{n-1}{n^2(n+2)}\ .
\end{eqnarray}

We will also require the moment
\begin{eqnarray}
    \mathbb{E}[ w_1^2 v_2^2 ] &= \mathbb{E}[ \tilde w_1^2]\,\mathbb{E}[ v_2^2] - 2\mathbb{E}[ \tilde w_1 v_1 v_2^2 \sum_j v_j \tilde w_j ] + \mathbb{E}[ v_1^2 v_2^2 \sum_{jk} v_jv_k \tilde w_j \tilde w_k ] \nonumber \\
    &= \mathbb{E}[ \tilde w_1^2]\,\mathbb{E}[ v_2^2] - 2 \mathbb{E}[ \tilde w_1^2]\, \mathbb{E}[ v_1^2 v_2^2] + \sum_j \mathbb{E}[ \tilde w_j^2]\, \mathbb{E}[ v_1^2 v_2^2 v_j^2 ]\nonumber \\
    &= \frac{1}{n^2} - \frac{2}{n^2(n+2)} + \frac{n-2}{n^2(n+2)(n+4)} + \frac{6}{n^2(n+2)(n+4)}\nonumber \\
    &= \frac{1}{n^2} - \frac{1}{n^2(n+2)}\nonumber \\
    &= \frac{n+1}{n^2(n+2)}\ .
\end{eqnarray}
Finally, we will also need
\begin{eqnarray}
   \hspace*{-2cm} \mathbb{E}[ w_1 w_2 v_1 v_2 ] &= - \mathbb{E}[ \tilde w_1 v_2^2 v_1 \sum_j v_j \tilde w_j] - \mathbb{E}[ \tilde w_2 v_2 v_1^2 \sum_j v_j \tilde w_j] + \mathbb{E}[ v_1^2 v_2^2  \sum_{jk} v_jv_k \tilde w_j \tilde w_k ] \nonumber \\
    &= - \mathbb{E}[ \tilde w_1^2]\, \mathbb{E}[ v_1^2 v_2^2 ] - \mathbb{E}[ \tilde w_2^2]\, \mathbb{E}[ v_1^2 v_2^2 ] + \sum_j \mathbb{E}[ \tilde w_j^2]\, \mathbb{E}[ v_1^2 v_2^2 v_j^2 ]\nonumber \\
    &= - \frac{2}{n^2(n+2)} + \frac{n-2}{n^2(n+2)(n+4)} + \frac{6}{n^2(n+2)(n+4)}\nonumber \\
    &=  - \frac{1}{n^2(n+2)}\ .
\end{eqnarray}
For later calculations we actually require the moments between $\bm{v}$ and a normalised version of $\bm{w}$,
\begin{eqnarray}
    \bm{\hat w} &\equiv \frac{\bm{w}}{|\bm{w}|} = \frac{\bm{w}}{\sqrt{1 - (\bm{v}\cdot \bm{\tilde w})^2}}\ .
\end{eqnarray}
In the following we approximate such moments as
\begin{eqnarray}
    \mathbb{E}[ \hat w_i \hat w_j v_k v_l ] &= \mathbb{E}\left[ \frac{w_i w_j v_k v_l}{1 - (\bm{v}\cdot \bm{\tilde w})^2} \right] \approx \frac{\mathbb{E}[ w_i w_j v_k v_l ]}{1 - \mathbb{E}[(\bm{v}\cdot \bm{\tilde w})^2]}\ .
\end{eqnarray}
Corrections to this approximation will involve moments of at least 68 vector components. Correspondingly these corrections will be of higher order, and vanish in the $n\rightarrow \infty$ limit. Since $\mathbb{E}[(\bm{v}\cdot \bm{\tilde w})^2] = 1/n$, our moments of interest hence become
\begin{eqnarray}
\label{eq:moments_of_ortho_0}
    \mathbb{E}[ \hat w_1^2 v_1^2 ] &\approx \frac{n-1}{n^2(n+2)} \frac{1}{1 - 1/n} = \frac{1}{n (n+2)}\\
\label{eq:moments_of_ortho_1}
    \mathbb{E}[ \hat w_1^2 v_2^2 ] &\approx \frac{n+1}{n^2(n+2)} \frac{1}{1 - 1/n} = \frac{n+1}{(n-1) n (n+2)}\\
\label{eq:moments_of_ortho_2}
    \mathbb{E}[ \hat w_1 \hat w_2 v_1 v_2 ] &\approx - \frac{1}{n^2(n+2)} \frac{1}{1 - 1/n} = \frac{-1}{(n-1)n(n+2)}\ .
\end{eqnarray}
Note especially that
\begin{eqnarray}
    \mathbb{E}[ \hat w_1^2 v_1^2 ] &= \mathbb{E}[ \hat w_1^2 ]\, \mathbb{E}[ v_1^2 ] + \mathcal{O}\left[\frac{1}{n^3}\right]\\
    \mathbb{E}[ \hat w_1^2 v_2^2 ] &= \mathbb{E}[ \hat w_1^2 ]\, \mathbb{E}[ v_2^2] + \mathcal{O}\left[\frac{1}{n^3}\right] \\
    \mathbb{E}[ \hat w_1 \hat w_2 v_1 v_2 ] &= 0 + \mathcal{O}\left[\frac{1}{n^3}\right]\ .
\end{eqnarray}
Together with the fact that $\mathbb{E}[ v_i \hat w_j ] = 0$ we take this as justification to treat $\bm{v}$ and $\bm{\hat w}$ as independent random vectors in the large-$n$ limit, and in particular to ignore the mixed moments $\mathbb{E}[ \hat w_1 \hat w_2 v_1 v_2 ]$ in the following calculations.

\subsection{The Johnson-Lindenstrauss theorem}
\label{app:JL_theorem}

We prove here a slightly different version of the JL-theorem than the one considered by \cite{Dasgupta2003}:\\

\noindent \textbf{Theorem} \emph{Johnson-Lindenstrauss} -- Consider a $k\times n$ dimensional matrix $\bm{V} = (\bm{v}_1\, ,\, \dots\, ,\, \bm{v}_n)$, obtained by drawing each element from a standard normal distribution $\mathcal{N}(0,1)$ and then normalising the column vectors of $\bm{V}$ such that $|\bm{v}_i|^2 = 1$~. Now take any $n$ vectors $\bm{x}_1\, ,\, \dots\, ,\, \bm{x}_n \in \mathbb{R}^n$~. As long as
\begin{eqnarray}
    k \geq \frac{8}{\epsilon^2}\ln\left(\frac{n}{\sqrt{\delta}}\right)
\end{eqnarray}
all pairwise squared distances between the $\bm{x}_i$ are preserved within precision $\pm\epsilon$ with probability $\geq 1-\delta$~. More precisely,
\begin{eqnarray}
\label{eq:JL_statement}
    (1-\epsilon) |\bm{x}_i - \bm{x}_j|^2 \leq |\bm{Vx}_i - \bm{Vx}_j|^2 \leq (1+\epsilon) |\bm{x}_i - \bm{x}_j|^2
\end{eqnarray}
for all pairs $\bm{x}_i, \bm{x}_j$ with probability $\geq 1 - \delta$~.
\\

\noindent An important detail here is the statement that \eqnref{JL_statement} is holds for \emph{all} pairs. If it would only hold for an individual pair then there would be no reason why $k$ should depend on $n$, since the statement would just be one about random angles in $\mathbb{R}^k$~.

To prove this modified version of the JL theorem, we first quote \cite{Dasgupta2003} who showed that with the above choice of $\bm{V}$ the probability that \eqnref{JL_statement} is not satisfied for some individual pair of vectors is given by
\begin{eqnarray}
    P\left(\left|\frac{|\bm{x}_i - \bm{x}_j|^2 - |\bm{Vx}_i - \bm{Vx}_j|^2}{|\bm{x}_i - \bm{x}_j|^2}\right| > \epsilon\right) \leq 2 \exp\left(-\frac{k \epsilon^2}{4}\right)\ .
\end{eqnarray}
With our lower bound on $k$ this becomes
\begin{eqnarray}
    P\left(\left|\frac{|\bm{x}_i - \bm{x}_j|^2 - |\bm{Vx}_i - \bm{Vx}_j|^2}{|\bm{x}_i - \bm{x}_j|^2}\right| > \epsilon\right) &\leq 2 \exp\left(-2\ln\left(\frac{n}{\sqrt{\delta}}\right)\right)\nonumber \\
 &= \frac{2\delta}{n^2}\ .
\end{eqnarray}
Taking a union bound over all pairs of points the probability of having at least one pair violating \eqnref{JL_statement} is then given by
\begin{eqnarray}
    P\left(\left|\frac{|\bm{x}_i - \bm{x}_j|^2 - |\bm{Vx}_i - \bm{Vx}_j|^2}{|\bm{x}_i - \bm{x}_j|^2}\right| > \epsilon\ \mathrm{for\ at\ least\ on\ pair}\right) &\leq \left(\matrix{n \cr 2}\right)\frac{2\delta}{n^2}\nonumber \\
    &\leq \delta\ .
\end{eqnarray}
So \eqnref{JL_statement} is satisfied for all pairs with probability $\geq 1-\delta$~. $\square$

Of course the theorem stated above is only a statement about distances between pairs of vectors, while in \figref{epsilon_of_k} or \eqnref{CJL_bound_intro} we made statements about the scalar products between pairs of vectors. We simply quote here \cite{Engebretsen2002}, who have shown that these are two equivalent viewpoints on the JL-theorem. In particular, they showed that
\\

\noindent \textbf{Corollary} -- Consider the same situation as stated in our above version of the JL-theorem. Then 
\begin{eqnarray}
\label{eq:JL_statement_angles}
\bm{x}_i\cdot \bm{x}_j -\epsilon \leq \bm{Vx}_i \cdot \bm{Vx}_j \leq \bm{x}_i\cdot \bm{x}_j + \epsilon
\end{eqnarray}
for all pairs $\bm{x}_i, \bm{x}_j$ with probability $\geq 1 - \delta$~.\\

\noindent If the vectors $\bm{x}_i$ are chosen to be orthonormal, this in particular means that $ | \bm{Vx}_i \cdot \bm{Vx}_j | < \epsilon$~, and this is in fact the statement we have used in the main text. There, the anti-commutator between two different, overlapping operators $\hat c_{\bm{p}}, \hat c_{\bm{q}}$ in the same Fourier space shell was given by
\begin{eqnarray}
\lbrace \hat c_{\bm{p}}^\dagger , \hat c_{\bm{q}} \rbrace &= \frac{\braket{v_{\bm{p}}|v_{\bm{q}}}+\braket{w_{\bm{p}}|w_{\bm{q}}}+i\braket{v_{\bm{p}}|w_{\bm{q}}}-i\braket{w_{\bm{p}}|v_{\bm{q}}}}{2}\ ,
\end{eqnarray}
where $\lbrace \ket{v_{\bm{p}}}, \ket{w_{\bm{p}}} \rbrace_{\bm{p} \in s}$ is a set of $2 N_s$ normalised vectors in $\mathbb{R}^{2 n_s}$~, $n_s$ being the number of physical qubits present in the Fourier space shell $s$ and $N_s$ the number of Fourier modes in that shell.

Hence, if in our above formulation of the JL-theorem we set $k = 2n_s$ and $n = 2N_s$, and if we draw the $\lbrace \ket{v_{\bm{p}}}, \ket{w_{\bm{p}}} \rbrace$ from orthonormal $\lbrace\bm{x}_i\rbrace$ via the random matrix $\bm{V}$ described above, then we can derive that
\begin{eqnarray}
|\lbrace \hat c_{\bm{p}}^\dagger , \hat c_{\bm{q}} \rbrace| &= \frac{|\braket{v_{\bm{p}}|v_{\bm{q}}}+\braket{w_{\bm{p}}|w_{\bm{q}}}+i\braket{v_{\bm{p}}|w_{\bm{q}}}-i\braket{w_{\bm{p}}|v_{\bm{q}}}|}{2}\nonumber \\
 &\leq \frac{|\braket{v_{\bm{p}}|v_{\bm{q}}}|+|\braket{w_{\bm{p}}|w_{\bm{q}}}|+|\braket{v_{\bm{p}}|w_{\bm{q}}}| + |\braket{w_{\bm{p}}|v_{\bm{q}}}|}{2}\nonumber \\
 &\leq \epsilon 
\end{eqnarray}
for \emph{all} mode pairs $\bm{p},\bm{q}$ with probability $> 1-\delta$ as long as we choose
\begin{eqnarray}
    n_s \geq \frac{4}{(\epsilon/2)^2}\ln\left(\frac{2 N_s}{\sqrt{\delta}}\right) = \frac{16}{\epsilon^2}\ln\left(\frac{2 N_s}{\sqrt{\delta}}\right)\ .
\end{eqnarray}
For fix $n_s$, $N_s$ and $\delta$ we can turn this around and compute the smallest mode overlap achievable to be
\begin{eqnarray}
    \epsilon_{\mathrm{best}} = \sqrt{\frac{16}{n_s} \ln\left(\frac{2 N_s}{\sqrt{\delta}}\right)}\ .
\end{eqnarray}

\subsection{Angles between complex vectors}
\label{app:complex_angles}

Consider two complex vectors
\begin{eqnarray}
    \bm{v} = \bm{x} + i\bm{y}\ \ ,\ \ \bm{w} = \bm{a} + i\bm{b}\ ,
\end{eqnarray}
with $\bm{x},\bm{y},\bm{a},\bm{b}\in \mathbb{R}^n$ being Gaussian random vectors like we have considered before in this section. Normalising $\bm{v}$ and $\bm{w}$ yields the vectors
\begin{eqnarray}
    \bm{\hat v} = \frac{\bm{x} + i\bm{y}}{\sqrt{|\bm{x}|^2 + |\bm{y}|^2}}\ \ ,\ \ \bm{\hat w} = \frac{\bm{a} + i\bm{b}}{\sqrt{|\bm{a}|^2 + |\bm{b}|^2}}\ .
\end{eqnarray}
The Hermitian product between these vectors is
\begin{eqnarray}
    \bm{\hat v}^\dagger \bm{\hat w} &= \frac{[(\bm{x}^T \bm{a}) + (\bm{y}^T \bm{b})] + i[(\bm{x}^T \bm{b}) - (\bm{y}^T \bm{a})]}{\sqrt{|\bm{x}|^2 + |\bm{y}|^2}\sqrt{|\bm{a}|^2 + |\bm{b}|^2}}\ .
\end{eqnarray}
For our derivation of the lifetime of plane waves in the holographic Weyl field in \ref{app:Heisenberg}, we need to calculate the expectation value of $|\bm{\hat v}^\dagger \bm{\hat w}|^2$, which is given by
\begin{eqnarray}
\label{eq:complex_overlap}
    \mathbb{E}\left[|\bm{\hat v}^\dagger \bm{\hat w}|^2 \right] &= \frac{\mathbb{E}\left[\lbrace (\bm{x}^T \bm{a}) + (\bm{y}^T \bm{b})\rbrace^2 \right]}{(|\bm{x}|^2 + |\bm{y}|^2)(|\bm{a}|^2 + |\bm{b}|^2)} + \frac{\mathbb{E}\left[\lbrace (\bm{x}^T \bm{b}) - (\bm{y}^T \bm{a})\rbrace^2 \right]}{(|\bm{x}|^2 + |\bm{y}|^2)(|\bm{a}|^2 + |\bm{b}|^2)}\nonumber \\
\end{eqnarray}
Let us define the vectors
\begin{eqnarray}
    \bm{r} \equiv \pmatrix{\bm{x} \cr \bm{y}}\ \ ,\ \ \bm{s} \equiv \pmatrix{-\bm{y} \cr \bm{x}}\ \ ,\ \ \bm{c} \equiv \pmatrix{\bm{a} \cr \bm{b}}\ .
\end{eqnarray}
Then \eqnref{complex_overlap} becomes
\begin{eqnarray}
\label{eq:complex_overlap_2}
    \mathbb{E}\left[|\bm{\hat v}^\dagger \bm{\hat w}|^2 \right] &= \frac{\mathbb{E}\left[(\bm{r}^T \bm{c})^2 \right]}{|\bm{r}|^2|\bm{c}|^2} + \frac{\mathbb{E}\left[(\bm{s}^T \bm{c})^2 \right]}{|\bm{s}|^2|\bm{c}|^2}\ .
\end{eqnarray}
The terms on the right hand side of this equation are expectation values of products of uniform random vectors on the unit sphere in $\mathbb{R}^{2n}$. Our results from \ref{app:real_angles} then tell us that
\begin{eqnarray}
\label{eq:complex_overlap_3}
    \mathbb{E}\left[|\bm{\hat v}^\dagger \bm{\hat w}|^2 \right] &= \frac{2}{2n} = \frac{1}{n}\ .
\end{eqnarray}

\subsection{Moments of the overlaps \texorpdfstring{$\braket{z_{\bm{q}} | z_{\bm{p}}}$}{}}
\label{app:z_overlap}

Let $\bm{p}$ be a wave mode located in some Fourier space shell $s$, and remember that $\ket{z_{\bm{p}}} = \ket{v_{\bm{p}}} + i\ket{w_{\bm{p}}}$ with
\begin{eqnarray}
\braket{v_{\bm{p}}|v_{\bm{p}}} = 1 = \braket{w_{\bm{p}}|w_{\bm{p}}}\ \ ,\ \ \braket{v_{\bm{p}}|w_{\bm{p}}} = 0\ .
\end{eqnarray}
For another mode $\bm{q}$ in the same shell we have
\begin{eqnarray}
    \braket{z_{\bm{q}}|z_{\bm{p}}} &= \left(\, \bra{v_{\bm{q}}} - i\bra{w_{\bm{q}}}\, \right)\left(\, \ket{v_{\bm{p}}} + i\ket{w_{\bm{p}}}\, \right)\nonumber \\
    &= \braket{v_{\bm{q}}|v_{\bm{p}}} + \braket{w_{\bm{q}}|w_{\bm{p}}} + i \left(\,\braket{v_{\bm{q}}|w_{\bm{p}}} - \braket{w_{\bm{q}}|v_{\bm{p}}}\,\right)\ ,
\end{eqnarray}
while
\begin{eqnarray}
    \braket{z_{\bm{q}}^*|z_{\bm{p}}} &= \left(\, \bra{v_{\bm{q}}} + i\bra{w_{\bm{q}}}\, \right)\left(\, \ket{v_{\bm{p}}} + i\ket{w_{\bm{p}}}\, \right)\nonumber \\
    &= \braket{v_{\bm{q}}|v_{\bm{p}}} - \braket{w_{\bm{q}}|w_{\bm{p}}} + i \left(\,\braket{v_{\bm{q}}|w_{\bm{p}}} + \braket{w_{\bm{q}}|v_{\bm{p}}}\,\right)\ .
\end{eqnarray}
For $\bm{q}\neq\bm{p}$ all of the above products between the $\lbrace \ket{v_{\bm{p}}}, \ket{w_{\bm{p}}}, \ket{v_{\bm{q}}}, \ket{w_{\bm{q}}} \rbrace$ have a variance of $1/n_{s}$, where $n_{s}$ is the total number of modes in the shell $s$. At the same time, the covariance between these vector products is $0$. 

For the moments of the complex products $\braket{z_{\bm{q}}|z_{\bm{p}}},\ \braket{z_{\bm{q}}^*|z_{\bm{p}}}$ this means
\begin{eqnarray}
    \mathbb{E}\left[ \braket{z_{\bm{q}}|z_{\bm{p}}}^2 \right] &= \frac{1}{2 n_s} + \frac{1}{2 n_s} - \frac{1}{2 n_s} - \frac{1}{2 n_s}\nonumber \\
    &= 0\\
    \mathbb{E}\left[ \braket{z_{\bm{q}}^*|z_{\bm{p}}}^2 \right] &= 0 \\
    \mathbb{E}\left[ \braket{z_{\bm{q}}|z_{\bm{p}}}\braket{z_{\bm{q}}^*|z_{\bm{p}}} \right] &= \frac{1}{2 n_s} - \frac{1}{2 n_s} - \frac{1}{2 n_s} + \frac{1}{2 n_s}\nonumber \\
    &= 0\\
    \mathbb{E}\left[ |\braket{z_{\bm{q}}|z_{\bm{p}}}|^2 \right] &= \frac{2}{n_s}\ = \mathbb{E}\left[ |\braket{z_{\bm{q}}^*|z_{\bm{p}}}|^2 \right]\ .
\end{eqnarray}

\section{Our construction as a Heisenberg model}
\label{app:Heisenberg}

In this appendix we will show that the Hamiltonian $\hat H_s$ in a single Fourier space shell $s$ of our holographic Weyl field can be re-interpreted as that of a Heisenberg model with general non-local, two-site interactions. In a second step we then re-formulate that Heisenberg Hamiltonian to resemble the Hamiltonian of a set of non-interacting spins in a stochastic, spatially varying magnetic field. This will allow us to characterise the energy spectrum of our field, and in particular the average vacuum energy of our construction. Similar calculations will then also yield the typical lifetime of plane wave states.

 In \secref{JL_embedding} we had defined
\begin{eqnarray}
    \bm\sigma_{x,\bm{p}}^c \equiv \sum_{j=1}^{2n} \braket{e_j|v_{\bm{p}}} \bm{C}_j\ \ ,\ \ \bm\sigma_{y,\bm{p}}^c \equiv \sum_{j=1}^{2n} \braket{e_j|w_{\bm{p}}} \bm{C}_j\ \ ,\ \ \bm\sigma_{z,\bm{p}}^c &= -i \bm\sigma_{x,\bm{p}}^c \bm\sigma_{y,\bm{p}}^c\nonumber
\end{eqnarray}
as well as
\begin{eqnarray}
    \bm{c}_{\bm{p}} = \frac{1}{2}\left(\bm\sigma_{x,\bm{p}}^c + i\bm\sigma_{y,\bm{p}}^c\right)\ .\nonumber
\end{eqnarray}
(Note the difference with the original CRSV-embedding because we want to produce approximately anti-commuting sets of operators instead of approximately commuting ones.) Each $\bm{p}$ represents one of the overlapping qubits of the $c$-part (i.e.\ the particle-part) our field. But since the dimension of the Hilbert space of Fourier space shell $s$ is a power of two ($\dim\mathcal{H}_s = 2^{n_s}$) we can also decompose our construction into a set of non-overlapping qubits. Let us e.g.\ define the creation and annihilation operators of these qubits as
\begin{eqnarray}
    \hat{A}_j \equiv \frac{1}{2}\left(\bm{C}_{2j-1} + i\bm{C}_{2j}\right)\ ,\ \hat{A}_j^\dagger \equiv \frac{1}{2}\left(\bm{C}_{2j-1} - i\bm{C}_{2j}\right)\nonumber\\
    \Rightarrow \bm{C}_{2j-1} = \hat{A}_j + \hat{A}_j^\dagger\ ,\ \bm{C}_{2j} = -i(\hat{A}_j - \hat{A}_j^\dagger)\ .\nonumber
\end{eqnarray}
If we furthermore set
\begin{eqnarray}
    f_{\bm{p}, j} \equiv \braket{e_{2j-1}|v_{\bm{p}}} + i \braket{e_{2j}|v_{\bm{p}}}\ ,\ g_{\bm{p}, j} \equiv \braket{e_{2j-1}|w_{\bm{p}}} + i \braket{e_{2j}|w_{\bm{p}}}\nonumber
\end{eqnarray}
then we have
\begin{eqnarray}
    &\ \ \ \ i\bm\sigma_{z,\bm{p}}^c\nonumber \\
    &=  \bm\sigma_{x,\bm{p}}^c \bm\sigma_{y,\bm{p}}^c \nonumber \\
    &= \sum_{j=1}^n \sum_{k=1}^n \left(\braket{e_{2j-1}|v_{\bm{p}}} \bm{C}_{2j-1} + \braket{e_{2j}|v_{\bm{p}}} \bm{C}_{2j} \right)\left(\braket{e_{2k-1}|w_{\bm{p}}} \bm{C}_{2k-1} + \braket{e_{2k}|w_{\bm{p}}} \bm{C}_{2k} \right)\nonumber \\
    &= \sum_{j=1}^n \sum_{k=1}^n \left( f_{\bm{p}, j}^* \hat{A}_j + f_{\bm{p}, j} \hat{A}_j^\dagger  \right) \left(g_{\bm{p}, k}^* \hat{A}_k + g_{\bm{p}, k} \hat{A}_k^\dagger   \right)\nonumber \\
    &= \sum_{j=1}^n \sum_{k=1}^n \left( f_{\bm{p}, j}^* g_{\bm{p}, k}^* \hat{A}_j\hat{A}_k + f_{\bm{p}, j}^* g_{\bm{p}, k} \hat{A}_j\hat{A}_k^\dagger  + f_{\bm{p}, j}g_{\bm{p}, k}^* \hat{A}_j^\dagger \hat{A}_k  + f_{\bm{p}, j}g_{\bm{p}, k} \hat{A}_j^\dagger \hat{A}_k^\dagger  \right)\nonumber \\
\end{eqnarray}
If we define the matrices $\bm{M}$ and $\bm{X}$ via
\begin{eqnarray}
    M_{jk} \equiv -i\sum_{\bm{p}\in s} f_{\bm{p}, j}^* g_{\bm{p}, k}^*\ ,\ X_{jk} \equiv -i\sum_{\bm{p}\in s} f_{\bm{p}, j}^* g_{\bm{p}, k}
\end{eqnarray}
then the Hamiltonian in the Fourier space shell $s$ (with $\Delta_s \ll k_s$) becomes
\begin{eqnarray}
\label{eq:Heisenberg_Hamiltonian_0}
    \hat H_s &\approx \frac{k_s}{2}\sum_{\bm{p}\in s} \bm\sigma_{z,\bm{p}}^c \nonumber \\
    &=\frac{k_s}{2}\sum_{j,k=1}^n \left( M_{jk} \hat{A}_j\hat{A}_k  + X_{jk} \hat{A}_j \hat{A}_k^\dagger  - X_{jk}^* \hat{A}_j^\dagger \hat{A}_k  - M_{jk}^* \hat{A}_j^\dagger \hat{A}_k^\dagger  \right)\ .
\end{eqnarray}
So the Hamiltonian in each Fourier space shell of our construction is equivalent to that of a general, non-local Heisenberg model with 2-site interactions. 

\subsection{Energy spectrum of the holographic field}
\label{app:energy_spectrum}

Using the anti-commutation relations between the (non-overlapping) operators $\hat{A}_j, \hat{A}_k, \hat{A}_j^\dagger, \hat{A}_k^\dagger$ we can reformulate \eqnref{Heisenberg_Hamiltonian_0} as
\begin{eqnarray}
\label{eq:Heisenberg_Hamiltonian_1}
    \hat H_s &= \frac{k_s}{2}\sum_{j,k=1}^n  \left\lbrace \frac{M_{jk} - M_{kj}}{2} \hat{A}_j\hat{A}_k -  \frac{M_{jk}^* - M_{kj}^*}{2} \hat{A}_j^\dagger\hat{A}_k^\dagger \right\rbrace\nonumber\\
    &\ \ \ + \frac{k_s}{2}\sum_{j,k=1}^n  \left\lbrace \frac{1}{2} X_{jk} \hat{A}_j \hat{A}_k^\dagger  + \frac{1}{2} X_{jk} \delta_{jk} - \frac{1}{2} X_{jk} \hat{A}_k^\dagger \hat{A}_j  \right\rbrace\nonumber\\
    &\ \ \ - \frac{k_s}{2}\sum_{j,k=1}^n  \left\lbrace \frac{1}{2} X_{jk}^* \hat{A}_j^\dagger \hat{A}_k + \frac{1}{2} X_{jk}^* \delta_{jk} - \frac{1}{2} X_{jk}^* \hat{A}_k \hat{A}_j^\dagger \right\rbrace\ ,
\end{eqnarray}
Note that 
\begin{eqnarray}
    \hspace*{-1.5cm}\sum_{j=1}^n X_{jj} &= -i \sum_{\bm{p}\in s} \sum_{j=1}^n \left\lbrace\braket{e_{2j}|v_{\bm{p}}} - i \braket{e_{2j+1}|v_{\bm{p}}}\right\rbrace\left\lbrace\braket{e_{2j}|w_{\bm{p}}} + i \braket{e_{2j+1}|w_{\bm{p}}}\right\rbrace\nonumber \\
    &=  -i \sum_{\bm{p}\in s} \left[ \braket{v_{\bm{p}}|w_{\bm{p}}} + i \sum_{j=1}^n \left\lbrace \braket{e_{2j}|v_{\bm{p}}}\braket{e_{2j+1}|w_{\bm{p}}} - \braket{e_{2j+1}|v_{\bm{p}}}\braket{e_{2j}|w_{\bm{p}}} \right\rbrace  \right] \nonumber \\
    &= \sum_{\bm{p}\in s}\sum_{j=1}^n \left[ \braket{e_{2j}|v_{\bm{p}}}\braket{e_{2j+1}|w_{\bm{p}}} - \braket{e_{2j+1}|v_{\bm{p}}}\braket{e_{2j}|w_{\bm{p}}} \right\rbrace
\end{eqnarray}
which is real, such that the terms proportional to $X_{jk} \delta_{jk}$ and $X_{jk}^* \delta_{jk}$ in the second line of \eqnref{Heisenberg_Hamiltonian_1} cancel each other. We can then write the shell-Hamiltonian as
\begin{eqnarray}
\label{eq:Heisenberg_Hamiltonian_2}
     \hspace*{-2cm} \hat H_s = \frac{k_s}{2}\sum_{j,k=1}^n  \left\lbrace \frac{M_{jk} - M_{kj}}{2} \hat{A}_j\hat{A}_k +  \frac{M_{kj}^* - M_{jk}^*}{2} \hat{A}_j^\dagger\hat{A}_k^\dagger \right\rbrace\nonumber \\
    \hspace*{-1.8cm}\ \ \ + \frac{k_s}{2}\sum_{j,k=1}^n  \left\lbrace \frac{X_{jk} + X_{kj}^*}{2}  \hat{A}_j \hat{A}_k^\dagger  - \frac{X_{kj} + X_{jk}^*}{2}  \hat{A}_j^\dagger \hat{A}_k  \right\rbrace\nonumber \\
    \hspace*{-1.4cm}\equiv \frac{k_s}{2} \left(\hat{A}_1^\dagger,\ \dots\ ,\ \hat{A}_n^\dagger,\ \hat{A}_1,\ \dots\ ,\ \hat{A}_n\right) \pmatrix{ -\frac{\displaystyle \bm{X}^T + \bm{X}^*}{\displaystyle 2} & \frac{\displaystyle \bm{M}^\dagger - \bm{M}^*}{\displaystyle 2} \cr & \cr \frac{\displaystyle \bm{M} - \bm{M}^T}{\displaystyle 2} & \frac{\displaystyle \bm{X} + \bm{X}^\dagger}{\displaystyle 2}} \pmatrix{\hat{A}_1 \cr \dots \cr \hat{A}_n \cr \hat{A}_1^\dagger \cr \dots \cr \hat{A}_n^\dagger}\nonumber \\
    \hspace*{-1.5cm}= \frac{k_s}{2} \left(\hat{A}_1^\dagger,\ \dots\ ,\ \hat{A}_n^\dagger,\ \hat{A}_n,\ \dots\ ,\ \hat{A}_1\right) \pmatrix{ -\frac{\displaystyle \bm{X}^T + \bm{X}^*}{\displaystyle 2} & \frac{\displaystyle \bm{M}^\dagger - \bm{M}^*}{\displaystyle 2}\bm{R}_n \cr & \cr \bm{R}_n\frac{\displaystyle \bm{M} - \bm{M}^T}{\displaystyle 2} & \bm{R}_n\frac{\displaystyle \bm{X} + \bm{X}^\dagger}{\displaystyle 2}\bm{R}_n} \pmatrix{\hat{A}_1 \cr \dots \cr \hat{A}_n \cr \hat{A}_n^\dagger \cr \dots \cr \hat{A}_1^\dagger}\nonumber\ , \\
\end{eqnarray}
where we have grouped all creation and annihilation operators into one operator valued vector $\left(\hat{A}_1^\dagger,\ \dots\ ,\ \hat{A}_n^\dagger,\ \hat{A}_1,\ \dots\ ,\ \hat{A}_n\right)$ and where in the last line we have reversed the order of the 2nd half of that vector by multiplication with the matrix $\bm{R}_n$ (the $n\times n$ matrix whose entries are 1 on the cross-diagonal and 0 everywhere else).

The $2n\times 2n$ matrix that appears in the last line of \eqnref{Heisenberg_Hamiltonian_2} is both Hermitian and anti-persymmetric, i.e.\ it switches sign when it is flipped along the cross-diagonal. As we show in \ref{app:persymmetric_matrices}, the eigenvalues of such a matrix come in pairs $\pm \lambda$. And if $\bm{v}$ is an eigenvector with eigenvalue $\lambda$, then $\bm{R}_{2n}\bm{v}^*$, i.e.\ the complex conjugate of $\bm{v}$ multiplied by $\bm{R}_{2n}$ (which reverses the ordering of the elements of $\bm{v}^*$), is an eigenvector with eigenvalue $-\lambda$. So we can express the above matrix as
\begin{eqnarray}
    \hspace*{-2cm}\pmatrix{ -\frac{\displaystyle \bm{X}^T + \bm{X}^*}{\displaystyle 2} & \frac{\displaystyle \bm{M}^\dagger - \bm{M}^*}{\displaystyle 2}\bm{R}_n \cr & \cr \bm{R}_n\frac{\displaystyle \bm{M} - \bm{M}^T}{\displaystyle 2} & \bm{R}_n\frac{\displaystyle \bm{X} + \bm{X}^\dagger}{\displaystyle 2}\bm{R}_n} = \bm{U}^\dagger \pmatrix{\lambda_1 & & & & & \cr  & \ddots & & & & \cr  & & \lambda_n & & & \cr  & & & -\lambda_n & & \cr  & & &  & \ddots & \cr  & & &  & & -\lambda_1} \bm{U}\nonumber\ , \\
\end{eqnarray}
where without loss of generality we can assume that all $\lambda_i > 0$ and $\lambda_{i+1} \geq \lambda_i$, and where the unitary matrix $\bm{U}^\dagger$ consists of the columns
\begin{eqnarray}
    \bm{U}^\dagger = \pmatrix{\bm{v}_1 ,\ \dots\ ,\ \bm{v}_n,\  \bm{R}_{n}\bm{v}_n^*,\ \dots\ ,\ \bm{R}_{n}\bm{v}_1^*}\ .
\end{eqnarray}
We can then define new creation and annihilation operators via
\begin{eqnarray}
    \pmatrix{\hat{B}_1 \cr \dots \cr \hat{B}_n \cr \hat{B}_n^\dagger \cr \dots \cr \hat{B}_1^\dagger} \equiv \bm{U} \cdot \pmatrix{\hat{A}_1 \cr \dots \cr \hat{A}_n \cr \hat{A}_n^\dagger \cr \dots \cr \hat{A}_1^\dagger}\ .
\end{eqnarray}
This is well defined because
\begin{eqnarray}
    \hspace*{-2cm}\left[\hat{B}_i\right]^\dagger = \left[\bm{v}_i^\dagger \cdot \pmatrix{\hat{A}_1 \cr \dots \cr \hat{A}_n \cr \hat{A}_n^\dagger \cr \dots \cr \hat{A}_1^\dagger}\right]^\dagger\nonumber &= \left(v_i^{1*} \hat{A}_1 + \dots + v_i^{n*} \hat{A}_n + v_i^{n+1*} \hat{A}_n^\dagger + \dots + v_i^{2n*} \hat{A}_1^\dagger \right)^\dagger\nonumber \\
    &= \left(v_i^{1} \hat{A}_1^\dagger + \dots + v_i^{n} \hat{A}_n^\dagger + v_i^{n+1} \hat{A}_n + \dots + v_i^{2n} \hat{A}_1 \right)\nonumber \\
    &= \left(\bm{R}_{2n} \bm{v}_i^*\right)^\dagger \cdot \pmatrix{\hat{A}_1 \cr \dots \cr \hat{A}_n \cr \hat{A}_n^\dagger \cr \dots \cr \hat{A}_1^\dagger}\, =\, \hat{B}_i^\dagger\ .
\end{eqnarray}
It is also straight forward to show that $\lbrace \hat{B}_i , \hat{B}_j \rbrace = 0 = \lbrace \hat{B}_i^\dagger , \hat{B}_j^\dagger \rbrace$ and that $\lbrace \hat{B}_i , \hat{B}_j^\dagger \rbrace = \delta_{ij}$~. The shell-Hamiltonian thus becomes
\begin{eqnarray}
    \hat H_s &= \frac{k_s}{2}\sum_{i=1}^n \lambda_i \left( \hat{B}_i^\dagger \hat{B}_i - \hat{B}_i \hat{B}_i^\dagger  \right)\nonumber \\
    &= \frac{k_s}{2}\sum_{i=1}^n \lambda_i \left( 2\hat{B}_i^\dagger \hat{B}_i - 1 \right)\ .
\end{eqnarray}
So we have reduced our Hamiltonian to that of a set of decoupled spins with varying coupling $\lambda_i$ to an external field. Since we have (without loss of generality) assumed that all $\lambda_i > 0\,$, the minimum energy eigenstate of that Hamiltonian is the one that has occupation number $0$ in each mode $i$~. Hence, the minimum energy eigenvalue is
\begin{eqnarray}
\label{eq:minimum_E_from_lambdas}
    E_{\min } = -\frac{k_s}{2} \sum_{i=1}^n \lambda_i\ .
\end{eqnarray}
To derive an estimate of this energy we will use a number of results from random matrix theory \cite{Reid1997, Anderson_Zeitouni2004, Erdos2010a, Erdos2010b}. Let us first note that the matrix
\begin{eqnarray}
    \bm{K} \equiv \pmatrix{ -\frac{\displaystyle \bm{X}^T + \bm{X}^*}{\displaystyle 2} & \frac{\displaystyle \bm{M}^\dagger - \bm{M}^*}{\displaystyle 2}\bm{R}_n \cr & \cr \bm{R}_n\frac{\displaystyle \bm{M} - \bm{M}^T}{\displaystyle 2} & \bm{R}_n\frac{\displaystyle \bm{X} + \bm{X}^\dagger}{\displaystyle 2}\bm{R}_n}
\end{eqnarray}
can be interpreted as a draw from a random ensemble of matrices, since the vector components $\braket{e_{j}|v_{\bm{p}}}$ and $\braket{e_{j}|w_{\bm{p}}}$ appearing in the definition of $\bm{K}$ where chosen at random according to the JL-embedding scheme described in \secref{JL_embedding_basics}~. Clearly, the ensemble from which $\bm{K}$ is drawn belongs to the Hermitian matrices. Hence there will be an eigenvalue distribution associated with this ensemble, i.e.\ there will be a function $p(\lambda)$ such that $p(\lambda)\mathrm{d}\lambda$ is the probability of finding a randomly chosen eigenvalue of a matrix in the ensemble within the interval $[\lambda,\, \lambda+\dd\lambda]\,$. The minimum energy of \eqnref{minimum_E_from_lambdas} can then be approximated as
\begin{eqnarray}
\label{eq:minimum_E_from_lambdas_2}
    E_{\min } \approx - k_s n \int_{\lambda \geq 0} \dd \lambda\ \lambda\ p(\lambda)\ .
\end{eqnarray}
So \eqnref{minimum_E_from_lambdas} can be viewed as \say{Monte-Carlo}-integration of \eqnref{minimum_E_from_lambdas_2}, and the factor of $1/2$ in \eqnref{minimum_E_from_lambdas} disappears in \eqnref{minimum_E_from_lambdas_2} because we are only integrating over half of the probability density function $p(\lambda)$. Note especially, that $p(\lambda) = p(-\lambda)$ because the eigenvalues of $\bm{K}$ come in pairs $\pm\lambda$ as discussed above.

So in order to obtain a theoretical expectation value for $E_{\min}$ we need to determine the eigenvalue density $p(\lambda)$ of the random matrix ensemble from which $\bm{K}$ is drawn. We will use the results of \cite{Anderson_Zeitouni2004, Erdos2010a} to do so. First, let us calculate the sum of the variances of the elements in the $k$th row of $\bm{K}$. With our results from \ref{app:random_angles} one can show that this is given by
\begin{eqnarray}
\label{eq:row_variance}
        \sum_{l} \mathbb{E}\left[ |K_{kl}|^2 \right] &= \frac{1}{4} \sum_{l} \left\lbrace\mathbb{E}\left[ |X_{kl}^* + X_{lk}|^2 \right] + \mathbb{E}\left[ |M_{lk}^* - M_{kl}^*|^2 \right] \right\rbrace \nonumber \\
        &= \frac{1}{4} \sum_{l} \left\lbrace\mathbb{E}\left[ |X_{kl}|^2 \right] + \mathbb{E}\left[ |X_{lk}|^2 \right] + \mathbb{E}\left[ |M_{kl}|^2 \right] + \mathbb{E}\left[ |M_{lk}|^2 \right] \right\rbrace \nonumber \\
        &\approx \frac{1}{2} \sum_{l}\sum_{\bm{p}} \left\lbrace\mathbb{E}\left[ |f_{\bm{p}, k}|^2 \right]\mathbb{E}\left[ |g_{\bm{p}, l}|^2 \right] + \mathbb{E}\left[ |f_{\bm{p}, l}|^2 \right]\mathbb{E}\left[ |g_{\bm{p}, k}|^2 \right] \right\rbrace \nonumber \\
        &= \frac{1}{2} \sum_{l}\sum_{\bm{p}} \left\lbrace\frac{2}{2n_s}\frac{2}{2n_s} + \frac{2}{2n_s}\frac{2}{2n_s} \right\rbrace \nonumber \\
        &= \frac{N_s}{n_s}\ .
\end{eqnarray}
If we can interpret $\bm{K}$ as drawn from the generalised Wigner ensembles considered by \cite{Anderson_Zeitouni2004, Erdos2010a}, then we can conclude from \eqnref{row_variance} that in the asymptotic limit $n_s \rightarrow \infty$ the eigenvalue distribution $p(\lambda)$ is given by the semi-circle distribution

\begin{eqnarray}
    p(\lambda)\approx \left\lbrace\matrix{\frac{\displaystyle n_s}{\displaystyle 2\pi N_s} \sqrt{4 \frac{\displaystyle N_s}{\displaystyle n_s} - \lambda^2} \ \ \ \ \mathrm{for}\ |\lambda| < 2 \sqrt{\frac{\displaystyle N_s}{\displaystyle n_s}} \cr \cr 0 \ \ \ \mathrm{else}} \right.\ \ \ \ .
\end{eqnarray}
For the minimum eigenvalue of the shell Hamiltonian this gives
\begin{eqnarray}
\label{eq:minimum_E_from_lambdas_3}
    E_{\min } &\approx - \frac{ n_s^2 k_s}{ 2\pi N_s}  \int_{\lambda = 0}^{2 \sqrt{ N_s/ n_s}} \dd \lambda\ \lambda\ \sqrt{4 \frac{\displaystyle N_s}{\displaystyle n_s} - \lambda^2}\nonumber \\
    &= - \frac{ n_s^2 k_s}{ 2\pi N_s}  \left[- \frac{1}{3} \left(4 \frac{\displaystyle N_s}{\displaystyle n_s} - \lambda^2\right)^{\frac{3}{2}} \right]_{\lambda = 0}^{2 \sqrt{ N_s/ n_s}}\nonumber \\
    &= -\frac{N_s k_s}{2}\cdot\left(\frac{8}{3\pi} \sqrt{\frac{\displaystyle n_s}{\displaystyle N_s}}\right)\ .
\end{eqnarray}
So the vacuum energy seems to be suppressed \wrt the standard, non-overlapping Weyl field by a factor of $\frac{8}{3\pi} \sqrt{\frac{\displaystyle n_s}{\displaystyle N_s}}$~. The only caveat to this result is the fact that $\bm{K}$ indeed satisfies all the properties of generalised Wigner ensembles considered by \cite{Anderson_Zeitouni2004, Erdos2010a} \emph{except} that it has the additional property of anti-persymmetry. A priori, one might expect that this can change the width of $p(\lambda)$ by a factor of $\mathcal{O}(1)$. We do not repeat the proofs of \cite{Anderson_Zeitouni2004, Erdos2010a} in our modified situation and instead simply check in numerical experiments that the width of $p(\lambda)$ is \emph{exactly} the one predicted by those authors, \ie the impact of anti-persymmetry is solely that the eigenspectrum is symmetrised around $0$ without changing the asymptotic width of the eigenvalue distribution.

\subsection{Lifetime of plane waves}
\label{app:lifetime}

Let $\bm{k}\in s$ be a wave mode in the Fourier space shell $s$, and let us define the state corresponding to a plane wave excitation $\ket{\bm{k}}$ of that mode as in \secref{dynamics}. We will again decompose the shell-Hamiltonian as
\begin{eqnarray}
\label{eq:Heisenberg_Hamiltonian_3}
    \hat H_s = \frac{k_s}{2} \left(\hat{A}_1^\dagger,\ \dots\ ,\ \hat{A}_n^\dagger,\ \hat{A}_n,\ \dots\ ,\ \hat{A}_1\right) \bm{K} \pmatrix{\hat{A}_1 \cr \dots \cr \hat{A}_n \cr \hat{A}_n^\dagger \cr \dots \cr \hat{A}_1^\dagger}\nonumber\ , \\
\end{eqnarray}
as we had done in the previous section. Note that the properties of our modified Weyl field do not depend on the exact vector pairs $\bm{v}_{\bm{p}},\,\bm{w}_{\bm{p}}$ used to construct the different field modes $\hat c_{\bm{p}}\,$, but only on the scalar products between the vectors in different vector pairs. Thus we can always rotate all embedding vectors in such a way that the vectors $\bm{v}_{\bm{k}}$ and $\bm{w}_{\bm{k}}$ defining the mode $\hat c_{\bm{k}}$ are given by
\begin{eqnarray}
    \bm{v}_{\bm{k}} = (1,\ 0,\ 0,\ \dots)\ \ ,\ \ \bm{w}_{\bm{k}} = (0,\ 1,\ 0,\ \dots)\ .
\end{eqnarray}
This means that we can without loss of generality assume that
\begin{eqnarray}
    \hat N_{\bm{k}} = \hat{A}_1^\dagger\hat{A}_1\ 
\end{eqnarray}
or equivalently, that $\hat c_{\bm{k}} \equiv \hat{A}_1\,$.

In \secref{dynamics} we estimated the lifetime of the plane wave excitation as
\begin{eqnarray}
\label{eq:Tscramble_appendix}
    \frac{1}{T_{\mathrm{scramble}}^2} \approx -\mathbb{E}\left\lbrace\frac{\mathrm{d}^2}{\mathrm{d}t^2} \braket{\hat N_{\bm{k}}}\right\rbrace = \mathbb{E}\left\lbrace\bra{\bm{k}}\left[\hat{H}_s,\left[\hat{H}_s, \hat N_{\bm{k}}\right]\right]\ket{\bm{k}}\right\rbrace\ ,
\end{eqnarray}
where the expectation value $\mathbb{E}$ is taken \wrt the random vectors that are used in the CRSV-embedding. Note that the first time derivative of $\braket{\hat N_{\bm{k}}}$ vanishes since $\ket{\bm{k}}$ is an eigenstate of $\hat N_{\bm{k}}$. Hence, we consider the second time derivative to estimate a characteristic time scale. To evaluate the double-commutator between $\hat N_{\bm{k}}$ and $\hat H_s$, let us first decompose the latter as
\begin{eqnarray}
    \hat H_s = \frac{k_s}{2} \left\lbrace K_{11}\, (2\hat N_{\bm{k}}-1) + 2\hat{A}_1 \hat B - 2\hat{A}_1^\dagger \hat B^\dagger\right\rbrace + \hat H_B
\end{eqnarray}
where the operator $\hat B^\dagger$ (which should not be confused with the operators $\hat B_i^\dagger$ from the previous subsection) is given by 
\begin{eqnarray}
    \hat B^\dagger = \sum_{j>1}^{n_s} \left\lbrace K_{1,j} \hat{A}_j + K_{1,2n_s + 1 - j} \hat{A}_j^\dagger \right\rbrace
\end{eqnarray}
and where $\hat H_B$ is the part of $\hat H_s$ that doesn't contain any factors $\hat{A}_1$ or $\hat{A}_1^\dagger$. It is clear that
\begin{eqnarray}
    \left[\hat H_B, \hat N_{\bm{k}}\right] = 0 = \left[ \hat B^\dagger, \hat N_{\bm{k}}\right]\ ,
\end{eqnarray}
because $\hat{A}_1$ and $\hat{A}_j$ represent non-overlapping qbits for $j>1$ (and because $\hat N_{\bm{k}}$ contains the two factors $\hat{A}_1$ and $\hat{A}_1^\dagger$ such that anti-commutation becomes commutation). Hence the first commutator of $\hat H_{s}$ and $\hat N_{\bm{k}}$ becomes
\begin{eqnarray}
    \left[\hat H_s, \hat N_{\bm{k}}\right] &= k_s \left\lbrace \left[\hat{A}_1, \hat N_{\bm{k}}\right] \hat B - \left[\hat{A}_1^\dagger, \hat N_{\bm{k}}\right] \hat B^\dagger\right\rbrace\nonumber \\
    &= k_s \left\lbrace \hat{A}_1 \hat B + \hat{A}_1^\dagger \hat B^\dagger\right\rbrace\ .
\end{eqnarray}
Note that $\bra{\bm{k}} \hat{A}_1 \ket{\bm{k}} = 0 = \bra{\bm{k}} \hat{A}_1^\dagger \ket{\bm{k}}$. Hence, when applying the commutator with $\hat H_s$ a second time, only terms without single factors $\hat{A}_1$ or $\hat{A}_1^\dagger$ will contribute to \eqnref{Tscramble_appendix}. This leads us to
\begin{eqnarray}
    \frac{1}{T_{\mathrm{scramble}}^2} &= k_s^2\, \mathbb{E}\left\lbrace\bra{\bm{k}}\left\lbrace \left[\hat{A}_1 \hat B, \hat{A}_1^\dagger \hat B^\dagger\right] - \left[ \hat{A}_1^\dagger \hat B^\dagger,\hat{A}_1 \hat B\right]\right\rbrace\ket{\bm{k}}\right\rbrace \nonumber \\
    &= 2k_s^2\, \mathbb{E}\left\lbrace\bra{\bm{k}}\hat N_{\bm{k}} \lbrace \hat B, \hat B^\dagger\rbrace - \hat B \hat B^\dagger\ket{\bm{k}}\right\rbrace\nonumber\\
    &= 2k_s^2\, \mathbb{E}\left\lbrace\bra{\bm{k}}\hat B^\dagger \hat B\ket{\bm{k}}\right\rbrace\ .
\end{eqnarray}
We can think of the operators $\hat B$ and $\hat A_1$ as living on different tensor factors of the shell Hilbert space, \ie we can understand them as tensor products of the form $\mathbb{I}_1\otimes_{JW} \hat B$ and $\hat A_1 \otimes_{JW}\mathbb{I}_{2-n}\,$, where the subscript $JW$ in $\otimes_{JW}$ signifies that appropriate Jordan-Wigner strings have to be added to the naive tensor product in order to ensure anti-commutation instead of commutation (\cf our discussion in \ref{app:JordanWigner}). The state $\ket{\bm{k}}$ can then be thought of as a product state
\begin{eqnarray}
    \ket{\bm{k}} = \ket{1}_A \otimes_{JW} \ket{b_{\bm{k}}}
\end{eqnarray}
where $\ket{b_{\bm{k}}}$ only lives on the Hilbert space factor corresponding to $\hat B$ (and again the subscript $JW$ is a reminder that the naive tensor product structure has to be augmented by appropriate Jordan-Wigner strings). The scrambling time thus becomes
\begin{eqnarray}
\label{eq:T_scramble_app_1}
    \frac{1}{T_{\mathrm{scramble}}^2} &= 2k_s^2\, \mathbb{E}\left\lbrace\bra{b_{\bm{k}}}\hat B^\dagger \hat B\ket{b_{\bm{k}}}\right\rbrace\ .
\end{eqnarray}
Now in \secref{dynamics} we had defined $\ket{\bm{k}}$ such that it maximizes $T_{\mathrm{scramble}}$, while also having $\bra{\bm{k}} \hat N_{\bm{k}} \ket{\bm{k}} = 1$. So $\ket{b_{\bm{k}}}$ needs to be an eigenvector of $\hat B^\dagger \hat B$ corresponding to its minimum eigenvalue.

We can express $\hat B$ as
\begin{eqnarray}
    \hat B = \left(K_{1,2},\ \dots ,\ K_{1,n},\ K_{1,n+1},\ \dots\ ,\ K_{1,2n-1}\right) \pmatrix{\hat{A}_2 \cr \dots \cr \hat{A}_n \cr \hat{A}_n^\dagger \cr \dots \cr \hat{A}_2^\dagger} \equiv \bm{a}^\dagger \cdot \pmatrix{\hat{A}_2 \cr \dots \cr \hat{A}_n \cr \hat{A}_n^\dagger \cr \dots \cr \hat{A}_2^\dagger}\ ,\nonumber \\
\end{eqnarray}
where we have defined $\bm{a} \equiv (K_{1,2}^*,\ \dots ,\ K_{1,2n-1}^*)^T$. So we need to find the minimum eigenvalue of
\begin{eqnarray}
    \hat B^\dagger \hat B = \left(\hat{A}_2^\dagger,\ \dots\ ,\ \hat{A}_n^\dagger,\ \hat{A}_n,\ \dots\ ,\ \hat{A}_2\right) [\bm{a}\cdot \bm{a}^\dagger] \pmatrix{\hat{A}_2 \cr \dots \cr \hat{A}_n \cr \hat{A}_n^\dagger \cr \dots \cr \hat{A}_2^\dagger}\ ,
\end{eqnarray}
where the dyadic product $\bm{a}\cdot \bm{a}^\dagger$ is a $(2n_s-2)\times (2n_s-2)$-matrix. Similarly to how we proceeded in \ref{app:energy_spectrum}, we can use the anti-commutation relations between $\hat{A}_j, \hat{A}_k, \hat{A}_j^\dagger, \hat{A}_k^\dagger$ to symmetrize the above expression. This yields
\begin{eqnarray}
\hspace*{-1.5cm}\hat B^\dagger \hat B = \frac{|\bm{a}|^2}{2} + \left(\hat{A}_2^\dagger,\ \dots\ ,\ \hat{A}_n^\dagger,\ \hat{A}_n,\ \dots\ ,\ \hat{A}_1\right) \left[\frac{1}{2}\bm{a}\cdot \bm{a}^\dagger - \frac{1}{2} \bm{b}\cdot \bm{b}^\dagger\right] \pmatrix{\hat{A}_1 \cr \dots \cr \hat{A}_n \cr \hat{A}_n^\dagger \cr \dots \cr \hat{A}_2^\dagger}\ ,\nonumber \\
\end{eqnarray}
where the vector $\bm{b}$ is given by
\begin{eqnarray}
    \bm{b} \equiv \left(\bm{R}_{2n_s - 2} \cdot \bm{a}\right)^*\ ,
\end{eqnarray}
i.e.\ it is the complex conjugate of the vector obtained by reversing the order of $\bm{a}$. 

Note that the matrix $[\bm{a}\cdot \bm{a}^\dagger - \bm{b}\cdot \bm{b}^\dagger]$ is Hermitian and anti-persymmetric (\ie it changes sign, when flipped along the cross-diagonal; \cf \ref{app:persymmetric_matrices}). So similarly to what we did in \ref{app:energy_spectrum} we can calculate the eigenspectrum of $\hat B^\dagger \hat B$ by determining the eigenspectrum of $[\bm{a}\cdot \bm{a}^\dagger - \bm{b}\cdot \bm{b}^\dagger]$. The latter is a rank two matrix, and a standard result shows that its only non-trivial eigenvalues are
\begin{eqnarray}
    \lambda_\pm[\bm{a}\cdot \bm{a}^\dagger - \bm{b}\cdot \bm{b}^\dagger] &= \frac{|\bm{a}|^2 - |\bm{b}|^2}{2} \pm \frac{1}{2} \sqrt{(|\bm{a}|^2 + |\bm{b}|^2)^2 - 4|(\bm{b}^\dagger\cdot \bm{a})|^2}\nonumber \\
    &= \pm |\bm{a}|^2 \sqrt{1 - \frac{|(\bm{b}^\dagger\cdot \bm{a})|^2}{|\bm{a}|^4}}\ ,
\end{eqnarray}
where in the second line we used the fact that in our case $|\bm{a}|^2 = |\bm{b}|^2$. The plane wave life time is then given by
\begin{eqnarray}
\label{eq:T_scramble_app_2}
    \frac{1}{T_{\mathrm{scramble}}^2} &= 2k_s^2\, \mathbb{E}\left\lbrace \frac{|\bm{a}|^2}{2} - \frac{1}{2}  \lambda_-[\bm{a}\cdot \bm{a}^\dagger - \bm{b}\cdot \bm{b}^\dagger] \right\rbrace\nonumber \\
    &= k_s^2\, \mathbb{E}\left\lbrace |\bm{a}|^2 -  |\bm{a}|^2\sqrt{1 - \frac{|(\bm{b}^\dagger\cdot \bm{a})|^2}{|\bm{a}|^4}} \right\rbrace\ .
\end{eqnarray}
To evaluate this further, let us write the $(2n_s - 2)$-dimensional vector $\bm{a}$ as
\begin{eqnarray}
    \bm{a} \equiv \pmatrix{\bm{v}^\dagger \cr \bm{R}_{n_s - 1} \bm{w}}\ \Rightarrow\ \bm{b} = \pmatrix{\bm{w} \cr \bm{R}_{n_s - 1} \bm{v}^\dagger}\ ,
\end{eqnarray}
where $\bm{v}, \bm{w}$ are random vectors $\in \mathbb{C}^{n_s-1}$. The term appearing in the square root of \eqnref{T_scramble_app_2} then becomes
\begin{eqnarray}
    \frac{|(\bm{b}^\dagger\cdot \bm{a})|^2}{|\bm{a}|^4} &= \frac{4|(\bm{v}^\dagger\cdot \bm{w})|^2}{|\bm{a}|^4}\nonumber \\
    &= 4\frac{|(\bm{v}^\dagger\cdot \bm{w})|^2}{|\bm{v}|^2|\bm{w}|^2}\cdot\frac{|\bm{v}|^2}{|\bm{v}|^2 + |\bm{w}|^2}\cdot\frac{|\bm{w}|^2}{|\bm{v}|^2 + |\bm{w}|^2}\ .
\end{eqnarray}
Using e.g.\ Lemma 2.2 of \cite{Dasgupta2003} one can show that $|\bm{v}|^2 / (|\bm{v}|^2 + |\bm{w}|^2) \in \left[1/2 - 1/(2n_s -2)^{1/4}, 1/2 + 1/(2n_s -2)^{1/4}\right]$ with a probability $\geq 1 - \sqrt{2n_s - 2}/ 8$. This means that as $n_s \rightarrow \infty$ the value of $|\bm{v}|^2 / (|\bm{v}|^2 + |\bm{w}|^2)$ (and similarly, of $|\bm{w}|^2 / (|\bm{v}|^2 + |\bm{w}|^2)$) will be concentrated in an increasingly narrow interval around $1/2$ with probability increasingly closer to $1$. We can hence conclude that in the high-$n_s$ limit
\begin{eqnarray}
    \frac{|(\bm{b}^\dagger\cdot \bm{a})|^2}{|\bm{a}|^4} &\approx \frac{|(\bm{v}^\dagger\cdot \bm{w})|^2}{|\bm{v}|^2|\bm{w}|^2}\ .
\end{eqnarray}
Our results from \ref{app:real_angles} and \ref{app:complex_angles} will then indicate that $|(\bm{b}^\dagger\cdot \bm{a})|^2 / |\bm{a}|^4$ is $\sim 1/(n_s-1)$, with fluctuations around this only of the order $1/(n_s-1)^2$. So in the high-$n_s$ limit it will be warranted to expand the square root in \eqnref{T_scramble_app_2} at linear order, which yields
\begin{eqnarray}
\label{eq:T_scramble_app_3}
    \frac{1}{T_{\mathrm{scramble}}^2} &\approx  \frac{k_s^2}{2}\, \mathbb{E}\left\lbrace(|\bm{v}|^2 + |\bm{w}|^2)\frac{|(\bm{v}^\dagger\cdot \bm{w})|^2}{|\bm{v}|^2|\bm{w}|^2}\right\rbrace\nonumber \\
    &= \frac{k_s^2}{2} \mathbb{E}\left\lbrace |\bm{v}|^2 + |\bm{w}|^2\right\rbrace \mathbb{E}\left\lbrace\frac{|(\bm{v}^\dagger\cdot \bm{w})|^2}{|\bm{v}|^2|\bm{w}|^2}\right\rbrace\nonumber \\
    &= \frac{k_s^2}{2} \frac{N_s (n_s-1)}{n_s^2} \frac{1}{n_s-1}\ ,
\end{eqnarray}
where in the second line we used the fact that the angle between $\bm{v}$ and $\bm{w}$ will be uncorrelated to the absolute value of these vectors, and in the third line we used our results from \ref{app:complex_angles} as well as a modified version of \eqnref{row_variance}. Using our definitions from \secref{JL_embedding} we finally arrive at
\begin{eqnarray}
    T_{\mathrm{scramble}} \approx 2\pi^2 \sqrt{\alpha} \left(\frac{\Lambda_{\mathrm{Planck}}}{\Lambda_{\mathrm{UV}}}\right)^2 \frac{N_{\mathrm{dof},\psi}}{N_{\mathrm{total}}} \sqrt{\frac{L}{k_s}}\ ,
\end{eqnarray}
where the quantities appearing in this equation were
\begin{itemize}
    \item \underline{$k_s$:} central radius of the Fourier space shell $s$,
    \item \underline{$\alpha$:} relative width $\Delta_s/k_s$ of the Fourier space shell $s$,
    \item \underline{$\Lambda_{\mathrm{UV}}/\Lambda_{\mathrm{Planck}}$:} UV-cutoff of our field in Planck units,
    \item \underline{$N_{\mathrm{dof},\psi}$:} number of degrees of freedom (\say{qbits}) present in our field
    \item \underline{$N_{\mathrm{total}}$:} total number of degrees of freedom present in the Universe,
    \item \underline{$L$:} IR-cutoff.
\end{itemize}

\subsection{Eigenvalue structure of Hermitian and anti-persymmetric matrices}
\label{app:persymmetric_matrices}

Let $\bm{R}_n$ be the $n\times n$-matrix such that
\begin{eqnarray}
    (\bm{R}_n)_{ij} = \left\lbrace\matrix{1 & \mathrm{if}\ i+j = n+1 \cr 0 & \mathrm{else}}\right.\ ,
\end{eqnarray}
\ie it is the skewed unit matrix with ones on the cross-diagonal. We define the \emph{skew transposed} $\bm{M}^S$ of a matrix $\bm{M}$ as
\begin{eqnarray}
    \bm{M}^S \equiv \bm{R}_n \bm{M}^T \bm{R}_n\ ,
\end{eqnarray}
where ${}^T$ is regular transposition. We follow \cite{Reid1997} in calling $\bm{M}$ \emph{persymmetric} if $\bm{M}^S = \bm{M}$. And we will call it \emph{anti-persymmetric} if $\bm{M}^S = -\bm{M}$. In analogy to situations considered by \cite{Reid1997} we now prove the following statement. \\

\noindent \textbf{Theorem:} \emph{Let $\bm{M}$ be a Hermitian and anti-persymmetric matrix (\ie $\bm{M} = \bm{M}^\dagger$ and $\bm{M}^S = -\bm{M}$) of dimension $n\times n$. Then
\begin{itemize}
    \item[-] if $\lambda$ is an eigenvalue of $\bm{M}$ then $-\lambda$ is also an eigenvalue;
    \item[-] if $\bm{v}$ is an eigenvector to the eigenvalue $\lambda$, then $\bm{R}_{n}\bm{v}^*$, i.e.\ the complex conjugate of $\bm{v}$ multiplied by $\bm{R}_{n}$ (which reverses the ordering of the elements of $\bm{v}^*$), is an eigenvector to the eigenvalue $-\lambda$.
\end{itemize}
}

\noindent \textbf{Proof:}

\noindent Let $\bm{v}$ be an eigenvector of $\bm{M}$ with eigenvalue $\lambda$, i.e.
\begin{eqnarray}
    \bm{M}\bm{v} &=  \lambda\bm{v}\ .
\end{eqnarray}
We will now do three transformations of this equation:
\begin{itemize}
    \item[-] we will take its complex conjugate,
    \item[-] we will multiply it from the left with $\bm{R}_{n}$,
    \item[-] we will insert a $\bm{I} = \bm{R}_{n} \bm{R}_{n}$ .
\end{itemize}
This gives
\begin{eqnarray}
    \lambda \bm{R}_{n}\bm{v}^* &= \bm{R}_{n}\bm{M}^* \bm{R}_{n} \bm{R}_{n} \bm{v}^* \nonumber \\
    &= \left[ \bm{R}_{n}\bm{M}^T \bm{R}_{n} \right]^\dagger \bm{R}_{n} \bm{v}^* \nonumber \\
    &= \left[ \bm{M}^S \right]^\dagger \bm{R}_{n} \bm{v}^* \nonumber \\
    &= -\bm{M} \bm{R}_{n}\bm{v}^*
\end{eqnarray}
\begin{eqnarray}
    \Rightarrow \bm{M} (\bm{R}_{n}\bm{v}^*) &= -\lambda (\bm{R}_{n}\bm{v}^*)\ .\ \ \square
\end{eqnarray}

\noindent \textbf{Corollary:} \emph{For odd $n$ the Matrix $\bm{M}$ must have at least one eigenvalue $\lambda = 0$.
}

\noindent This corollary is however irrelevant for our work, since our results of \ref{app:Heisenberg} only involve matrices of even dimension.

\section{Jordan-Wigner strings to construct anti-commuting creation and annihilation operators}
\label{app:JordanWigner}

For each mode $\bm{p}$ in a narrow Fourier space shell $s$, the CRSV-embedding we described in \secref{JL_embedding} is generating creation and annihilation operators $\hat c_{\bm{p}}, \hat c_{\bm{p}}^\dagger$ (and $\hat d_{\bm{p}}, \hat d_{\bm{p}}^\dagger$ for anti-particles) whose anti-commutator deviates from the standard anti-commutators of the non-overlapping field by only some small value $\epsilon$ (with high probability). We then took different shells $s_1, s_2$ to be non-overlapping, i.e.\ operators $\hat c_{\bm{p}_1}, \hat c_{\bm{p}_1}^\dagger$ and $\hat c_{\bm{p}_2}, \hat c_{\bm{p}_2}^\dagger$ for modes $\bm{p}_1\in s_1$ and $\bm{p}_2\in s_2$ should be anti-commuting. If $\mathcal{H}_1, \mathcal{H}_2$ are the Hilbert spaces of the two shells, then their combined Hilbert space will be isomorphic to the tensor product
\begin{eqnarray}
    \mathcal{H}_{12} \simeq \mathcal{H}_1 \otimes \mathcal{H}_2\ .
\end{eqnarray}
However, we cannot simply take the mode operators in this combined space to be the tensor products of operators in the individual spaces such as
\begin{eqnarray}
    \hat c_{\bm{p}_1} \otimes \bm{1}\ \ ,\ \ \bm{1}\otimes \hat c_{\bm{p}_2}\ \ \mathrm{etc.}\nonumber
\end{eqnarray}
because such a construction would make the operators of different shells \emph{commute} instead of anti-commute. To embed the operators of the individual shells into the combined Hilbert space in a manner that makes them anti-commute we will have to incorporate so-called Jordan-Wigner strings. This is best done with the help of the non-overlapping creation and annihilation operators we introduced in \ref{app:Heisenberg},
\begin{eqnarray}
    \hat{A}_{s,j} \equiv \frac{1}{2}\left(\bm{C}_{s,2j} + i\bm{C}_{s,2j+1}\right)\ ,\ \hat{A}_{s,j}^\dagger \equiv \frac{1}{2}\left(\bm{C}_{s,2j} - i\bm{C}_{s,2j+1}\right)\ ,\nonumber
\end{eqnarray}
where $\bm{C}_{s,j}$ are the generators of the Clifford algebra we chose for each shell $s$. We can now sort the operators $\hat{A}_{s,j}$ from all shells $s$ into a single list of operators, e.g.\ in the form
\begin{eqnarray}
\label{eq:ordering_the_a_operators}
    \left( \hat a_1 ,\  \hat a_2,\  \hat a_3,\ \dots  \right)\ \equiv \left(\hat{A}_{s_1,1},\ \hat{A}_{s_1,2},\ \dots,\ \hat{A}_{s_2,1},\ \hat{A}_{s_2,2},\ \dots,\ \hat{A}_{s_3,1},\ \dots\right)\ ,\nonumber\\
\end{eqnarray}
where we understand the operators on the right-hand side to be embedded into the combined Hilbert space by means of the naive tensor products mentioned above. In particular, operators between different different shells are still only commuting at this point. We can now devise new, anti-commuting operators $\left( \hat b_1 ,\  \hat b_2,\  \hat b_3,\ \dots  \right)$ as follows: Let the index $j$ correspond to some shell $s$. We then define
\begin{eqnarray}
    \hat b_j \equiv \exp\left(-i\pi \sum_{s' < s} \sum_{j' \in s'} \hat a_{j'}^\dagger \hat a_{j'} \right)\hat a_j\ .
\end{eqnarray}
Here the criterion $s' < s$ means that we are summing over all shells $s'$ that come before the shell $s$ in the order of shells implied by \eqnref{ordering_the_a_operators}. The new operators $\left( \hat b_1 ,\  \hat b_2,\  \hat b_3,\ \dots  \right)$ as well as their Hermitian conjugates are now anti-commuting between the different shells. At the same time, within each shell the above transformation is just a constant unitary, i.e.\ the anti-commutation relations within one shell are unchanged. We can then apply the same transformation also to the overlapping mode operators $\hat c_{\bm{p}}, \hat c_{\bm{p}}^\dagger$, and since they can be expressed as linear combinations of the non-overlapping $\hat A_{i}, \hat A_{i}^\dagger$ they will also anti-commute between different shells. This construction can also be extended to incorporate the anti-particle Hilbert space in a straight forward manner. Note especially that none of our previous calculations are impacted by the above transformation and this appendix is only meant to close a conceptual gap in our presentation.

\end{document}

%% file: journaldef.tex
\def\aj{AJ}%
\def\araa{ARA\&A}%
\def\apj{ApJ}%
\def\apjl{ApJ}%
\def\apjs{ApJS}%
\def\ao{Appl.~Opt.}%
\def\apss{Ap\&SS}%
\def\aap{A\&A}%
\def\aapr{A\&A~Rev.}%
\def\aaps{A\&AS}%
\def\azh{AZh}%
\def\baas{BAAS}%
\def\jrasc{JRASC}%
\def\memras{MmRAS}%
\def\mnras{MNRAS}%
\def\pra{Phys.~Rev.~A}%
\def\prb{Phys.~Rev.~B}%
\def\prc{Phys.~Rev.~C}%
\def\prd{Phys.~Rev.~D}%
\def\pre{Phys.~Rev.~E}%
\def\prl{Phys.~Rev.~Lett.}%
\def\pasp{PASP}%
\def\pasj{PASJ}%
\def\qjras{QJRAS}%
\def\skytel{S\&T}%
\def\solphys{Sol.~Phys.}%
\def\sovast{Soviet~Ast.}%
\def\ssr{Space~Sci.~Rev.}%
\def\zap{ZAp}%
\def\nat{Nature}%
\def\iaucirc{IAU~Circ.}%
\def\aplett{Astrophys.~Lett.}%
\def\apspr{Astrophys.~Space~Phys.~Res.}%
\def\bain{Bull.~Astron.~Inst.~Netherlands}%
\def\fcp{Fund.~Cosmic~Phys.}%
\def\gca{Geochim.~Cosmochim.~Acta}%
\def\grl{Geophys.~Res.~Lett.}%
\def\jcap{JCAP}%
\def\jcp{J.~Chem.~Phys.}%
\def\jgr{J.~Geophys.~Res.}%
\def\jqsrt{J.~Quant.~Spec.~Radiat.~Transf.}%
\def\memsai{Mem.~Soc.~Astron.~Italiana}%
\def\nphysa{Nucl.~Phys.~A}%
\def\physrep{Phys.~Rep.}%
\def\physscr{Phys.~Scr}%
\def\planss{Planet.~Space~Sci.}%
\def\procspie{Proc.~SPIE}%